\documentclass[11pt]{article}
\usepackage[margin=1in]{geometry}
\usepackage{amsmath,amssymb}
\usepackage{booktabs,multirow,longtable,array}
\usepackage{graphicx}
\usepackage[hidelinks]{hyperref}
\usepackage[round,authoryear]{natbib}
\usepackage{xcolor}
\newcommand{\stcmd}[1]{\texttt{#1}}
\newcommand{\E}{\mathrm{E}}
\newcommand{\pr}{\Pr}
\title{Ratio-of-mediator-probability weighting for causal mediation analysis: estimation, extensions, and practical guidance}
\author{
Ariel Linden, DrPH \\
University of California, San Francisco \\
Department of Medicine \\
Division of Clinical Informatics \& Digital Transformation (DoC-IT) \\
ariel.linden@ucsf.edu
}
\date{}

\begin{document}
\maketitle

\begin{abstract}

\textbf{Background:} Ratio-of-mediator-probability weighting (RMPW) estimates natural direct and indirect effects by reweighting mediator distributions across treatment conditions rather than modeling the outcome, thereby accommodating treatment-by-mediator interaction. RMPW has evolved into a broad family of methods, but the available approaches and evidence about their practical performance are scattered across the literature.

\textbf{Methods:} Existing RMPW methods are brought together within a common estimation framework that includes a heteroskedastic normal model for continuous mediators and extends two-step variance estimation to multiple parametric weight components. The framework is illustrated using a national health survey to decompose disparities in cigarette smoking by sexual-minority status and a randomized welfare-to-work experiment to examine pathways from program assignment through employment to maternal depressive symptoms. Targeted simulations evaluate model specification, weighting strategies, treatment-selection adjustment, weight stability, variance estimation, and continuous treatments.

\textbf{Results:} Parametric RMPW performed well when models were adequately specified and weights were stable, but correct specification alone did not prevent unstable estimation from concentrated weights. The empirical applications similarly illustrate the importance of weight construction and stability. Stratification-based weighting provided some robustness to model misspecification. Standard errors that treated estimated weights as fixed were unreliable, while two-step and bootstrap inference performed well when used under appropriate conditions.

\textbf{Conclusions:} Successful application of RMPW requires attention not only to model specification but also to weight stability and inference. The empirical illustrations and simulation evidence provide practical guidance for selecting weighting strategies, diagnosing instability, and conducting inference. The methods are implemented in the community-contributed Stata command \texttt{mmwsmed}.
\end{abstract}

\noindent\textbf{Keywords:} causal mediation; ratio-of-mediator-probability weighting; natural direct and indirect effects; propensity score; inverse probability of treatment weighting; simulation

\section{Introduction}

Evaluations of health care interventions typically focus on whether an intervention changes an outcome, but an estimate of the overall treatment effect does not explain how that effect was
produced. Many interventions are explicitly designed to operate through intermediate processes, for example, by changing patient behavior, treatment adherence, engagement with care, or other modifiable mechanisms that subsequently affect health outcomes. Mediation analysis examines these causal pathways by decomposing the overall effect of a treatment into components operating through a mediator of interest and through alternative pathways.
Understanding these mechanisms can help confirm that an effective intervention operates as intended and, when an intervention fails, identify where the hypothesized causal pathway breaks down and how the intervention might be improved
\citep{lindenkarlson2013,schuler2025}.

Methods for conducting mediation analysis have expanded substantially and can be viewed as arising from three broad analytic conventions. Parametric regression and structural equation models estimate mediated effects through models for the mediator and outcome, commonly using products or differences of regression coefficients
\citep{baronkenny1986,mackinnon2007,vanderweele2015}. The potential-outcomes framework instead defines direct and indirect effects as contrasts between counterfactual outcomes and makes
explicit the assumptions required for their causal identification
\citep{robinsgreenland1992,pearl2001,imai2010}. Within the potential-outcomes framework itself, a large collection of estimators has developed from combinations of weighting,
model-based prediction, and hybrid strategies, with different modeling requirements and robustness properties \citep{nguyen2023}. More recently, flexible nonparametric and
machine-learning approaches have provided alternatives that reduce reliance on prespecified functional forms, permit data-adaptive estimation of nuisance functions, and can accommodate
complex nonlinearities and interactions \citep{lindenyarnold2018,farbmacher2022,hejazi2022}. Thus, the
proliferation of mediation methods reflects not only alternative estimation strategies but also differences in the causal structures, identifying assumptions, and scientific questions they are designed to address.

Within the important setting in which natural direct and indirect effects are identified, \citet{hong2010} and \citet{hongdeutschhill2015} developed ratio-of-mediator-probability
weighting (RMPW). The premise is to estimate counterfactual potential-outcome means by reweighting the observed mediator distribution under one treatment condition to resemble the
distribution it would have had under another. For example, the counterfactual outcome that treated individuals would have had if their mediator had followed its distribution under control
is estimated by weighting treated individuals by the ratio of the mediator's probability under control to its probability under treatment. Because the outcome itself is not modeled, RMPW
accommodates treatment-by-mediator interaction without requiring that interaction to be specified in an outcome model and can be used with different types of outcomes \citep{hong2015book}. Hong subsequently developed a broad framework encompassing
parametric and stratification-based nonparametric mediator weights, multivalued and nonrandom treatments, moderated mediation, and multiple mediators \citep{hong2015book}. Other work has extended the available estimation tools: \citet{bein2018} developed two-step variance estimation that accounts for estimation of the weights; mediator-probability ratios can alternatively be expressed through treatment odds \citep{tchetgen2013,huber2014}; and generalized propensity-score weighting extends mediation weighting to continuous treatments
\citep{hsu2020}.

This increasingly broad framework gives investigators considerable flexibility, but it also creates consequential choices in applied analyses. Mediator weights may be constructed from a parametric model, from stratification, or through treatment odds; nonrandom treatment requires an additional treatment-selection model; continuous mediators require specification of a conditional density; and continuous treatments require local weighting and a bandwidth choice. Inference must additionally account for the fact that the weights themselves are estimated. These choices involve tradeoffs between model specification, robustness, and weight stability that are not apparent from asymptotic identification results alone. The simulations of \citet{hongdeutschhill2015} demonstrated good performance of RMPW in the original binary-treatment, binary-mediator setting and showed that parametric mediator weights can be sensitive to model misspecification. However, the RMPW framework has since come to encompass a much broader range of mediator, outcome, and treatment types and weighting strategies, for which finite-sample performance has received less systematic evaluation. This is particularly important for weighting estimators because a correctly specified model can nevertheless generate highly variable weights and unstable estimation
\citep{colehernan2008,austinstuart2015}.

This article makes three contributions. First, it brings these developments together within a unified estimation framework. For discrete treatments, the basic estimator and its extensions share
a single data-duplication estimator and differ primarily in how the component weights are constructed. Second, it incorporates a heteroskedastic normal mediator model that jointly estimates
the conditional mean and variance of a continuous mediator, rather than obtaining them from separately fitted conditional moments. It also extends the two-step variance estimator of \citet{bein2018} to propagate uncertainty from all parametric models used to construct mediator and treatment-selection weights. Third, targeted simulation experiments evaluate the finite-sample
consequences of the principal choices available within the framework. The experiments examine mediator-model misspecification, alternative weighting strategies, treatment confounding, weight instability, continuous treatments, and variance estimation. The resulting evidence is used to
develop practical guidance for choosing models and weights, diagnosing instability, selecting a variance estimator, and reporting an RMPW analysis. The methods are implemented in
\texttt{mmwsmed}, a community-contributed Stata command \citep{linden2026mmwsmed}.

Section~\ref{sec:methods} defines the causal estimands and identification assumptions and develops RMPW
estimation and its extensions. Section~\ref{sec:example} illustrates the framework using a national health survey
and a randomized welfare-to-work experiment; Section~\ref{sec:simulation} describes the targeted simulation
experiments, and Section~\ref{sec:results} reports their results; Section~\ref{sec:discussion} translates the findings into
practical guidance for investigators and discusses limitations; and Section~\ref{sec:conclusion} concludes.

\section{Methods}
\label{sec:methods}

\subsection{Causal estimands and identification}
\label{sec:estimands}

Let $T$ denote treatment, with levels $0,1,\ldots,L-1$, one of which, $c$, is the control condition; let $M$ denote the mediator, $Y$ the outcome, and $\mathbf{X}$ a vector of pretreatment covariates. For each individual, $M(t)$ denotes the potential value of the mediator under treatment level $t$, and $Y(t,m)$ the potential outcome under treatment $t$ and mediator
value $m$. The composite potential outcome $Y(t,M(t'))$ is the outcome under treatment $t$ when the mediator takes the value it would have taken under treatment $t'$
\citep{robinsgreenland1992,pearl2001,imai2010,vanderweele2015,nguyen2022}.
The causal effects considered here are linear contrasts of population potential-outcome means $E\{Y(a,M(b))\}$.

For a treatment level $t \ne c$, four potential-outcome means define the effects considered in this article. The means $E\{Y(t,M(t))\}$ and $E\{Y(c,M(c))\}$ are the mean outcomes under
treatment and control, respectively. The cross-world mean $E\{Y(t,M(c))\}$ is the mean outcome if treatment were set to $t$ but the mediator followed the distribution it would have
under control, and $E\{Y(c,M(t))\}$ is defined analogously with treatment set to control and the mediator following its distribution under treatment. Table~\ref{tab:effects} summarizes the resulting effects.

For a treatment level $t \ne c$ compared with the control condition, the total effect is
\[
\mathrm{TE}=E\{Y(t,M(t))\}-E\{Y(c,M(c))\}.
\]
The natural indirect effect and natural direct effect are
\[
\mathrm{NIE}=E\{Y(t,M(t))\}-E\{Y(t,M(c))\},
\]
and
\[
\mathrm{NDE}=E\{Y(t,M(c))\}-E\{Y(c,M(c))\},
\]
so that $\mathrm{TE}=\mathrm{NIE}+\mathrm{NDE}$. The alternative decomposition uses the pure natural indirect effect
\[
\mathrm{PNIE}=E\{Y(c,M(t))\}-E\{Y(c,M(c))\},
\]
and the total natural direct effect
\[
\mathrm{TNDE}=E\{Y(t,M(t))\}-E\{Y(c,M(t))\},
\]
so that $\mathrm{TE}=\mathrm{PNIE}+\mathrm{TNDE}$. The natural treatment-by-mediator
interaction is
\[
\mathrm{INT}=\mathrm{NIE}-\mathrm{PNIE}
            =\mathrm{TNDE}-\mathrm{NDE},
\]
which measures the extent to which the indirect effect depends on treatment, or equivalently the direct effect depends on the mediator distribution
\citep{hongdeutschhill2015,hong2015book,vanderweele2014fourway}. These alternative decompositions are useful when treatment and mediator interact in their effects on the outcome, because the natural indirect and direct effects then depend on the treatment level or mediator distribution at which they are defined.

For a multivalued treatment, these effects are defined separately for every treatment level $t$ relative to the control level $c$. For a continuous treatment, $t$ and $c$ are replaced by the
evaluation dose $d_t$ and control dose $d_0$. Extensions of RMPW to multiple mediators, moderated mediation, and controlled direct effects are described in the Supplementary Material; their finite-sample performance is not evaluated here.

Identification of natural direct and indirect effects requires assumptions linking these counterfactual quantities to the observed data \citep{robinsgreenland1992,imai2010,vanderweele2015,nguyen2022}.
Conditional on pretreatment covariates $\mathbf{X}$, these require that (i) treatment assignment is independent of the potential mediator and outcomes; (ii) the mediator is independent of the
potential outcomes conditional on treatment; and (iii) no posttreatment variable confounds the mediator--outcome relationship. The last condition is particularly important because natural direct and indirect effects are generally not identified
when a variable affected by treatment also confounds the mediator--outcome relationship \citep{vanderweele2014confounder}.

Positivity requires a positive probability or density of the relevant treatment and mediator values at covariate patterns represented in the target population, and consistency requires the observed mediator and outcome to equal their corresponding potential values under the treatment actually received. When treatment is randomized, the treatment ignorability condition holds by design, but mediator assignment remains nonrandom;
$\mathbf{X}$ must therefore still contain pretreatment confounders of the mediator--outcome relationship \citep{imai2010,hong2015book,schuler2025}. When treatment is observational, $\mathbf{X}$ must additionally contain sufficient covariates to control treatment confounding.

\subsection{Ratio-of-mediator-probability weighting}
\label{sec:rmpw}

Under these assumptions, $E\{Y(t,M(t))\}$ and $E\{Y(c,M(c))\}$ are identified from the observed treatment groups, with additional treatment-selection weighting when treatment is not randomized. The cross-world mean $E\{Y(t,M(c))\}$ is estimated from individuals who received treatment $t$ using the ratio-of-mediator-probability weight \begin{equation}
\omega_{t\rightarrow c}
=
\frac{\Pr(M=m\mid T=c,\mathbf{X})}
     {\Pr(M=m\mid T=t,\mathbf{X})},
\label{eq:rmpw}
\end{equation}
where $m$ is the individual's observed mediator value
\citep{hong2010,hongdeutschhill2015,hong2015book}. For a continuous mediator, the probabilities in Equation~\ref{eq:rmpw} are replaced by conditional densities. The weight transforms the mediator distribution among individuals receiving treatment $t$ into the distribution it would have had under control, conditional on $\mathbf{X}$. The reverse weight $\omega_{c\rightarrow t}$ applied to the control group identifies $E\{Y(c,M(t))\}$, which is
required for the PNIE, TNDE, and INT.

The essential attraction of RMPW is that this reweighting leaves the observed outcome-generating process within each treatment group intact. Consequently, estimation of the natural effects does not require an outcome model and permits the relation between the mediator and outcome to differ by treatment
\citep{hongdeutschhill2015,hong2015book,nguyen2023}.

Following \citet{hong2015book}, estimation proceeds by data duplication. For each requested potential-outcome mean $E\{Y(a,M(b))\}$, the observations at treatment level $a$ are copied and assigned weight $\omega_{a\rightarrow b}$, equal to one when $a=b$, multiplied by any treatment-selection weight. The required copies are stacked, and a single weighted outcome
regression on indicators for the copies estimates the potential-outcome means jointly. The causal effects are linear contrasts of these means.

More generally, the weight assigned to individual $i$ in copy $c$ can be written as
\begin{equation}
W_{ic}=u_i\,\tau_i\prod_k\omega_{ick},
\label{eq:weight}
\end{equation}
where $u_i$ is a sampling weight, $\tau_i$ is a treatment-selection weight, and $\omega_{ick}$ is the mediator-probability or density ratio for mediator $k$ required by that
copy. When sampling weights are required for population inference, they should also be incorporated
when estimating the models used to construct the mediation weights \citep{coffman2024}.
Components that are not required for a particular analysis equal one. Thus, the discrete-treatment RMPW extensions considered here share the same data-duplication estimator and differ primarily in how the component weights are constructed; the continuous-treatment estimator of Section~\ref{sec:continuous} replaces these weights with kernel-localized generalized propensity-score weights. This common representation also provides the basis for the extension of two-step variance estimation described in Section~\ref{sec:variance}.

\subsection{Extensions and estimation choices}
The basic RMPW construction requires choices about how the mediator distribution is represented, how treatment selection is handled when treatment is not randomized, and how uncertainty in
the estimated weights is incorporated into inference. Continuous treatments require a related generalized propensity-score construction. Table~\ref{tab:contrib} summarizes the methodological basis for these components, the developments introduced here, and the aspects evaluated in the simulation experiments. These choices are described below independently of the simulation designs used later to evaluate them.

\subsubsection{Mediator weights}

\paragraph{Parametric ratio weights.}
For the RMPW estimator in Equation~\ref{eq:rmpw}, the conditional distribution of the mediator is modeled separately within each treatment level \citep{hong2010,hongdeutschhill2015,hong2015book}. The fitted models provide, for every observation, the probability or density of its observed mediator value under each treatment level, and ratios of these predictions form the mediator weights. The model therefore depends on the scale and distribution of the mediator: binary and ordinal mediators require probability models,
count mediators may be represented by an appropriate count distribution, and continuous mediators require a conditional density. With a multivalued treatment, a separate mediator model
is fitted within each treatment level.

For a continuous mediator, a normal linear model commonly assumes constant conditional variance within each treatment level. Hong's construction allows the variance to depend on covariates by estimating $E(M\mid\mathbf{X})$ and $E(M^2\mid\mathbf{X})$ separately and obtaining the conditional variance from
$E(M^2\mid\mathbf{X})-\{E(M\mid\mathbf{X})\}^2$
\citep[sec.~12.1.2]{hong2015book}. Because separately fitted mean equations need not imply a positive variance for every observation, this construction can yield negative fitted variances.

Here I instead use a heteroskedastic normal model in which the conditional mean and variance are estimated jointly:
\[
M\mid\mathbf{X},T=t
\sim
N\!\left\{\mathbf{x}'\boldsymbol{\beta}_t,
          \exp(\mathbf{z}'\boldsymbol{\gamma}_t)\right\},
\]
where $\mathbf{x}$ contains covariates of the conditional mean and $\mathbf{z}$ contains
covariates of the conditional variance. The mean and variance equations are estimated jointly by maximum likelihood; for treatment level $t$, the log-likelihood is
$\ell_t(\boldsymbol{\beta}_t,\boldsymbol{\gamma}_t)=-\tfrac12\sum_{i:T_i=t}w_i\{\log(2\pi)+\mathbf{z}_i'\boldsymbol{\gamma}_t+(M_i-\mathbf{x}_i'\boldsymbol{\beta}_t)^2/\exp(\mathbf{z}_i'\boldsymbol{\gamma}_t)\}$,
where $w_i$ is any sampling or treatment-selection weight. The exponential variance specification guarantees a positive fitted variance and permits the mediator density, and therefore the RMPW density ratio, to reflect treatment-specific and covariate-dependent heteroskedasticity.

\paragraph{Odds-ratio weights.}
An alternative avoids modeling the mediator distribution directly. By Bayes' rule, the mediator-probability ratio can be expressed through two treatment models, one conditional on the
mediator and one without it \citep{tchetgen2013,huber2014}:
\begin{equation}
\omega_{a\rightarrow b}
=
\frac{\Pr(T=b\mid M,\mathbf{C})\Pr(T=a\mid\mathbf{C})}
     {\Pr(T=a\mid M,\mathbf{C})\Pr(T=b\mid\mathbf{C})},
\label{eq:odds}
\end{equation}
where $\mathbf{C}$ denotes the mediator-model covariates. This representation can be useful when the mediator distribution is difficult to specify, particularly for nonstandard continuous or
count mediators. It does not eliminate modeling assumptions; rather, it transfers them from the mediator distribution to the treatment distribution conditional on the mediator and covariates
\citep{tchetgen2013,huber2014,nguyen2023}.

\paragraph{Stratification-based nonparametric ratio weights.}
The nonparametric RMPW estimator replaces fitted mediator probabilities with observed proportions within propensity-score strata \citep{hongdeutschhill2015}. Observations are first
stratified on the linear predictor of the mediator model under treatment and then, within those strata, on the corresponding predictor under control. Within each cross-stratum, the mediator
probability is estimated by its observed proportion at the relevant treatment level. This marginal mean weighting through stratification (MMWS) construction \citep{hong2012,linden2014}
reduces reliance on the functional form of the mediator model, at the cost of residual differences in mediator propensity within strata and increasingly sparse cross-strata as the number of strata
increases.

\subsubsection{Nonrandom treatment}

When treatment is not randomized, mediator weighting alone does not address confounding of treatment with the mediator or outcome. Hong therefore combines the mediator weight with a treatment-selection weight \citep[sec.~12.2.1]{hong2015book}. One approach uses the stabilized inverse-probability-of-treatment weight (IPTW)
\[
\frac{\Pr(T=z)}
     {\Pr(T=z\mid\mathbf{X})}
\]
for an individual observed at treatment level $z$. The treatment model may be binary or multivalued, as appropriate to the treatment. As with inverse-probability weighting more generally, performance depends on adequate treatment-model specification and sufficient overlap in the estimated treatment probabilities \citep{colehernan2008,austinstuart2015}.

As an alternative, MMWS treatment weights \citep{hong2012,linden2014} stratify observations
on the estimated treatment propensity and weight observations within each stratum to reproduce the marginal treatment distribution. IPTW therefore relies directly on the fitted propensity score, whereas MMWS replaces individual inverse-probability weights with stratum-specific weights. The latter can reduce the influence of extreme estimated propensities but may retain residual treatment confounding within strata \citep{hong2012,linden2014}.

\subsubsection{Continuous treatment}
\label{sec:continuous}

For a continuous treatment, exact treatment levels have probability zero and the discrete-treatment RMPW construction cannot be applied directly. I therefore use the generalized
propensity-score (GPS) weighting estimator of \citet{hsu2020}. Two conditional treatment densities are estimated: $f_1(d\mid\mathbf{X})$, conditional on pretreatment covariates, and
$f_2(d\mid M,\mathbf{X})$, additionally conditional on the mediator. Let $K_d(T)$ denote an Epanechnikov kernel in $(T-d)/h$. The potential-outcome means at evaluation dose $d_t$ and
control dose $d_0$ are estimated as normalized weighted means with weights
\[
E\{Y(d_t,M(d_t))\}:
\quad
\frac{K_{d_t}(T)}
     {f_1(d_t\mid\mathbf{X})},
\]
and
\[
E\{Y(d_t,M(d_0))\}:
\quad
\frac{K_{d_t}(T)f_2(d_0\mid M,\mathbf{X})}
     {f_2(d_t\mid M,\mathbf{X})f_1(d_0\mid\mathbf{X})},
\]
with symmetric expressions for $E\{Y(d_0,M(d_0))\}$ and
$E\{Y(d_0,M(d_t))\}$. To limit the influence of extreme weights, observations whose normalized weight $w_i/\sum_j w_j$ exceeds a threshold (0.1 in all analyses here) in any of the potential-outcome means are removed from all means, and the means are then computed from the remaining observations with renormalized weights.
The bandwidth $h$ determines how strongly observations are localized around the evaluation dose: smaller bandwidths make the comparison more local but reduce the effective information contributing to each weighted mean \citep{hsu2020}. Inference for this estimator is obtained by bootstrap.

\subsubsection{Variance estimation}
\label{sec:variance}

RMPW and the related weighting estimators considered here are multistep estimators: models used to construct mediator or treatment-selection weights are estimated before the final weighted
outcome model. A conventional robust variance estimator applied only to the final outcome model treats the estimated weights as fixed and therefore does not propagate sampling uncertainty from
the models used to construct them \citep{bein2018}.

\citet{bein2018} developed a two-step variance estimator for RMPW that combines the estimating equations for the weight model and weighted outcome model and obtains an M-estimation sandwich
variance. I extend this construction to propagate uncertainty from all parametric models used to construct mediator and treatment-selection weights by stacking their estimating equations with
those of the weighted outcome model. The resulting variance estimator accommodates ratio and odds-ratio mediator weights, including Poisson and heteroskedastic normal mediator models; binary and multivalued treatment-selection models; and the final weighted outcome model. This permits the same two-step framework to be used when both mediator and treatment-selection weights are
estimated. The stacked estimating equations, sandwich variance, and delta-method transformation to the causal effects are given in Supplementary Section~S2.

The derivative-based two-step estimator does not cover stratification-based RMPW or MMWS weights, which are step functions of estimated propensity scores. Bootstrap inference provides a general alternative for these estimators and can also be used with parametric weights when finite-sample performance of the sandwich approximation is a concern. All models, strata, and
weights are re-estimated within each bootstrap resample.

\subsubsection{Weight diagnostics}
\label{sec:diagnostics}

Estimated treatment or mediator probabilities near the boundaries can cause a small number of
observations to receive a large share of the total weight \citep{colehernan2008,austinstuart2015}.
I therefore consider descriptive diagnostics of two kinds.
The first concerns concentration of the weights. Within each weighted copy, I examine the largest
observation's share of the total weight and Kish's effective sample size,
\[
\mathrm{ESS}
=
\frac{\left(\sum_i w_i\right)^2}
     {\sum_i w_i^2},
\]
\citep{kish1965}, expressed as a percentage of the number of observations in the copy. For an
analysis involving several weighted copies, the smallest ESS across copies provides a summary of
the least stable component of the estimator.

The second concerns whether the weights accomplish what they are intended to do. Each
cross-world copy is weighted to reproduce the mediator distribution of the other treatment group,
so the weighted distribution of the mediator in that copy---the proportion at each level of a
categorical mediator, or the mean of a continuous or count mediator---can be compared with the
corresponding distribution in the targeted treatment group, weighted by any treatment-selection
weights \citep[p.~290]{hong2015book}. Substantial discrepancies indicate a misspecified mediator
model (or, for odds-ratio weights, treatment model) or limited overlap. Covariate balance after
weighting can be assessed in the same spirit \citep{hong2015book,austinstuart2015}. All of these
quantities are used as warning indicators rather than formal criteria for estimator validity, and
satisfactory diagnostics do not establish the identification assumptions.

\section{Empirical illustrations}
\label{sec:example}

Two applications illustrate the framework. The first, based on a national health survey, combines
an observational exposure, treatment-selection weighting, survey sampling weights, and
alternative constructions of the mediator weights, and shows how weight diagnostics and
inference operate together in a realistic analysis. The second, a randomized welfare-to-work
experiment, complements it by illustrating alternative representations of the mediator and a
continuous outcome.

\subsection{National Survey on Drug Use and Health}
\label{sec:nsduh}

Sexual-minority adults experience disparities in cigarette smoking and other substance-use
outcomes relative to heterosexual adults, with particularly pronounced disparities among women
\citep{schuler2018}. Earlier smoking initiation has been identified as one pathway contributing
to these disparities in adulthood \citep{schulercollins2019}. I use the publicly available National
Survey on Drug Use and Health (NSDUH) data analyzed by \citet{coffman2024} and distributed
with the \texttt{twangMediation} package for R \citep{mccaffrey2026}, comprising 40,293 adult
women, of whom 4,130 identified as lesbian, gay, or bisexual (LGB). The exposure
is sexual-minority status (LGB vs.\ heterosexual), the mediator is smoking initiation before age
15, and the outcome is current (past-month) cigarette smoking.

Sexual-minority status is not an assignable treatment, so the natural effects are interpreted here
as a decomposition of the covariate-adjusted smoking disparity rather than as effects of a
manipulable intervention. Under the identification assumptions of Section~\ref{sec:estimands}, the NIE represents
the component of the disparity associated with differences in early smoking initiation, whereas
the NDE represents the disparity remaining after equalizing its distribution.

Because the exposure is observational, stabilized IPTW weights were estimated from a logit
model containing age, race and ethnicity, education, household income, and employment status.
The mediator models used the same covariates. Analyses were conducted both without and with
the NSDUH sampling weights; when used, the sampling weights entered all fitted models and the
potential-outcome means. Parametric RMPW with logit mediator models was compared with
odds-ratio weights and stratification-based nonparametric RMPW, and robustness to a probit
treatment model and restriction to common support was examined (Table~\ref{tab:nsduh}).

Incorporating the NSDUH sampling weights reduced the estimated smoking disparity and both
of its components while substantially increasing weight concentration. Parametric and
nonparametric RMPW nevertheless produced very similar decompositions both with and without
sampling weights. Odds-ratio weights agreed closely with these estimators without sampling
weights but assigned somewhat more of the weighted disparity to the indirect pathway. The mediator distribution check (Section~\ref{sec:diagnostics}) is informative here. With
sampling weights, parametric and nonparametric RMPW reproduced the target distributions more
closely than odds-ratio weighting, favoring the mediator-probability weights in this application. The estimated treatment-by-mediator
interaction was small throughout. Results were essentially unchanged when the treatment model
was changed from logit to probit or the analysis was restricted to common support.

For parametric RMPW, two-step, PSU-clustered two-step, and PSU-cluster bootstrap standard
errors were similar. For nonparametric RMPW, however, treating the step-function weights as
known understated uncertainty relative to the bootstrap, consistent with the simulation results
in Section~\ref{sec:simvar}. The analyses incorporate the NSDUH sampling weights and PSU
clustering but not survey strata and therefore do not constitute full design-based survey
inference. Overall, the example illustrates that population weighting can affect both the estimated
decomposition and weight stability, that diagnostics can help distinguish among alternative
mediator-weight constructions, and that inference should reflect how the weights are estimated.

\subsection{Riverside welfare-to-work experiment}
\label{sec:riverside}

The second illustration complements the first: treatment was randomized, the outcome is
continuous, and the mediator can be represented in several ways. It uses data from the Riverside
site of the National Evaluation of Welfare-to-Work Strategies (NEWWS), analyzed by
\citet{hongdeutschhill2015} and \citet[ch.~11]{hong2015book} and distributed with the
\texttt{rmpw} package for R \citep{rmpwR}. Mothers with at least one child aged 3--5 years were
randomly assigned to a labor force attachment (LFA) program emphasizing rapid entry into
employment ($n=208$) or to a control condition ($n=486$). The outcome is maternal depressive
symptoms two years after randomization, and the mediator is employment during those two years.
LFA increased employment but had no statistically significant overall effect on depressive
symptoms. \citet{hongdeutschhill2015} proposed that this overall effect could conceal opposing
pathways---program-induced employment might reduce depressive symptoms, while other
features of the program, such as the stress associated with the threat of sanctions, could operate
in the opposite direction---and that the relationship between employment and depression might
differ by treatment assignment. Employment can be represented as a binary indicator, a
three-category measure, or the number of quarters employed (0--8). Mediator models adjusted for
nine baseline covariates (prior employment, four measures of depressive symptoms, two barriers
to work, Hispanic ethnicity, and marital status), which were also included in the outcome model
for precision. Because treatment was randomized, no treatment-selection weights were required
(Table~\ref{tab:example}).

With employment represented as a binary mediator, parametric RMPW indicated opposing
pathways: LFA increased employment (65\% vs.\ 40\% employed), the indirect effect through
employment was beneficial (NIE $=-0.87$), the direct effect was detrimental (NDE $=1.28$), and
the natural treatment-by-mediator interaction was negative (INT $=-1.19$), consistent with the
hypothesis that employment was more beneficial under LFA than under the control condition.
Representing employment by three ordered categories, modeled by ordered or multinomial logit,
gave nearly identical estimates. Bootstrap standard errors of the NIE and the interaction were 25\%--30\% larger than
two-step standard errors for the binary mediator, and with bootstrap inference the interaction was no
longer statistically significant. The number of quarters employed is bounded, with 53\% of
observations equal to zero, making normal and Poisson mediator models unattractive. Odds-ratio
weights avoided modeling this distribution directly and gave a smaller NIE ($-0.62$) and
interaction ($-0.38$), with stable weights and similar two-step and bootstrap standard errors.
This illustrates both the attraction and the tradeoff of odds-ratio weighting: it avoids specifying a
difficult mediator distribution but shifts the modeling assumptions to the treatment model
conditional on the mediator and covariates.

\section{Simulation evaluation}
\label{sec:simulation}

The empirical illustrations demonstrate several choices that arise in applying RMPW, including
how treatment selection is addressed, how the mediator is represented and modeled, how mediator
weights are constructed, and how weight stability and uncertainty are assessed. Targeted simulation
experiments were therefore used to evaluate the finite-sample consequences of these and related
choices under conditions in which the true causal effects are known. The experiments broaden the
settings evaluated by \citet{hongdeutschhill2015} beyond the original binary-treatment, binary-mediator
design and examine mediator-model specification, alternative weighting strategies, treatment-selection
adjustment, weight instability, variance estimation, and continuous-treatment weighting. The
simulations were designed and reported following the ADEMP framework---aims, data-generating
mechanisms, estimands, methods, and performance measures---of \citet{morris2019}. Complete
data-generating mechanisms, true estimand values, models, sample sizes, and methods for all designs
are given in Supplementary Table~\ref{tab:simdesign}.

\subsection{Simulation designs}

The simulations begin by anchoring the evaluation to \citet{hongdeutschhill2015}. A replication
experiment reproduces their three randomized binary-treatment, binary-mediator scenarios. A
misspecification experiment reconstructs their mediator-model misspecification analysis by adding
either a nonlinear squared term or a nonadditive product term to the mediator model and omitting
that term from the fitted model. Because the original article did not report the covariate
coefficients or complete misspecification data-generating mechanisms, these components were
reconstructed to preserve the reported marginal mediator probabilities and true effects. These
experiments compare parametric RMPW with correctly and incorrectly specified mediator models
and stratification-based nonparametric RMPW.

A second set of designs examines whether the weighting framework continues to perform well
beyond the original binary-mediator, continuous-outcome setting. The designs include continuous
and count mediators, binary and count outcomes, and a three-level treatment. A
continuous-mediator design also allows covariate effects and residual variance to differ between
treatment groups, providing a setting in which the parametric density-ratio model is correctly
specified but a simple treatment model linear in the mediator, as used for the basic odds-ratio
weight, is misspecified.

Nonrandom treatment assignment is examined in designs that compare adjustment for mediator
selection alone with IPTW and MMWS treatment-selection weights and examine omission of a
treatment--outcome confounder from the treatment model. A further design uses an ordinal
three-level treatment and compares alternative multivalued treatment models.

Two sets of designs examine the relation between model specification and weight stability. In
the first, treatment is generated from logit, probit, and heteroskedastic probit models, and each
model is fitted. In the second, a continuous mediator has a conditional variance that depends on
treatment and covariates; increasing degrees of heteroskedasticity are used to compare the
conventional constant-variance density-ratio model, the heteroskedastic normal model described
above, and odds-ratio weights. Because increasing heteroskedasticity can also make the mediator
densities under the two treatment groups increasingly dissimilar, this design provides a direct
assessment of the relation between model specification, weight concentration, and estimator
performance.

Inference is examined by reusing representative binary-mediator and three-level-treatment designs
with nonrandom treatment at $N=800$ to compare two-step and bootstrap inference for IPTW and
bootstrap inference for MMWS, nonparametric RMPW, and odds-ratio mediator weights. Additional
two-step-versus-bootstrap comparisons are made in the heteroskedastic continuous-mediator
design.

Finally, the GPS estimator of \citet{hsu2020} is evaluated for a continuous treatment. A linear
Gaussian treatment model is evaluated at two treatment contrasts, and the default bandwidth,
$\mathrm{sd}(T)\,2.34/N^{1/4}$, is compared with one-half of that value to examine the
bias--variance consequences of more localized kernel weighting.

Across the simulations, the principal estimands are the NIE and NDE for each treatment level
relative to control: level 1 versus 0 for binary treatments, levels 1 and 2 versus 0 for three-level
treatments, and doses 2 and 0 versus 1 for the continuous treatment. True values were obtained
analytically when possible and otherwise by Gauss--Hermite integration over the covariate
distribution. Unless otherwise stated, mediator and treatment models include the covariates as
linear additive terms. An unadjusted estimator obtained from a mediator model without
covariates provides the reference for the proportion of bias removed in the randomized-treatment
and nonrandom-treatment designs.

\subsection{Performance measures}

For each estimand, performance is summarized by bias, empirical standard error, mean model
standard error, coverage of nominal 95\% confidence intervals, root mean squared error, and the
proportion of replications that fail to produce an estimate. To facilitate comparison with
\citet{hongdeutschhill2015}, the randomized- and nonrandom-treatment designs also report the proportion of the naive estimator's
absolute bias removed,
\[
1-\frac{|\mathrm{bias}|}{|\mathrm{bias}_{\mathrm{naive}}|}.
\]
The accuracy of model-based inference is assessed by the relative error of the mean estimated
standard error,
\[
\frac{\overline{\widehat{\mathrm{SE}}}}{\mathrm{EmpSE}}-1,
\]
where $\mathrm{EmpSE}$ is the empirical standard error of the estimates across replications.
A value of zero indicates agreement with the empirical sampling variability, whereas negative
and positive values indicate underestimation and overestimation, respectively. For designs specifically examining weight instability,
the smallest Kish ESS across weighted copies, the largest observation's share of a copy's total
weight, and the proportion of replications in which at least one copy has ESS below 5\% are also
recorded.

Each design is evaluated at $N=800$ and $N=5{,}000$, except the bootstrap-inference designs, which use $N=800$.
The simulations use 1,000 replications except for the two-step-versus-bootstrap comparisons, which use 500. Bootstrap standard errors are based on 200 resamples.
Monte Carlo standard errors are computed for all performance measures following
\citet{morris2019}, with the largest value reported in each table's notes. With 1,000 replications,
the Monte Carlo standard error of 95\% coverage is approximately 0.007 and that of estimated
bias is 3.2\% of the empirical standard error; with 500 replications, the corresponding values are
0.010 and 4.5\%.

All simulation analyses were performed with the community-contributed package \stcmd{mmwsmed}
\citep{linden2026mmwsmed} in Stata version 19.

\section{Simulation results}
\label{sec:results}

\subsection{Mediator-model specification and alternative weights}

The replication of \citet{hongdeutschhill2015} confirmed the performance of RMPW in the
original binary-treatment, binary-mediator setting (Supplementary Table~\ref{tab:study1}). The
naive estimator was biased in every scenario, with NIE coverage of at most 0.18. Parametric
RMPW removed 98\%--99\% of the bias, while stratification-based nonparametric RMPW
removed 81\%--84\% with $3\times3$ strata and 78\%--90\% with $4\times4$ strata. These
results closely reproduced the proportions reported by \citet{hongdeutschhill2015}
(Supplementary Table~\ref{tab:hongcompare}). At $N=800$, increasing from $3\times3$ to
$4\times4$ strata did not consistently reduce bias. Coverage for the nonparametric estimators,
whose default standard errors treat the weights as known, was below nominal in several
settings.

Mediator-model misspecification produced a clear difference between parametric and
stratification-based weighting (Supplementary Table~\ref{tab:study1b}). When a nonlinear or
nonadditive term was omitted from the mediator model, the proportion of NIE bias removed by
parametric RMPW declined as the strength of the omitted term increased. In the nonlinear
designs, bias removal declined from 51\%--53\% at $\delta=0.25$ to about 20\% at
$\delta=1$; in the nonadditive designs, it declined from 79\%--81\% to 61\%--65\%.
At $N=5{,}000$, NIE coverage was as low as 0.41. Nonparametric RMPW removed
65\%--88\% of the NIE bias across these designs and had the smallest NIE root mean squared
error in every design and at both sample sizes, although its NIE coverage remained below
nominal (0.65--0.85). Correctly specifying the parametric mediator model reduced bias but
could substantially increase sampling variability; in the most extreme designs, its empirical
standard error was as much as 2.8 times that of the misspecified model.

Across different mediator, outcome, and treatment types, parametric RMPW was essentially
unbiased when its mediator model was adequately specified (Supplementary Table~\ref{tab:study2}).
Absolute bias was at most 0.006 for continuous and count mediators, binary and count outcomes,
and a three-level treatment. Odds-ratio weights performed similarly when a continuous mediator
had common covariate effects and residual variance across treatment groups. When these features
differed by treatment group, however, odds-ratio weights removed only 44\%--46\% of the bias,
leaving bias of approximately 0.06 for both the NIE and NDE at both sample sizes.
Stratification-based nonparametric RMPW removed 82\%--97\% of the bias in the designs in
which it was evaluated.

\subsection{Treatment-selection adjustment}

With nonrandom treatment, mediator weighting alone was insufficient to recover both natural
effects (Supplementary Table~\ref{tab:study3a}). Adjusting only for mediator selection removed
80\%--100\% of the NIE bias but left NDE bias of 0.11--0.55, exceeding the bias of the
naive estimator in every design. Adding IPTW treatment-selection weights reduced the absolute
NDE bias to at most 0.009. MMWS treatment weights retained some NDE bias, ranging from
0.012 to 0.058 with five strata and from 0.005 to 0.029 with ten strata. When a
treatment--outcome confounder was omitted from the treatment model, absolute NDE bias ranged
from 0.015 to 0.087, whereas NIE bias remained at most 0.007.

For an ordinal three-level treatment, IPTW based on multinomial logit, ordered logit, and
ordered probit treatment models was essentially unbiased, with absolute bias at most 0.013 and
coverage of 0.930--0.956 (Supplementary Table~\ref{tab:study3c}). Odds-ratio mediator weights, which rely on a model for treatment conditional on the mediator,
had NDE bias as large as $-0.036$ and coverage as low as 0.859.

\subsection{Correct specification does not ensure stable weights}

Correct model specification did not ensure stable weighting or good finite-sample performance.
When treatment was generated by conventional logit or probit models, the alternative treatment
models performed similarly (Supplementary Table~\ref{tab:study3b}). Under strong
heteroskedasticity ($g=0.50$), however, the correctly specified heteroskedastic probit model
performed worse than the simpler logit model, with greater bias, sampling variability, and
undercoverage. This deterioration coincided with greater weight concentration: at $N=5{,}000$,
the mean smallest effective sample size was 19.2\% for the heteroskedastic probit model versus
approximately 30\% for the logit model, and very low effective sample sizes became more
frequent. Both the estimation problems and weight concentration diminished as heteroskedasticity
weakened.

The same pattern was more pronounced with a heteroskedastic continuous mediator (Supplementary Table~\ref{tab:study4}). Assuming constant variance produced increasing bias and severe
undercoverage as heteroskedasticity increased. Modeling the conditional variance substantially
reduced the bias but did not prevent large sampling variability and poor coverage under strong
heteroskedasticity; odds-ratio weights also deteriorated as heteroskedasticity increased. These
problems again coincided with increasingly concentrated weights: at $c=0.40$ and
$N=5{,}000$, the mean smallest effective sample size was 15.5\% with the heteroskedastic
mediator model and 9.6\% with odds-ratio weights. In both experiments, deterioration in estimator
performance coincided with deterioration in the weight diagnostics.

\subsection{Variance estimation}
\label{sec:simvar}

Standard errors that treated estimated weights as fixed were unreliable in both directions
(Supplementary Tables~\ref{tab:study1} and \ref{tab:study1b}--\ref{tab:study3a}). They understated sampling
variability for several mediator-weighting estimators but overstated it for some treatment-selection
estimators. Two-step standard errors generally performed well in large samples when the weights
were stable, but could substantially understate sampling variability when the weights were highly
variable, including in the mediator-model misspecification and heteroskedastic continuous-mediator
designs (Supplementary Tables~\ref{tab:study1b} and \ref{tab:study4boot}).
In the nonrandom-treatment designs, where both mediator and treatment-selection weights were
estimated, the generalized two-step standard errors produced coverage close to the nominal level
when the weights were stable (Supplementary Table~\ref{tab:study3a}).

Bootstrap inference generally more closely reproduced empirical sampling variability at
$N=800$ (Supplementary Table~\ref{tab:study6}). Coverage was near nominal for IPTW, MMWS,
and nonparametric RMPW; poorer coverage for some odds-ratio estimates coincided with
appreciable estimator bias. Bootstrap inference also improved coverage relative to two-step
inference in the heteroskedastic continuous-mediator design at $N=800$.

Bootstrap inference did not, however, remedy severe weight instability. At $N=5{,}000$ under
strong mediator heteroskedasticity, both two-step and bootstrap standard errors substantially
underestimated the empirical sampling variability
(Supplementary Table~\ref{tab:study4boot}). Estimation failures were uncommon overall:
failures occurred in only 17 simulation cells, with at most 11 of 1,000 replications failing in
any cell, and bootstrap fitting failures were rare.

\subsection{Continuous-treatment weighting}

GPS weighting with the default bandwidth produced little bias and coverage near the nominal
level in the continuous-treatment design (Supplementary Table~\ref{tab:study5}). Halving the
bandwidth increased sampling variability and, at $N=800$, produced appreciable NDE bias and
lower coverage.

\section{Discussion}
\label{sec:discussion}

The empirical illustrations and simulation experiments together highlight three distinct
considerations in applying RMPW: whether the causal effect is identified, whether the models
used to construct the weights are sufficiently well specified and produce stable weights, and
whether inference adequately accounts for estimation of those weights. These considerations are
related but not interchangeable. No weighting strategy can overcome failure of the identification
assumptions; correct specification of a weight model does not guarantee good finite-sample
performance when the resulting weights are highly concentrated; and a better variance estimator
cannot repair an unstable point estimator.

Parametric RMPW performed well across different mediator, outcome, and treatment types when
the models producing the weights were adequately specified and the weights were stable. At the
same time, the simulations clarify an important tradeoff between specification and robustness.
Stratification-based nonparametric RMPW was less sensitive to nonlinear and nonadditive
misspecification of the mediator model but retained some residual bias, whereas correctly specified
parametric models could reduce bias at the cost of substantially greater sampling variability.
Alternative weighting formulations likewise changed rather than eliminated the modeling
assumptions: odds-ratio weighting avoided direct specification of the mediator distribution but
instead required an adequate model for treatment conditional on the mediator and covariates
\citep{tchetgen2013,huber2014}. The NSDUH illustration reinforces this point: alternative
mediator-weight constructions gave similar results without sampling weights but differed more
once sampling weights were incorporated, and reproduction of the target mediator distribution
favored the mediator-probability weights.

Perhaps the most consequential finding is that model specification and weight stability should be
treated as separate properties of a weighting analysis. In the treatment-model experiment, under
strong heteroskedasticity, the correctly specified heteroskedastic probit model generated more
concentrated weights and poorer finite-sample performance than the simpler logit model; in the continuous-mediator experiment, modeling the conditional
variance correctly reduced bias but still produced concentrated weights and poor coverage under
strong heteroskedasticity. The NSDUH analysis provides a complementary empirical example:
incorporating sampling weights changed the target population estimates while sharply reducing
effective sample size. The problem was especially pronounced in the simulations for continuous
mediators with strongly treatment- or covariate-dependent variances. Ratios of continuous densities
can become extremely variable when the treatment groups have limited overlap in their mediator
distributions, so increasing sample size need not provide the stabilization expected from conventional
large-sample arguments. The effective-sample-size diagnostics deteriorated in the same settings in
which bias, sampling variability, and coverage deteriorated. This does not imply that investigators
should deliberately misspecify their models. Rather, it means that goodness of fit and substantive
plausibility should be considered together with the behavior of the weights implied by those models.

The variance results lead to a related conclusion. Standard errors that treated estimated weights
as fixed were unreliable in both directions. The generalized two-step estimator, building on the
approach of \citet{bein2018}, performed well for many parametric weighting estimators when samples
were sufficiently large and weights were stable, while bootstrap inference generally performed
better at moderate sample sizes and provides a natural approach for the step-function weights used
by nonparametric RMPW and MMWS. The empirical illustrations showed the same general pattern:
two-step and bootstrap inference were similar for parametric weights, whereas bootstrap inference
was more consequential for the nonparametric weights in NSDUH and for some effects in Riverside.
Neither approach, however, produced reliable inference in the simulations when the point estimator
itself was severely destabilized by extreme weights. Variance estimation and weight diagnostics
should therefore be viewed as complementary components of an RMPW analysis rather than as
substitutes for one another.

\subsection{Practical guidance}

These findings lead to several practical recommendations, summarized in
Table~\ref{tab:recommendations}.

\paragraph{Begin with the causal estimand and identification assumptions.}
The choice of weighting estimator comes after the causal question has been defined. Natural
direct and indirect effects require assumptions about treatment assignment, mediator--outcome
confounding, positivity, and consistency
\citep{imai2010,vanderweele2015,nguyen2022}. Pretreatment covariates used to model the
mediator should therefore include confounders of the mediator--outcome relationship even when
treatment is randomized, because the mediator itself is not randomized. When treatment is
observational, treatment confounding must additionally be addressed. Treatment-induced
mediator--outcome confounding requires particular care because the natural effects considered
here are generally not identified under the standard assumptions
\citep{vanderweele2014confounder}.

\paragraph{Choose the mediator-weighting strategy to match the modeling problem.}
When a credible parametric model is available for the mediator distribution, parametric ratio
weights provide a natural starting point. For binary and ordinal mediators,
stratification-based nonparametric RMPW can provide a useful robustness analysis for
functional-form specification, although increasingly fine cross-stratification can create sparse
cells and need not improve performance in modest samples. Count and continuous mediators
require particular attention to their conditional distributions. For continuous mediators, the
conditional variance should be examined as well as the conditional mean; imposing constant
variance when substantial heteroskedasticity is present can produce biased density ratios.
Odds-ratio weighting may be attractive when direct modeling of the mediator distribution is
difficult, as in the bounded, zero-heavy count mediator in the Riverside example, but the
required model for treatment conditional on the mediator and covariates must itself be credible.

\paragraph{Address treatment selection separately when treatment is not randomized.}
Mediator weights address mediator selection; they do not in general remove confounding of
treatment with the mediator and outcome. A treatment-selection weight is therefore required
when treatment is observational under the identification strategy considered here
\citep{hong2015book}. IPTW performed well when the treatment model included the relevant
confounders. Stratification-based treatment weights reduced reliance on individual inverse
probabilities but retained residual confounding when strata were too coarse. More flexible
treatment models should be considered when substantively justified, but flexibility should not
be pursued without examining the stability of the resulting weights.

\paragraph{Examine the weights before interpreting the effects.}
Investigators should examine overlap and covariate balance and assess whether weighting
reproduces the intended target distribution \citep{colehernan2008,austinstuart2015}. For RMPW,
effective sample size and the largest observation's share of the total weight indicate whether a
small part of the sample dominates a weighted potential-outcome mean. The
simulations do not establish universal thresholds for these diagnostics. Unusually concentrated
weights should instead prompt investigation of overlap, model specification, and whether the
chosen weighting formulation is appropriate for the observed treatment and mediator
distributions. Restricting an analysis to a prespecified region of adequate common support may
be informative when lack of overlap is responsible for instability, but doing so changes the
population to which the resulting causal effect applies.

\paragraph{Match inference to the way the weights are estimated.}
Standard errors that treat estimated weights as fixed should generally be avoided. For
parametric weights, the two-step estimator provides a computationally convenient way to
propagate uncertainty from estimation of the weight models and performed well in sufficiently
large samples when the weights were reasonably stable. Bootstrap inference provides an
alternative in moderate samples and is required for the stratification-based estimators considered
here because their step-function weights are not
covered by the derivative-based variance calculation. All models, strata, and weights should be
re-estimated within each bootstrap resample. Repeated fitting failures should be treated as
evidence of sparse data or unstable models rather than simply discarded as a computational
inconvenience.

\paragraph{Use additional care with continuous treatments.}
For continuous treatments, the evaluation doses should lie in regions with adequate support,
and the bandwidth should be regarded as a bias--variance choice rather than a tuning parameter
for which smaller is necessarily better. In the designs considered here, the default bandwidth
used by the GPS estimator of \citet{hsu2020} performed well, whereas halving it increased
sampling variability and finite-sample bias. Because only two bandwidths and a single
data-generating treatment model were examined, these results support caution rather than a
general bandwidth-selection rule.

\paragraph{Report the information needed to evaluate the weighting analysis.}
An RMPW analysis should report the causal estimands and identification assumptions; the
mediator and, when applicable, treatment models and their covariates; the weighting strategy;
the variance estimator; and relevant balance and weight-stability diagnostics. Bootstrap
analyses should describe the resampling procedure and report fitting failures. Restrictions based
on overlap or common support should be reported together with their implications for the target
population. When alternative weighting strategies encode meaningfully different modeling
assumptions, presenting those analyses as robustness checks can help readers assess how strongly
the conclusions depend on a particular specification.

\subsection{Limitations}

The simulation experiments were designed to isolate several practically important choices in
RMPW implementation rather than exhaust the full range of data-generating mechanisms,
estimators, and applications. They considered representative mediator, outcome, and treatment
types at principally two sample sizes, and only selected forms of model misspecification and
weight instability. Extensions such as multiple mediators and moderated mediation were not
evaluated, and the continuous-treatment experiment considered a single treatment model and
two bandwidths. The results should therefore be interpreted as evidence about the methods under
the conditions examined rather than as rankings that necessarily generalize to other applications.

The simulations also assume that the natural effects are identified and therefore do not address
violations such as unmeasured mediator--outcome confounding; weighting-based sensitivity
analysis provides a complementary approach to this problem \citep{hongqinyang2018}. Finally, although
weight instability was examined directly, the simulations did not compare remedies such as
weight truncation or alternative weighting targets. Such modifications can alter the
bias--variance tradeoff and, in some cases, the population to which the estimand applies, and
therefore require separate evaluation.

\section{Conclusion}
\label{sec:conclusion}

Ratio-of-mediator-probability weighting provides a flexible approach to causal mediation analysis that has evolved into a broad family of weighting methods for different mediator and treatment types. This article brings these methods together within a common estimation framework for discrete treatments, incorporates a heteroskedastic normal model for continuous mediators, extends two-step variance estimation to multiple parametric weight components, and broadens the finite-sample evidence available to guide their application. The results show that successful application requires more than correct specification of the models used to construct the weights: investigators must also assess whether those models produce stable weights and use an inference procedure that appropriately accounts for weight estimation. These considerations---identification, model specification and weight stability, and inference---provide a practical framework for applying and evaluating RMPW analyses. The estimators, variance procedures, and diagnostics examined here are implemented in the community-contributed Stata command \texttt{mmwsmed} \citep{linden2026mmwsmed}.

\bibliographystyle{plainnat}
\bibliography{references}

@article{hongdeutschhill2015,
  author  = {Hong, Guanglei and Deutsch, Jonah and Hill, Heather D.},
  title   = {Ratio-of-mediator-probability weighting for causal mediation analysis in the presence of treatment-by-mediator interaction},
  journal = {Journal of Educational and Behavioral Statistics},
  year    = {2015},
  volume  = {40},
  number  = {3},
  pages   = {307--340},
  doi       = {10.3102/1076998615583902}
}

@book{hong2015book,
  author    = {Hong, Guanglei},
  title     = {Causality in a Social World: Moderation, Mediation and Spill-over},
  publisher = {Wiley},
  address   = {Chichester, UK},
  year      = {2015},
  doi       = {10.1002/9781119030638}
}

@inproceedings{hong2010,
  author    = {Hong, Guanglei},
  title     = {Ratio of mediator probability weighting for estimating natural direct and indirect effects},
  booktitle = {JSM Proceedings, Biometrics Section},
  pages     = {2401--2415},
  publisher = {American Statistical Association},
  address   = {Alexandria, VA},
  year      = {2010}
}

@article{hong2012,
  author  = {Hong, Guanglei},
  title   = {Marginal mean weighting through stratification: a generalized method for evaluating multivalued and multiple treatments with nonexperimental data},
  journal = {Psychological Methods},
  year    = {2012},
  volume  = {17},
  pages   = {44--60},
  doi       = {10.1037/a0024918}
}

@article{bein2018,
  author  = {Bein, Edward and Deutsch, Jonah and Hong, Guanglei and Porter, Kristin E. and Qin, Xu and Yang, Cheng},
  title   = {Two-step estimation in ratio-of-mediator-probability weighted causal mediation analysis},
  journal = {Statistics in Medicine},
  year    = {2018},
  volume  = {37},
  number  = {8},
  pages   = {1304--1324},
  doi       = {10.1002/sim.7581}
}

@article{morris2019,
  author  = {Morris, Tim P. and White, Ian R. and Crowther, Michael J.},
  title   = {Using simulation studies to evaluate statistical methods},
  journal = {Statistics in Medicine},
  year    = {2019},
  volume  = {38},
  number  = {11},
  pages   = {2074--2102},
  doi       = {10.1002/sim.8086}
}

@article{hsu2020,
  author  = {Hsu, Yu-Chin and Huber, Martin and Lee, Ying-Ying and Lettry, Layal},
  title   = {Direct and indirect effects of continuous treatments based on generalized propensity score weighting},
  journal = {Journal of Applied Econometrics},
  year    = {2020},
  volume  = {35},
  number  = {7},
  pages   = {814--840},
  doi       = {10.1002/jae.2765}
}

@article{tchetgen2013,
  author  = {Tchetgen Tchetgen, Eric J.},
  title   = {Inverse odds ratio-weighted estimation for causal mediation analysis},
  journal = {Statistics in Medicine},
  year    = {2013},
  volume  = {32},
  pages   = {4567--4580},
  doi       = {10.1002/sim.5864}
}

@article{linden2014,
  author  = {Linden, Ariel},
  title   = {Combining propensity score-based stratification and weighting to improve causal inference in the evaluation of health care interventions},
  journal = {Journal of Evaluation in Clinical Practice},
  year    = {2014},
  volume  = {20},
  pages   = {1065--1071},
  doi       = {10.1111/jep.12254}
}

@book{kish1965,
  author    = {Kish, Leslie},
  title     = {Survey Sampling},
  publisher = {Wiley},
  address   = {New York},
  year      = {1965}
}

@article{imai2010,
  author  = {Imai, Kosuke and Keele, Luke and Tingley, Dustin},
  title   = {A general approach to causal mediation analysis},
  journal = {Psychological Methods},
  year    = {2010},
  volume  = {15},
  pages   = {309--334},
  doi       = {10.1037/a0020761}
}

@article{lindenkarlson2013,
  author  = {Linden, Ariel and Karlson, Kristian Bernt},
  title   = {Using mediation analysis to identify causal mechanisms in disease management interventions},
  journal = {Health Services and Outcomes Research Methodology},
  year    = {2013},
  volume  = {13},
  pages   = {86--108},
  doi     = {10.1007/s10742-013-0106-5}
}

@article{huber2014,
  author  = {Huber, Martin},
  title   = {Identifying causal mechanisms (primarily) based on inverse probability weighting},
  journal = {Journal of Applied Econometrics},
  year    = {2014},
  volume  = {29},
  pages   = {920--943},
  doi       = {10.1002/jae.2341}
}

@manual{rmpwR,
  author = {Qin, Xu and Hong, Guanglei and Yang, Fan},
  title  = {rmpw: Causal Mediation Analysis Using Weighting Approach},
  note   = {R package version 0.0.7},
  url    = {https://CRAN.R-project.org/package=rmpw},
  year   = {2026}
}

@article{lindenyarnold2018,
  author  = {Linden, Ariel and Yarnold, Paul R.},
  title   = {Identifying causal mechanisms in health care interventions using classification tree analysis},
  journal = {Journal of Evaluation in Clinical Practice},
  year    = {2018},
  volume  = {24},
  pages   = {353--361},
  doi     = {10.1111/jep.12848}
}

@article{nguyen2023,
  author  = {Nguyen, Trang Quynh and Ogburn, Elizabeth L. and Schmid, Ian and Sarker, Elizabeth B. and Greifer, Noah and Koning, Ina M. and Stuart, Elizabeth A.},
  title   = {Causal mediation analysis: From simple to more robust strategies for estimation of marginal natural (in)direct effects},
  journal = {Statistics Surveys},
  year    = {2023},
  volume  = {17},
  pages   = {1--41},
  doi     = {10.1214/22-SS140}
}

@article{schuler2025,
  author  = {Schuler, Megan S. and Coffman, Donna L. and Stuart, Elizabeth A. and Nguyen, Trang Q. and Vegetabile, Brian and McCaffrey, Daniel F.},
  title   = {Practical challenges in mediation analysis: a guide for applied researchers},
  journal = {Health Services and Outcomes Research Methodology},
  year    = {2025},
  volume  = {25},
  number  = {1},
  pages   = {57--84},
  doi     = {10.1007/s10742-024-00327-4}
}

@article{baronkenny1986,
  author  = {Baron, Reuben M. and Kenny, David A.},
  title   = {The Moderator--Mediator Variable Distinction in Social Psychological Research: Conceptual, Strategic, and Statistical Considerations},
  journal = {Journal of Personality and Social Psychology},
  year    = {1986},
  volume  = {51},
  number  = {6},
  pages   = {1173--1182},
  doi     = {10.1037/0022-3514.51.6.1173}
}

@article{mackinnon2007,
  author  = {MacKinnon, David P. and Fairchild, Amanda J. and Fritz, Matthew S.},
  title   = {Mediation Analysis},
  journal = {Annual Review of Psychology},
  year    = {2007},
  volume  = {58},
  pages   = {593--614},
  doi     = {10.1146/annurev.psych.58.110405.085542}
}

@book{vanderweele2015,
  author    = {VanderWeele, Tyler J.},
  title     = {Explanation in Causal Inference: Methods for Mediation and Interaction},
  publisher = {Oxford University Press},
  address   = {New York},
  year      = {2015}
}

@article{robinsgreenland1992,
  author  = {Robins, James M. and Greenland, Sander},
  title   = {Identifiability and Exchangeability for Direct and Indirect Effects},
  journal = {Epidemiology},
  year    = {1992},
  volume  = {3},
  number  = {2},
  pages   = {143--155},
  doi     = {10.1097/00001648-199203000-00013}
}

@inproceedings{pearl2001,
  author    = {Pearl, Judea},
  title     = {Direct and Indirect Effects},
  booktitle = {Proceedings of the Seventeenth Conference on Uncertainty in Artificial Intelligence},
  year      = {2001},
  pages     = {411--420},
  publisher = {Morgan Kaufmann},
  address   = {San Francisco, CA}
}

@article{farbmacher2022,
  author  = {Farbmacher, Helmut and Huber, Martin and Laff{\'e}rs, Luk{\'a}{\v{s}} and Langen, Henrika and Spindler, Martin},
  title   = {Causal Mediation Analysis with Double Machine Learning},
  journal = {The Econometrics Journal},
  year    = {2022},
  volume  = {25},
  number  = {2},
  pages   = {277--300},
  doi     = {10.1093/ectj/utac003}
}

@article{hejazi2022,
  author  = {Hejazi, Nima S. and Rudolph, Kara E. and D{\'i}az, Iv{\'a}n},
  title   = {\texttt{medoutcon}: Nonparametric Efficient Causal Mediation Analysis with Machine Learning in \texttt{R}},
  journal = {Journal of Open Source Software},
  year    = {2022},
  volume  = {7},
  number  = {69},
  pages   = {3979},
  doi     = {10.21105/joss.03979}
}

@article{colehernan2008,
  author  = {Cole, Stephen R. and Hern{\'a}n, Miguel A.},
  title   = {Constructing Inverse Probability Weights for Marginal Structural Models},
  journal = {American Journal of Epidemiology},
  year    = {2008},
  volume  = {168},
  number  = {6},
  pages   = {656--664},
  doi     = {10.1093/aje/kwn164}
}

@article{austinstuart2015,
  author  = {Austin, Peter C. and Stuart, Elizabeth A.},
  title   = {Moving towards best practice when using inverse probability of treatment weighting ({IPTW}) using the propensity score to estimate causal treatment effects in observational studies},
  journal = {Statistics in Medicine},
  year    = {2015},
  volume  = {34},
  number  = {28},
  pages   = {3661--3679},
  doi     = {10.1002/sim.6607}
}

@article{hongqinyang2018,
  author  = {Hong, Guanglei and Qin, Xu and Yang, Fan},
  title   = {Weighting-based sensitivity analysis in causal mediation studies},
  journal = {Journal of Educational and Behavioral Statistics},
  year    = {2018},
  volume  = {43},
  number  = {1},
  pages   = {32--56},
  doi     = {10.3102/1076998617749561}
}

@article{nguyen2022,
  author  = {Nguyen, Trang Quynh and Schmid, Ian and Ogburn, Elizabeth L. and Stuart, Elizabeth A.},
  title   = {Clarifying Causal Mediation Analysis: Effect Identification via Three Assumptions and Five Potential Outcomes},
  journal = {Journal of Causal Inference},
  year    = {2022},
  volume  = {10},
  number  = {1},
  pages   = {246--279},
  doi     = {10.1515/jci-2021-0049}
}

@article{vanderweele2014confounder,
  author  = {VanderWeele, Tyler J. and Vansteelandt, Stijn and Robins, James M.},
  title   = {Effect Decomposition in the Presence of an Exposure-Induced Mediator-Outcome Confounder},
  journal = {Epidemiology},
  year    = {2014},
  volume  = {25},
  number  = {2},
  pages   = {300--306},
  doi     = {10.1097/EDE.0000000000000034}
}

@article{vanderweele2014fourway,
  author  = {VanderWeele, Tyler J.},
  title   = {A Unification of Mediation and Interaction: A Four-Way Decomposition},
  journal = {Epidemiology},
  year    = {2014},
  volume  = {25},
  number  = {5},
  pages   = {749--761},
  doi     = {10.1097/EDE.0000000000000121}
}

@misc{linden2026mmwsmed,
  author       = {Linden, Ariel},
  title        = {{MMWSMED}: {Stata} module for conducting causal mediation analysis using ratio-of-mediator-probability weighting with parametric or nonparametric weights},
  howpublished = {Statistical Software Components S460014, Boston College Department of Economics},
  year         = {2026}
}

@article{coffman2024,
  author  = {Coffman, Donna L. and Zhou, Haoyu and Castellano, Katherine E. and Schuler, Megan S. and McCaffrey, Daniel F.},
  title   = {Sampling weighting strategies in causal mediation analysis},
  journal = {BMC Medical Research Methodology},
  year    = {2024},
  volume  = {24},
  pages   = {133},
  doi     = {10.1186/s12874-024-02262-x}
}

@article{schuler2018,
  author  = {Schuler, Megan S. and Rice, Cara E. and Evans-Polce, Rebecca J. and Collins, Rebecca L.},
  title   = {Disparities in substance use behaviors and disorders among adult sexual minorities by age, gender, and sexual identity},
  journal = {Drug and Alcohol Dependence},
  year    = {2018},
  volume  = {189},
  pages   = {139--146},
  doi     = {10.1016/j.drugalcdep.2018.05.008}
}

@article{schulercollins2019,
  author  = {Schuler, Megan S. and Collins, Rebecca L.},
  title   = {Early alcohol and smoking initiation: a contributor to sexual minority disparities in adult use},
  journal = {American Journal of Preventive Medicine},
  year    = {2019},
  volume  = {57},
  number  = {6},
  pages   = {808--817},
  doi     = {10.1016/j.amepre.2019.07.020}
}

@manual{mccaffrey2026,
  author = {McCaffrey, Dan and Castellano, Katherine and Coffman, Donna and Vegetabile, Brian and Schuler, Megan and Zhou, Haoyu},
  title  = {twangMediation: Twang Causal Mediation Modeling via Weighting},
  note   = {R package version 1.2.1},
  url    = {https://CRAN.R-project.org/package=twangMediation},
  year   = {2026}
}

\clearpage

\begin{table}[htbp]
\centering
\caption{Components of the RMPW framework: previous methodological basis, methodological development in this article, and finite-sample evaluation in this article.}
\label{tab:contrib}
\footnotesize
\begin{tabular}{>{\raggedright\arraybackslash}p{2.6cm}>{\raggedright\arraybackslash}p{3.4cm}>{\raggedright\arraybackslash}p{4.0cm}>{\raggedright\arraybackslash}p{4.7cm}}
\toprule
Component & Previous basis & Developed here & Evaluated here \\
\midrule
Basic RMPW & \citet{hong2010}; \citet{hongdeutschhill2015}; \citet{hong2015book} & Common weight representation (Equation~\ref{eq:weight}) & Replication of \citet{hongdeutschhill2015}; continuous mediators, binary and count outcomes, and multivalued treatments \\
\addlinespace
Nonparametric RMPW & \citet{hongdeutschhill2015}; \citet{hong2015book} & --- & Number of strata and residual bias; misspecification designs extending those of \citet{hongdeutschhill2015}; bootstrap inference \\
\addlinespace
Continuous mediator & Separately estimated conditional moments \citep[sec.~12.1.2]{hong2015book} & Heteroskedastic normal model: joint maximum-likelihood mean and log-variance equations, with guaranteed positive fitted variance & Constant-variance, heteroskedastic, and odds-ratio weights under increasing heteroskedasticity \\
\addlinespace
Odds-ratio mediator weights & \citet{tchetgen2013}; \citet{huber2014} & --- & Comparison with ratio weights for continuous mediators and ordinal treatments; bootstrap inference \\
\addlinespace
Nonrandom treatment & \citet[sec.~12.2.1]{hong2015book}; \citet{hong2012}; \citet{linden2014} & --- & IPTW versus MMWS; omitted treatment-selection adjustment; binary, multivalued, and ordinal treatments; heteroskedastic treatment assignment \\
\addlinespace
Continuous treatment & \citet{hsu2020} & --- & Default versus halved bandwidth \\
\addlinespace
Two-step variance & \citet{bein2018} & Extension to stacked estimating equations for all parametric weight components (mediator, treatment-selection, marginal-share, and odds-ratio models) & Comparison with known-weight and bootstrap standard errors across designs \\
\addlinespace
Weight diagnostics & \citet{kish1965}; general weighting literature \citep{colehernan2008,austinstuart2015} & --- & Correspondence of effective sample size and weight concentration with estimator instability \\
\bottomrule
\end{tabular}

\vspace{2pt}\begin{minipage}{0.97\textwidth}\footnotesize IPTW, inverse-probability-of-treatment weighting; MMWS, marginal mean weighting through stratification. The derivation of the generalized two-step variance estimator is given in Supplementary Section~S2.\end{minipage}
\end{table}

\clearpage
\begin{table}[htbp]
\centering
\caption{Causal effects considered for treatment level $t$ compared with the control level $c$.}
\label{tab:effects}
\small
\begin{tabular}{>{\raggedright\arraybackslash}p{1.9cm}>{\raggedright\arraybackslash}p{5.6cm}>{\raggedright\arraybackslash}p{7.3cm}}
\toprule
Effect & Definition & Interpretation \\
\midrule
TE & $Y_{tt} - Y_{cc}$ & total effect of the treatment \\
NIE & $Y_{tt} - Y_{tc}$ & effect transmitted through the mediator, under treatment \\
NDE & $Y_{tc} - Y_{cc}$ & effect not transmitted through the mediator, with the mediator at its control value \\
PNIE & $Y_{ct} - Y_{cc}$ & effect transmitted through the mediator, under control \\
TNDE & $Y_{tt} - Y_{ct}$ & effect not transmitted through the mediator, with the mediator at its treated value \\
INT & $\text{NIE} - \text{PNIE} = \text{TNDE} - \text{NDE}$ & natural treatment-by-mediator interaction \\
\bottomrule
\end{tabular}

\vspace{2pt}\begin{minipage}{0.95\textwidth}\footnotesize $Y_{ab} = \E[Y(a, M(b))]$. With a continuous treatment, $t$ and $c$ are the evaluation dose $d_t$ and the control dose $d_0$.\end{minipage}
\end{table}

\clearpage
\begin{table}[htbp]
\centering
\caption{NSDUH data: decomposition of the disparity in current smoking by sexual-minority status into components through and not through early smoking initiation; estimates (standard errors) by weighting method, without and with the NSDUH sampling weights.}
\label{tab:nsduh}
\footnotesize\setlength{\tabcolsep}{4pt}
\begin{tabular}{>{\raggedright\arraybackslash}p{4.4cm}rrrrr}
\toprule
Weighting method & NIE & NDE & TE & INT & ESS (\%) \\
\midrule
\multicolumn{6}{l}{\emph{Without NSDUH sampling weights}} \\
Parametric RMPW            & 0.028 (0.003) & 0.097 (0.008) & 0.125 (0.008) & $-$0.005 (0.003) & 60.3 \\
Odds-ratio weights         & 0.028 (0.003) & 0.097 (0.008) & 0.125 (0.008) & $-$0.001 (0.003) & 58.7 \\
Nonparametric RMPW, $3\times 3$ & 0.028 (0.003) & 0.097 (0.009) & 0.125 (0.009) & $-$0.002 (0.002) & 59.9 \\
Parametric RMPW, probit treatment model & 0.028 (0.003) & 0.097 (0.008) & 0.124 (0.008) & $-$0.005 (0.003) & 59.4 \\
\addlinespace
\multicolumn{6}{l}{\emph{With NSDUH sampling weights}} \\
Parametric RMPW            & 0.018 (0.004) & 0.080 (0.013) & 0.098 (0.014) & $-$0.009 (0.005) & 15.7 \\
Odds-ratio weights         & 0.024 (0.004) & 0.075 (0.014) & 0.098 (0.014) & 0.001 (0.004) & 14.3 \\
Nonparametric RMPW, $3\times 3$ & 0.018 (0.006) & 0.080 (0.014) & 0.098 (0.014) & $-$0.006 (0.006) & 15.6 \\
Parametric RMPW, probit treatment model & 0.018 (0.004) & 0.079 (0.013) & 0.098 (0.013) & $-$0.009 (0.005) & 15.5 \\
\bottomrule
\end{tabular}

\vspace{4pt}
\begin{minipage}{0.97\textwidth}\footnotesize
$N = 40{,}293$ adult women (4,130 LGB, 36,163 heterosexual). Exposure: LGB vs.\ heterosexual; mediator: smoking
initiation before age 15; outcome: past-month cigarette smoking (effects are differences in prevalence). Treatment-selection
(IPTW) and mediator models adjust for age, race and ethnicity, education, household income, and employment status; the
treatment model is logit unless stated, and the mediator models are logit. With sampling weights, the weights enter every
model and the potential-outcome means. Standard errors are two-step, except for nonparametric RMPW (bootstrap, 1,000
resamples of the 100 primary sampling units). ESS: smallest Kish effective sample size of the four weighted copies, as a
percentage of the copy's observations.
\end{minipage}
\end{table}

\clearpage
\begin{table}[htbp]
\centering
\caption{NEWWS Riverside data: estimates (standard errors) by mediator representation, weighting method, and variance estimator.}
\label{tab:example}
\footnotesize\setlength{\tabcolsep}{4pt}
\begin{tabular}{>{\raggedright\arraybackslash}p{3.3cm}lrrrrr}
\toprule
Mediator model / weights & Std.\ err. & NIE & NDE & PNIE & INT & ESS (\%) \\
\midrule
\multicolumn{7}{l}{\emph{Binary mediator} (\texttt{emp}), logit model} \\
Parametric RMPW      & two-step  & $-$0.871 (0.475) & 1.277 (0.861) & 0.318 (0.275) & $-$1.189 (0.555) & 68.8 \\
                     & bootstrap & $-$0.871 (0.620) & 1.277 (0.942) & 0.318 (0.302) & $-$1.189 (0.691) & \\
\addlinespace
\multicolumn{7}{l}{\emph{Categorical mediator} (\texttt{emp\_cat})} \\
Ordered logit        & two-step  & $-$0.865 (0.394) & 1.269 (0.801) & 0.134 (0.270) & $-$0.999 (0.470) & 67.1 \\
                     & bootstrap & $-$0.865 (0.456) & 1.269 (0.816) & 0.134 (0.298) & $-$0.999 (0.546) & \\
Multinomial logit    & two-step  & $-$0.886 (0.473) & 1.282 (0.843) & 0.071 (0.367) & $-$0.957 (0.595) & 59.3 \\
\addlinespace
\multicolumn{7}{l}{\emph{Count mediator} (\texttt{emp\_cont}, quarters employed)} \\
Odds ratio           & two-step  & $-$0.616 (0.280) & 1.018 (0.738) & $-$0.234 (0.190) & $-$0.382 (0.325) & 72.2 \\
                     & bootstrap & $-$0.616 (0.282) & 1.018 (0.718) & $-$0.234 (0.201) & $-$0.382 (0.331) & \\
\bottomrule
\end{tabular}

\vspace{4pt}
\begin{minipage}{0.97\textwidth}\footnotesize
$N = 694$ (486 control, 208 LFA). All analyses adjust for the nine covariates in the mediator and outcome
models. Two-step: standard errors that account for the estimation of the weights; bootstrap: 1,000 resamples. ESS: smallest
Kish effective sample size of the two weighted copies, as a percentage of the copy's observations. The total effect was
0.40--0.41 in every analysis.
\end{minipage}
\end{table}

\clearpage
\begin{table}[!htbp]
\centering
\caption{Practical recommendations for ratio-of-mediator-probability weighting analyses.}
\label{tab:recommendations}
\small
\begin{tabular}{>{\raggedright\arraybackslash}p{0.23\textwidth}>{\raggedright\arraybackslash}p{0.69\textwidth}}
\toprule
Issue & Recommendation \\
\midrule
Identification and covariates
& Define the causal estimand and assess the identification assumptions before choosing an
estimator. Adjust the mediator model for pretreatment mediator--outcome confounders even
with randomized treatment; with nonrandom treatment, additionally adjust for treatment
confounders. \\
Discrete mediator
& Use a suitable parametric mediator model when its functional form is credible. For binary
or ordinal mediators, compare with stratification-based nonparametric RMPW as a robustness
analysis for functional-form specification; avoid overly fine strata when sample sizes within
cross-strata are small. \\
Count mediator
& Use a count model when its distributional assumptions are plausible; otherwise consider
odds-ratio weighting, recognizing that this shifts the modeling assumptions to the treatment
model conditional on the mediator. \\
Continuous mediator
& Examine the conditional mean and variance of the mediator across treatment groups. Model
heteroskedasticity when present and inspect weight stability carefully; use odds-ratio weighting
only when the treatment model conditional on the mediator can be adequately specified. \\
Nonrandom treatment
& Model treatment selection explicitly. IPTW performs well with an adequately specified
treatment model; stratification-based treatment weights can reduce sensitivity to extreme
propensities but may retain residual confounding. Balance model flexibility against weight
stability. \\
Continuous treatment
& Use GPS weighting at evaluation doses with adequate support. Treat bandwidth as a
bias--variance choice rather than assuming that a smaller bandwidth is preferable. \\
Weight diagnostics
& Examine covariate balance, reproduction of the target mediator distribution, effective sample
size, and concentration of the weights. Treat unusually low effective sample size or highly
concentrated weights as warnings requiring investigation, not as formal validity thresholds. \\
Variance estimation
& Avoid standard errors that treat estimated weights as fixed. Use two-step inference for
parametric weights in sufficiently large samples with reasonably stable weights; use bootstrap
inference for step-function weights and when finite-sample approximation is a concern. Refit all
models and weights within each bootstrap resample. \\
Reporting
& Report the estimand and identification assumptions, models and covariates, weighting strategy,
variance estimator, balance and weight-stability diagnostics, bootstrap fitting failures, overlap
restrictions, and sensitivity to substantively different weighting approaches. \\
\bottomrule
\end{tabular}
\end{table}

\clearpage
\appendix
\section*{Supplementary material}
\setcounter{table}{0}
\renewcommand{\thetable}{S\arabic{table}}

\subsection*{S1. Additional RMPW estimands and identification conditions}

\paragraph{Two mediators.}
With two mediators $M_1$ and $M_2$, the potential-outcome means depend on their causal
ordering. In the parallel (concurrent) design, neither mediator affects the other, and define
\[
Y(z,a,b)=E\{Y(z,M_1(a),M_2(b))\}.
\]
The total effect $Y(t,t,t)-Y(c,c,c)$ is decomposed into the natural direct effect
$Y(t,c,c)-Y(c,c,c)$ and the joint indirect effect
$Y(t,t,t)-Y(t,c,c)$. The latter can be decomposed into the indirect effect through $M_1$
with $M_2$ at its control or treated value,
\[
\mathrm{NIE1}_0=Y(t,t,c)-Y(t,c,c), \qquad
\mathrm{NIE1}_1=Y(t,t,t)-Y(t,c,t),
\]
and the corresponding effects through $M_2$,
\[
\mathrm{NIE2}_0=Y(t,c,t)-Y(t,c,c), \qquad
\mathrm{NIE2}_1=Y(t,t,t)-Y(t,t,c).
\]
The mediated interaction is
\[
\mathrm{M1xM2}=\mathrm{NIE2}_1-\mathrm{NIE2}_0,
\]
which is nonzero when the effect of one mediator on the outcome depends on the value of the
other \citep[sec.~13.2]{hong2015book}.

In the sequential (consecutive) design, $M_1$ precedes and may affect $M_2$, and
$M_2(t,M_1(t'))$ denotes the potential value of the second mediator under treatment $t$ when
the first mediator takes its value under $t'$. The weight proposed by
\citet[eq.~13.16]{hong2015book} identifies potential-outcome means of the form
\[
Y(z,a,b)=E\{Y(z,M_1(a),M_2(b,M_1(a)))\},
\]
so that $M_2$ responds to the individual's own value of $M_1$. Using these means,
the NDE, joint NIE, indirect effects through $M_1$ and
$M_2$, their mediated interaction, the corresponding pure indirect effects,
and the treatment-by-mediator interactions can be defined as shown in
Table~\ref{tab:effectsS}. The NDE, joint NIE, and TE coincide with the
definitions of \citet[sec.~13.3.1]{hong2015book}.

\paragraph{Moderated mediation and controlled direct effects.}
For a pretreatment moderator $G$, the models, weights, and effects may be
estimated within each level $g$ of $G$, with differences in each effect
defined relative to a reference level of the moderator
\citep[sec.~13.1]{hong2015book}. Controlled direct effects can additionally
be estimated from a weighted outcome regression containing treatment,
mediator, and their interaction \citep{hong2015book}.

\paragraph{Identification with two mediators.}
With two parallel mediators, conditions (ii) and (iii) of Section~\ref{sec:estimands} apply to each mediator, and the mediators
must additionally be conditionally independent of each other given treatment and $\mathbf{X}$
\citep[sec.~13.2.2]{hong2015book}. Sequential mediators require stronger conditions
\citep[sec.~13.3.2]{hong2015book}: given treatment and $\mathbf{X}$, $M_1$ must be independent
not only of the potential outcomes but also of the potential values of $M_2$; and given
treatment, $M_1$, and $\mathbf{X}$, $M_2$ must be independent of the potential outcomes.
Because $M_1$ is itself a posttreatment variable, the $M_2$--outcome relationship may be
confounded by $M_1$, which the weights condition on, but by no other variable affected by the
treatment or by $M_1$.

\subsection*{S2. Generalized two-step variance estimator}
\label{sec:twostepS}

This section states the estimating equations stacked by the two-step variance estimator of
Section~\ref{sec:variance}, as implemented in \texttt{mmwsmed}.

\paragraph{Weights and outcome model.}
Estimation proceeds in the duplicated data (Section~\ref{sec:rmpw}). Copy $c=1,\ldots,C$ targets the
potential-outcome mean $E\{Y(a_c,M(b_c))\}$ and contains the individuals with $T_i=a_c$.
The weight of individual $i$ in copy $c$ is
\[
W_{ic}(\boldsymbol{\theta})
=
u_i\,\tau_i(\boldsymbol{\theta}_T,\boldsymbol{\pi})\prod_{k}\omega_{ick}(\boldsymbol{\theta}_{M}),
\]
where $u_i$ is a sampling weight (one if none); $\tau_i=\pi_{T_i}/\Pr(T=T_i\mid\mathbf{X}_i;\boldsymbol{\theta}_T)$
is the stabilized IPTW (one under randomization), with marginal treatment shares
$\boldsymbol{\pi}=(\pi_z)$ and treatment-model parameters $\boldsymbol{\theta}_T$; and
$\omega_{ick}=f_k(M_{ik}\mid b_c,\mathbf{X}_i;\boldsymbol{\theta}_{M})/f_k(M_{ik}\mid a_c,\mathbf{X}_i;\boldsymbol{\theta}_{M})$
is the mediator-probability (or density) ratio for mediator $k$, equal to one when $a_c=b_c$.
For odds-ratio weights, $\omega_{ick}$ is Equation~\ref{eq:odds}, and $\boldsymbol{\theta}_M$
contains the parameters of the treatment models with and without the mediator. Collect the
parameters of all weight components in $\boldsymbol{\theta}=(\boldsymbol{\theta}_1,\ldots,\boldsymbol{\theta}_J)$,
one block per fitted model (a mediator model for each treatment level, the treatment-selection
model, the marginal shares, or an odds-weight treatment model).

The outcome model is a generalized linear model
$\mu_{ic}=g^{-1}(\mathbf{d}_{ic}'\boldsymbol{\lambda})$, where $\mathbf{d}_{ic}$ contains
the copy indicators and, optionally, outcome covariates centered at their (sampling-weighted)
sample means, $\tilde{\mathbf{v}}_i$, so that $\boldsymbol{\lambda}=(\lambda_1,\ldots,\lambda_C,\boldsymbol{\lambda}_v')'$; $g$ is the identity, logit, probit, or log link. Its weighted estimating function
for individual $i$ is
\[
\boldsymbol{\psi}_{\lambda,i}(\boldsymbol{\lambda},\boldsymbol{\theta})
=
\sum_{c}W_{ic}(\boldsymbol{\theta})\,\mathbf{d}_{ic}\,\alpha_{ic}(Y_i-\mu_{ic}),
\]
with $\alpha_{ic}=1$ for canonical links and
$\alpha_{ic}=\phi(\eta_{ic})/[\Phi(\eta_{ic})\{1-\Phi(\eta_{ic})\}]$ for the probit link,
$\eta_{ic}=\mathbf{d}_{ic}'\boldsymbol{\lambda}$. The sum over copies reflects that all copies of
an individual derive from one observation.

\paragraph{Weight-model estimating functions.}
Each weight component $j$ is estimated by solving $\sum_i\boldsymbol{\psi}_{j,i}(\boldsymbol{\theta}_j)=\mathbf{0}$,
where $\boldsymbol{\psi}_{j,i}$ is the (weighted) maximum-likelihood score of that model. For the
marginal treatment shares, $\psi_{\pi_z,i}=u_i\{\mathbf{1}(T_i=z)-\pi_z\}$. The mediator models
are fitted with weights $u_i\tau_i$ and therefore depend on the treatment-selection parameters:
$\boldsymbol{\psi}_{M,i}=u_i\tau_i(\boldsymbol{\theta}_T,\boldsymbol{\pi})\,
\mathbf{s}_{M,i}(\boldsymbol{\theta}_M)$, with $\mathbf{s}_{M,i}$ the individual score. The full
stacked estimating function is
\[
\boldsymbol{\Psi}_i
=
\bigl(\boldsymbol{\psi}_{T,i}',\;\boldsymbol{\psi}_{\pi,i}',\;\boldsymbol{\psi}_{M,i}',\;
\boldsymbol{\psi}_{\lambda,i}'\bigr)',
\]
in which the treatment-selection and marginal-share blocks are absent under randomization and
the mediator block includes the odds-weight treatment models when those weights are used.

\paragraph{Sandwich variance.}
Let $\mathbf{A}_N=-\sum_i\partial\boldsymbol{\Psi}_i/\partial(\boldsymbol{\theta}',\boldsymbol{\lambda}')$
and $\mathbf{B}_N=\sum_q\bigl(\sum_{i\in q}\boldsymbol{\Psi}_i\bigr)\bigl(\sum_{i\in q}\boldsymbol{\Psi}_i\bigr)'$,
where $q$ indexes clusters (individuals in the absence of clustering). The variance of
$(\hat{\boldsymbol{\theta}},\hat{\boldsymbol{\lambda}})$ is estimated by
$\mathbf{A}_N^{-1}\mathbf{B}_N\mathbf{A}_N^{-\top}$, the empirical sandwich
$\mathbf{A}^{-1}\mathbf{B}\mathbf{A}^{-\top}/N$ written in sums. Because the stack is
block-lower-triangular (the weight models do not depend on $\boldsymbol{\lambda}$, and the
mediator models depend on $\boldsymbol{\theta}_T$ and $\boldsymbol{\pi}$ but not conversely),
the $\boldsymbol{\lambda}$ block can be computed through influence functions. For each weight
component, let $\mathbf{H}_j=-\sum_i\partial\boldsymbol{\psi}_{j,i}/\partial\boldsymbol{\theta}_j'$ and
\[
\mathbf{IF}_{j,i}=\mathbf{H}_j^{-1}\boldsymbol{\psi}_{j,i}
\quad\text{(treatment and marginal-share models)},
\]
\[
\mathbf{IF}_{M,i}=\mathbf{H}_M^{-1}\Bigl\{\boldsymbol{\psi}_{M,i}
+\sum_{j\in\{T,\pi\}}\Bigl(\sum_l\boldsymbol{\psi}_{M,l}\,
\frac{\partial\log\tau_l}{\partial\boldsymbol{\theta}_j'}\Bigr)\mathbf{IF}_{j,i}\Bigr\}
\quad\text{(mediator models)}.
\]
Then, with
\[
\mathbf{Q}=\sum_i\sum_c W_{ic}\,\mathbf{d}_{ic}\mathbf{d}_{ic}'
\Bigl\{\alpha_{ic}\frac{\partial\mu_{ic}}{\partial\eta_{ic}}
-(Y_i-\mu_{ic})\frac{\partial\alpha_{ic}}{\partial\eta_{ic}}\Bigr\},
\qquad
\mathbf{D}_j=\sum_i\sum_c W_{ic}\,\mathbf{d}_{ic}\,\alpha_{ic}(Y_i-\mu_{ic})
\frac{\partial\log W_{ic}}{\partial\boldsymbol{\theta}_j'},
\]
and
$\mathbf{U}_i=\boldsymbol{\psi}_{\lambda,i}+\sum_j\mathbf{D}_j\,\mathbf{IF}_{j,i}$,
\[
\widehat{\mathrm{Var}}(\hat{\boldsymbol{\lambda}})
=
\mathbf{Q}^{-1}\Bigl\{\sum_q\Bigl(\sum_{i\in q}\mathbf{U}_i\Bigr)\Bigl(\sum_{i\in q}\mathbf{U}_i\Bigr)'\Bigr\}\mathbf{Q}^{-1},
\]
which equals the $\boldsymbol{\lambda}$ block of $\mathbf{A}_N^{-1}\mathbf{B}_N\mathbf{A}_N^{-\top}$.
The derivatives of the log weight are
$\partial\log W_{ic}/\partial\boldsymbol{\theta}_j=h_{jc}\,\mathbf{G}_{j,i}$, where
$\mathbf{G}_{j,i}$ is the derivative of the log fitted probability or density of model $j$ for
individual $i$ and $h_{jc}$ indicates how that probability enters copy $c$'s weight: $-1$ for the
treatment-selection model (denominator of $\tau_i$); $+1$ for the marginal share; for the
mediator model at treatment level $t$, $+1$ if $t=b_c$, $-1$ if $t=a_c$, and $0$ otherwise
(so the mediator models do not contribute when $a_c=b_c$); and the corresponding signs of the
log-odds terms in Equation~\ref{eq:odds} for odds-ratio weights. For the heteroskedastic normal
mediator model, for example,
$\mathbf{G}_{i}=\bigl(\mathbf{x}_i' r_i/\sigma_i^2,\;
\mathbf{z}_i'(r_i^2/\sigma_i^2-1)/2\bigr)'$, with $r_i=M_i-\mathbf{x}_i'\boldsymbol{\beta}_t$
and $\sigma_i^2=\exp(\mathbf{z}_i'\boldsymbol{\gamma}_t)$.

\paragraph{Potential-outcome means and effects.}
For a linear outcome model with centered covariates, the copy coefficients in
$\boldsymbol{\lambda}$ are the potential-outcome means $\boldsymbol{\nu}$. For a nonlinear
outcome model, the means are standardized,
$\hat\nu_c=\sum_i u_i\,g^{-1}(\hat\lambda_c+\tilde{\mathbf{v}}_i'\hat{\boldsymbol{\lambda}}_v)/\sum_i u_i$,
and
$\widehat{\mathrm{Var}}(\hat{\boldsymbol{\nu}})=\mathbf{J}\,\widehat{\mathrm{Var}}(\hat{\boldsymbol{\lambda}})\,\mathbf{J}'$,
with $\mathbf{J}=\partial\hat{\boldsymbol{\nu}}/\partial\boldsymbol{\lambda}'$; in both cases the
empirical covariate distribution over which the means are averaged is treated as fixed. The causal
effects are linear contrasts $\boldsymbol{\delta}=\mathbf{L}\boldsymbol{\nu}$ (Table~\ref{tab:effects}),
with $\widehat{\mathrm{Var}}(\hat{\boldsymbol{\delta}})=\mathbf{L}\,\widehat{\mathrm{Var}}(\hat{\boldsymbol{\nu}})\,\mathbf{L}'$.

\paragraph{Relation to Bein et al.}
With a single mediator, randomized treatment, no sampling weights or outcome covariates, and a
linear outcome model, the treatment-selection and marginal-share blocks are absent and the stack
reduces to the mediator-model scores and the weighted estimating equations for the
potential-outcome means, the construction of \citet{bein2018}.

\paragraph{Coverage.}
The estimator applies whenever every weight component is a smooth function of parameters
estimated by solving estimating equations. As implemented, this includes logit, probit, Poisson,
ordered logit, multinomial logit, normal, and heteroskedastic normal mediator models; logit,
probit, heteroskedastic probit, ordered, and multinomial logit treatment-selection models with
estimated marginal shares; the treatment models underlying odds-ratio weights; and linear,
logit, probit, and Poisson outcome models. It does not apply to stratification-based
nonparametric RMPW or MMWS weights, which are step functions of estimated propensity scores:
the weights are not differentiable in the propensity-model parameters, so $\mathbf{D}_j$ is not
defined and stratum boundaries contribute variability that a first-order expansion does not
capture. Nor does it apply to the kernel-based GPS estimator for continuous treatment. Bootstrap
inference is used in these cases.

\clearpage
\begin{table}[htbp]
\centering
\caption{Effects for two mediators and moderated mediation, for treatment level $t$ compared with the control level $c$.}
\label{tab:effectsS}
\small
\begin{tabular}{>{\raggedright\arraybackslash}p{1.9cm}>{\raggedright\arraybackslash}p{5.6cm}>{\raggedright\arraybackslash}p{7.3cm}}
\toprule
Effect & Definition & Interpretation \\
\midrule
\multicolumn{3}{l}{\textit{Two parallel mediators}; $Y(z,a,b) = \E[Y(z, M_1(a), M_2(b))]$} \\
TE, NDE & $Y(t,t,t) - Y(c,c,c)$; $Y(t,c,c) - Y(c,c,c)$ & total and natural direct effects \\
NIE & $Y(t,t,t) - Y(t,c,c)$ & joint indirect effect through both mediators \\
NIE1\_0, NIE1\_1 & $Y(t,t,c) - Y(t,c,c)$; $Y(t,t,t) - Y(t,c,t)$ & indirect effect through $M_1$, with $M_2$ at its control or treated value \\
NIE2\_0, NIE2\_1 & $Y(t,c,t) - Y(t,c,c)$; $Y(t,t,t) - Y(t,t,c)$ & indirect effect through $M_2$, with $M_1$ at its control or treated value \\
M1xM2 & $\text{NIE2\_1} - \text{NIE2\_0}$ & indirect effect through the $M_1$-by-$M_2$ interaction \\
\addlinespace
\multicolumn{3}{l}{\textit{Two sequential mediators}; $Y(z,a,b) = \E[Y(z, M_1(a), M_2(b, M_1(a)))]$} \\
TE, NDE, NIE & as for parallel mediators & total, natural direct, and joint indirect effects \\
NIE1 & $Y(t,t,c) - Y(t,c,c)$ & indirect effect through $M_1$, including its effect on $M_2$ under control \\
NIE2 & $Y(t,c,t) - Y(t,c,c)$ & indirect effect through $M_2$, with $M_1$ at its control value \\
M1xM2 & $Y(t,t,t) - Y(t,t,c) - Y(t,c,t) + Y(t,c,c)$ & indirect effect through the $M_1$-by-$M_2$ interaction \\
PNIE1, PNIE2 & $Y(c,t,c) - Y(c,c,c)$; $Y(c,c,t) - Y(c,c,c)$ & indirect effects through $M_1$ and $M_2$, under control \\
ZxM1, ZxM2 & $\text{NIE1} - \text{PNIE1}$; $\text{NIE2} - \text{PNIE2}$ & treatment-by-mediator interactions \\
\addlinespace
\multicolumn{3}{l}{\textit{Other}} \\
\textit{effect}\_diff & effect at moderator level $g$ minus effect at the first level & difference in an effect between subpopulations \\
POmeans & $Y_{ab}$ or $Y(z,a,b)$ & potential-outcome means \\
\bottomrule
\end{tabular}

\vspace{2pt}\begin{minipage}{0.95\textwidth}\footnotesize $Y_{ab} = \E[Y(a, M(b))]$ for one mediator.\end{minipage}
\end{table}

\clearpage
\subsection*{S3. Simulation design and complete results}

Supplementary Table~\ref{tab:simdesign} gives the data-generating mechanisms, true values of the
estimands, and methods compared for every simulation design. Supplementary
Tables~\ref{tab:study1}--\ref{tab:study5} report the complete results for the designs summarized
in Section~\ref{sec:results}.

\begin{footnotesize}\setlength{\tabcolsep}{3pt}
\begin{longtable}{>{\raggedright\arraybackslash}p{1.9cm}>{\raggedright\arraybackslash}p{3.1cm}>{\raggedright\arraybackslash}p{4.0cm}>{\raggedright\arraybackslash}p{4.0cm}>{\raggedleft\arraybackslash}p{1.0cm}>{\raggedleft\arraybackslash}p{1.0cm}}
\caption{Simulation design: data-generating mechanisms, true values of the estimands, and methods compared, by simulation design \citep[ADEMP;][]{morris2019}.}\label{tab:simdesign}\\
\toprule
Design & Treatment $T$ & Mediator $M$ & Outcome $Y$ & NIE & NDE \\
\midrule
\endfirsthead
\caption[]{(continued)}\\
\toprule
Design & Treatment $T$ & Mediator $M$ & Outcome $Y$ & NIE & NDE \\
\midrule
\endhead
\midrule
\multicolumn{6}{r}{\emph{continued on next page}}\\
\endfoot
\bottomrule
\multicolumn{6}{p{16.2cm}}{\emph{Notes:} Covariates are independent: $X_1, X_2, X_3, W_1 \sim N(0,1)$ and $W_2 \sim \text{Bernoulli}(0.5)$
($W_1$ and $W_2$ only in designs with nonrandom treatment); $S = X_1 + X_2 + X_3$; $\varepsilon, U \sim N(0,1)$; $\Lambda$ and $\Phi$ are the
logistic and standard normal distribution functions. Treatment indices:
$v = 0.5X_1 - 0.5X_2 + 0.3X_3 + 0.8W_1 + 0.8(W_2 - 0.5)$ and $v_2 = -0.3X_1 + 0.5X_2 + 0.5X_3 + 0.5W_1 - 0.8(W_2 - 0.5)$.
Mediator intercepts $\alpha_t$ were calibrated so that $\E\{\pr(M = 1 \mid T = t)\}$ takes the stated values. True values are the NIE
and NDE of treatment level~1 vs.\ 0 (and~2 vs.\ 0 where two contrasts are shown); for binary and count outcomes they are on the
mean (probability or count) scale and were obtained by Gauss--Hermite integration over the covariates. The analysis models
used the covariates $X_1$--$X_3$ (plus $W_1$ for the mediator and $X_1$--$X_3$, $W_1$, $W_2$ for the treatment in designs with nonrandom
treatment) as linear additive terms, except where stated. Known: standard errors that treat the estimated weights as known; two-step:
standard errors accounting for estimation of the weights using the generalized two-step estimator described in Section~\ref{sec:variance} and Supplementary Section~S2; bootstrap: 200 resamples. RMPW: parametric
ratio-of-mediator-probability weights; NP $k \times k$: nonparametric RMPW with $k$ strata of each mediator propensity score;
odds: odds-ratio mediator weights; IPTW: inverse-probability-of-treatment weights; MMWS $k$: marginal mean weighting through
stratification with $k$ strata.}\\
\endlastfoot
\multicolumn{6}{l}{\textbf{Replication of \citet{hongdeutschhill2015}.} $N = 800$, $5{,}000$; 1,000 replications.}\\[2pt]
\multicolumn{6}{p{16.2cm}}{All designs: $T \sim \text{Bernoulli}(0.5)$;
$\text{logit}\,\pr(M = 1 \mid T = 0) = \alpha_0 + X_1 + 0.5X_2$, $\text{logit}\,\pr(M = 1 \mid T = 1) = \alpha_1 + 0.5X_2 + X_3$;
$Y = \mu_{TM} + 0.5S + \varepsilon$, with $\mu_{00} = 0$, $\mu_{10} = \theta_1$, $\mu_{01} = \theta_2$,
$\mu_{11} = \theta_1 + \theta_2 + \theta_3$.}\\[2pt]
(a) & Randomized & $\E(\pi_0, \pi_1) = (0.392, 0.661)$ & $\theta = (0.392, 1, -1)$ & 0 & 0 \\
(b) & Randomized & $\E(\pi_0, \pi_1) = (0.392, 0.661)$ & $\theta = (2.158, 0.5, -3.5)$ & $-0.807$ & 0.787 \\
(c) & Randomized & $\E(\pi_0, \pi_1) = (0.194, 0.807)$ & $\theta = (1.112, 0.5, -1.75)$ & $-0.766$ & 0.773 \\[2pt]
\multicolumn{6}{p{16.2cm}}{\emph{Methods:} naive (mediator model without covariates); RMPW (known, two-step); NP $3 \times 3$ and
$4 \times 4$ (known).}\\
\midrule
\multicolumn{6}{l}{\textbf{Mediator-model misspecification.} $N = 800$, $5{,}000$; 1,000 replications.}\\[2pt]
\multicolumn{6}{p{16.2cm}}{All designs: treatment and outcome of design (b); the mediator models of the replication designs plus a nonlinear
(q) or nonadditive (i) term of strength $\delta$, with $\alpha_0, \alpha_1$ recalibrated so that $\E(\pi_0, \pi_1) = (0.392, 0.661)$.}\\[2pt]
Nonlinear & Randomized & $T = 0$: $+\,\delta(X_1^2 - 1)$; $T = 1$: $+\,\delta(X_3^2 - 1)$; $\delta = 0.25, 0.50, 1$ & as (b) & $-0.807$ & 0.787 \\
Nonadditive & Randomized & $T = 0$: $+\,\delta X_1X_2$; $T = 1$: $+\,\delta X_2X_3$; $\delta = 0.25, 0.50, 1$ & as (b) & $-0.807$ & 0.787 \\[2pt]
\multicolumn{6}{p{16.2cm}}{\emph{Methods:} naive; RMPW with the linear additive (misspecified) mediator model (two-step); RMPW with
the correct mediator model (two-step); NP $3 \times 3$ and $4 \times 4$ (known).}\\
\midrule
\multicolumn{6}{l}{\textbf{Alternative mediator, outcome, and treatment types.} $N = 800$, $5{,}000$; 1,000 replications.}\\[2pt]
Continuous $M$, common & Bernoulli(0.5) & $M = 0.5T + 0.5S + U$ & $Y = 0.2T + 0.4M + 0.3TM + 0.5S + \varepsilon$ & 0.350 & 0.200 \\
Continuous $M$, arm-specific & Bernoulli(0.5) & $T = 0$: $M = 0.5X_1 + 0.25X_2 + U$; $T = 1$: $M = 0.5 + 0.25X_2 + 0.5X_3 + 1.2U$ & as above & 0.350 & 0.200 \\
Count $M$ & Bernoulli(0.5) & $M \sim \text{Poisson}(\lambda_T)$; $\ln\lambda_0 = 0.637 + 0.3X_1 + 0.15X_2$, $\ln\lambda_1 = 1.042 + 0.15X_2 + 0.3X_3$; $\E(M \mid T) = 2, 3$ & $Y = 0.2T + 0.3M + 0.1TM + 0.5S + \varepsilon$ & 0.400 & 0.400 \\
Binary $Y$ & Bernoulli(0.5) & Binary, as replication design (b) & $\pr(Y = 1) = \Lambda(-1 + 0.6T + 1.2M + 0.4TM + 0.4S)$ & 0.093 & 0.145 \\
Count $Y$ & Bernoulli(0.5) & as above & $Y \sim \text{Poisson}\{\exp(0.3T + 0.7M - 0.2TM + 0.2S)\}$ & 0.245 & 0.305 \\
Three-level $T$ & Uniform on $\{0, 1, 2\}$ & $\text{logit}\,\pr(M = 1 \mid T = t)$: $\alpha_0 + X_1 + 0.5X_2$; $\alpha_1 + 0.5X_2 + X_3$; $\alpha_2 + 0.5S$; $\E(\pi_0, \pi_1, \pi_2) = (0.40, 0.60, 0.75)$ & $Y = \beta_T + \gamma_TM + 0.5S + \varepsilon$; $\beta = (0, 0.3, 0.5)$, $\gamma = (0.6, 0.8, 1)$ & 0.160\newline 0.350 & 0.380\newline 0.660 \\[2pt]
\multicolumn{6}{p{16.2cm}}{\emph{Methods:} naive; RMPW (known, two-step); odds (known, two-step) for the continuous mediators; NP $4 \times 4$
for the binary and count outcomes and $3 \times 3$ for the three-level treatment (known).}\\
\midrule
\multicolumn{6}{l}{\textbf{Nonrandom treatment assignment.} $N = 800$, $5{,}000$; 1,000 replications.}\\[2pt]
Binary $M$ & $\pr(T = 1) = \Lambda(v)$ & $\text{logit}\,\pr(M = 1 \mid T)$ as replication design (b) $+\,0.5W_1$; $\E(\pi_0, \pi_1) = (0.392, 0.661)$ & $Y = 0.3T + (0.6 + 0.3T)M + 0.5S + 0.4W_1 + 0.4W_2 + \varepsilon$ & 0.242 & 0.418 \\
Continuous $M$ & as above & $M = T + 0.5S + 0.5W_1 + U$ & $\pr(Y = 1) = \Lambda\{-1 + 0.5T + (0.6 + 0.2T)M + 0.3S + 0.4W_1 + 0.4(W_2 - 0.5)\}$ & 0.133 & 0.092 \\
Three-level $T$ & Multinomial logit; log odds vs.\ $T = 0$: $v$ ($T = 1$), $v_2$ ($T = 2$) & as randomized three-level $T$ $+\,0.5W_1$ & as randomized three-level $T$ $+\,0.4W_1 + 0.4W_2$ & 0.160\newline 0.350 & 0.380\newline 0.660 \\[2pt]
\multicolumn{6}{p{16.2cm}}{\emph{Methods:} naive; mediator selection only (no treatment weights); IPTW (known, two-step); MMWS 5 and
10 (known); IPTW with a misspecified treatment model omitting $W_2$ (known).}\\
\midrule
\multicolumn{6}{l}{\textbf{Treatment-model specification and weight stability.} $N = 800$, $5{,}000$; 1,000 replications.}\\[2pt]
Logit & $\pr(T = 1) = \Lambda(v)$ & as nonrandom binary $M$ & as nonrandom binary $M$ & 0.242 & 0.418 \\
Probit & $\pr(T = 1) = \Phi(v/1.6)$ & as above & as above & 0.242 & 0.418 \\
Heterosk.\ probit & $\pr(T = 1) = \Phi\{v/(1.6e^{gW_1})\}$; $g = 0.50, 0.25, 0.10$ & as above & as above & 0.242 & 0.418 \\[2pt]
\multicolumn{6}{p{16.2cm}}{\emph{Methods:} IPTW from logit, probit, and heteroskedastic probit (variance covariate $W_1$) treatment
models (two-step).}\\
\midrule
\multicolumn{6}{l}{\textbf{Ordinal treatment.} $N = 800$, $5{,}000$; 1,000 replications.}\\[2pt]
Ordinal $T$ & Ordered logit: $T^* = 0.6v + $ logistic error; cutpoints $-0.4$, $0.8$ & as nonrandom three-level $T$ & as nonrandom three-level $T$ & 0.160\newline 0.350 & 0.380\newline 0.660 \\[2pt]
\multicolumn{6}{p{16.2cm}}{\emph{Methods:} IPTW from multinomial logit, ordered logit, and ordered probit treatment models; odds
from multinomial logit and ordered logit (two-step).}\\
\midrule
\multicolumn{6}{l}{\textbf{Heteroskedastic continuous mediator.} $N = 800$, $5{,}000$.}\\[2pt]
Four levels of $c$ & as nonrandom binary $M$ & $M = T + 0.5S + 0.5W_1 + \exp\{0.5(1.5cX_1 + cTX_2)\}U$; $c = 0.40, 0.20, 0.10, 0$ & $Y = 0.3T + (0.5 + 0.2T)M + 0.5S + 0.4W_1 + 0.4W_2 + \varepsilon$ & 0.700 & 0.300 \\[2pt]
\multicolumn{6}{p{16.2cm}}{\emph{Methods:} (a) 1,000 replications: constant-variance mediator model; heteroskedastic model with the
variance depending on all mediator covariates or on $X_1$ and $X_2$; odds (all with IPTW, two-step). (b) 500 replications:
heteroskedastic model (all $c$) and constant-variance model ($c = 0$), two-step vs.\ bootstrap on the same data sets.}\\
\midrule
\multicolumn{6}{l}{\textbf{Continuous treatment.} $N = 800$, $5{,}000$; 1,000 replications.}\\[2pt]
Continuous $T$ & $T = 1 + 0.5X_1 + 0.5X_2 + \varepsilon_T$ & $M = 0.5 + 0.6T + 0.4X_1 + 0.3X_2 + U$ & $Y = 0.3T + 0.5M + 0.2TM + 0.5X_1 + 0.5X_2 + \varepsilon$ & 0.540\newline $-0.300$ & 0.520\newline $-0.520$ \\[2pt]
\multicolumn{6}{p{16.2cm}}{Covariates $X_1$, $X_2$ only. Estimands: dose 2 vs.\ 1 and dose 0 vs.\ 1. \emph{Methods:} GPS weighting
\citep{hsu2020} with the default bandwidth and half of it (bootstrap).}\\
\midrule
\multicolumn{6}{l}{\textbf{Variance estimation.} $N = 800$; 500 replications.}\\[2pt]
Binary $M$; three-level $T$ & as nonrandom designs & as nonrandom designs & as nonrandom designs & & \\[2pt]
\multicolumn{6}{p{16.2cm}}{\emph{Methods:} IPTW (two-step, bootstrap); MMWS 5 and 10, NP $3 \times 3$, and odds (bootstrap).}\\
\end{longtable}
\end{footnotesize}

\clearpage
\begin{footnotesize}\setlength{\tabcolsep}{4pt}
\begin{longtable}{lllrrrrrrr}
\caption{Replication of \citet{hongdeutschhill2015}: randomized binary treatment, binary mediator, and continuous outcome; 1,000 replications per cell.}\label{tab:study1}\\
\toprule
Design & $N$ & Method & Bias & \%BR & EmpSE & ModSE & Cov & RMSE & Fail \\
\midrule
\endfirsthead
\multicolumn{10}{l}{\textit{(continued)}}\\
\toprule
Design & $N$ & Method & Bias & \%BR & EmpSE & ModSE & Cov & RMSE & Fail \\
\midrule
\endhead
\bottomrule
\endfoot
\multicolumn{10}{l}{\textit{A. Natural indirect effect}}\\\addlinespace[2pt]
(a) & 800 & Naive & 0.164 & 0.000 & 0.041 & 0.037 & 0.011 & 0.170 & 0 \\
 &  & RMPW, known weights & 0.003 & 0.982 & 0.079 & 0.071 & 0.946 & 0.079 & 0 \\
 &  & RMPW, two-step & 0.003 & 0.982 & 0.079 & 0.066 & 0.933 & 0.079 & 0 \\
 &  & Nonparametric $3\times3$ & 0.030 & 0.817 & 0.059 & 0.063 & 0.935 & 0.066 & 0 \\
 &  & Nonparametric $4\times4$ & 0.019 & 0.887 & 0.062 & 0.068 & 0.957 & 0.065 & 0 \\
\addlinespace[3pt]
(a) & 5,000 & Naive & 0.162 & 0.000 & 0.017 & 0.015 & 0.000 & 0.163 & 0 \\
 &  & RMPW, known weights & $-$0.002 & 0.991 & 0.029 & 0.030 & 0.961 & 0.029 & 0 \\
 &  & RMPW, two-step & $-$0.002 & 0.991 & 0.029 & 0.028 & 0.939 & 0.029 & 0 \\
 &  & Nonparametric $3\times3$ & 0.026 & 0.837 & 0.021 & 0.023 & 0.807 & 0.034 & 0 \\
 &  & Nonparametric $4\times4$ & 0.017 & 0.896 & 0.022 & 0.025 & 0.923 & 0.028 & 0 \\
\addlinespace[3pt]
(b) & 800 & Naive & 0.162 & 0.000 & 0.089 & 0.037 & 0.179 & 0.185 & 0 \\
 &  & RMPW, known weights & $-$0.004 & 0.977 & 0.132 & 0.105 & 0.867 & 0.132 & 0 \\
 &  & RMPW, two-step & $-$0.004 & 0.977 & 0.132 & 0.132 & 0.945 & 0.132 & 0 \\
 &  & Nonparametric $3\times3$ & 0.026 & 0.842 & 0.113 & 0.090 & 0.857 & 0.116 & 0 \\
 &  & Nonparametric $4\times4$ & 0.036 & 0.778 & 0.120 & 0.101 & 0.850 & 0.125 & 0 \\
\addlinespace[3pt]
(b) & 5,000 & Naive & 0.164 & 0.000 & 0.037 & 0.015 & 0.001 & 0.168 & 0 \\
 &  & RMPW, known weights & 0.001 & 0.994 & 0.055 & 0.044 & 0.878 & 0.055 & 0 \\
 &  & RMPW, two-step & 0.001 & 0.994 & 0.055 & 0.054 & 0.950 & 0.055 & 0 \\
 &  & Nonparametric $3\times3$ & 0.027 & 0.833 & 0.046 & 0.033 & 0.754 & 0.053 & 0 \\
 &  & Nonparametric $4\times4$ & 0.018 & 0.892 & 0.046 & 0.036 & 0.830 & 0.049 & 0 \\
\addlinespace[3pt]
(c) & 800 & Naive & 0.382 & 0.000 & 0.102 & 0.100 & 0.047 & 0.396 & 0 \\
 &  & RMPW, known weights & 0.007 & 0.981 & 0.168 & 0.160 & 0.939 & 0.168 & 0 \\
 &  & RMPW, two-step & 0.007 & 0.981 & 0.168 & 0.142 & 0.914 & 0.168 & 0 \\
 &  & Nonparametric $3\times3$ & 0.073 & 0.810 & 0.133 & 0.155 & 0.922 & 0.152 & 0 \\
 &  & Nonparametric $4\times4$ & 0.082 & 0.786 & 0.139 & 0.149 & 0.899 & 0.161 & 0 \\
\addlinespace[3pt]
(c) & 5,000 & Naive & 0.387 & 0.000 & 0.042 & 0.040 & 0.000 & 0.389 & 0 \\
 &  & RMPW, known weights & 0.004 & 0.989 & 0.069 & 0.071 & 0.946 & 0.069 & 0 \\
 &  & RMPW, two-step & 0.004 & 0.989 & 0.069 & 0.064 & 0.923 & 0.069 & 0 \\
 &  & Nonparametric $3\times3$ & 0.068 & 0.825 & 0.051 & 0.060 & 0.811 & 0.085 & 0 \\
 &  & Nonparametric $4\times4$ & 0.044 & 0.887 & 0.054 & 0.065 & 0.910 & 0.070 & 0 \\
\addlinespace[3pt]
\multicolumn{10}{l}{\textit{B. Natural direct effect}}\\\addlinespace[2pt]
(a) & 800 & Naive & $-$0.164 & 0.000 & 0.105 & 0.107 & 0.666 & 0.195 & 0 \\
 &  & RMPW, known weights & $-$0.003 & 0.983 & 0.129 & 0.124 & 0.952 & 0.129 & 0 \\
 &  & RMPW, two-step & $-$0.003 & 0.983 & 0.129 & 0.121 & 0.948 & 0.129 & 0 \\
 &  & Nonparametric $3\times3$ & $-$0.030 & 0.819 & 0.117 & 0.119 & 0.951 & 0.120 & 0 \\
 &  & Nonparametric $4\times4$ & $-$0.019 & 0.885 & 0.121 & 0.123 & 0.955 & 0.122 & 0 \\
\addlinespace[3pt]
(a) & 5,000 & Naive & $-$0.161 & 0.000 & 0.040 & 0.043 & 0.028 & 0.166 & 0 \\
 &  & RMPW, known weights & 0.003 & 0.984 & 0.049 & 0.050 & 0.948 & 0.049 & 0 \\
 &  & RMPW, two-step & 0.003 & 0.984 & 0.049 & 0.049 & 0.946 & 0.049 & 0 \\
 &  & Nonparametric $3\times3$ & $-$0.025 & 0.842 & 0.045 & 0.046 & 0.925 & 0.051 & 0 \\
 &  & Nonparametric $4\times4$ & $-$0.016 & 0.901 & 0.045 & 0.047 & 0.948 & 0.048 & 0 \\
\addlinespace[3pt]
(b) & 800 & Naive & $-$0.165 & 0.000 & 0.127 & 0.119 & 0.695 & 0.208 & 0 \\
 &  & RMPW, known weights & 0.001 & 0.993 & 0.148 & 0.155 & 0.961 & 0.148 & 0 \\
 &  & RMPW, two-step & 0.001 & 0.993 & 0.148 & 0.145 & 0.949 & 0.148 & 0 \\
 &  & Nonparametric $3\times3$ & $-$0.027 & 0.837 & 0.131 & 0.145 & 0.957 & 0.133 & 0 \\
 &  & Nonparametric $4\times4$ & $-$0.024 & 0.855 & 0.139 & 0.153 & 0.954 & 0.141 & 0 \\
\addlinespace[3pt]
(b) & 5,000 & Naive & $-$0.162 & 0.000 & 0.055 & 0.047 & 0.111 & 0.171 & 0 \\
 &  & RMPW, known weights & 0.001 & 0.992 & 0.061 & 0.063 & 0.951 & 0.061 & 0 \\
 &  & RMPW, two-step & 0.001 & 0.992 & 0.061 & 0.059 & 0.944 & 0.061 & 0 \\
 &  & Nonparametric $3\times3$ & $-$0.025 & 0.844 & 0.054 & 0.056 & 0.936 & 0.060 & 0 \\
 &  & Nonparametric $4\times4$ & $-$0.016 & 0.904 & 0.054 & 0.057 & 0.954 & 0.056 & 0 \\
\addlinespace[3pt]
(c) & 800 & Naive & $-$0.382 & 0.000 & 0.138 & 0.137 & 0.214 & 0.407 & 0 \\
 &  & RMPW, known weights & $-$0.007 & 0.981 & 0.188 & 0.187 & 0.945 & 0.188 & 0 \\
 &  & RMPW, two-step & $-$0.007 & 0.981 & 0.188 & 0.168 & 0.926 & 0.188 & 0 \\
 &  & Nonparametric $3\times3$ & $-$0.068 & 0.821 & 0.169 & 0.184 & 0.926 & 0.182 & 0 \\
 &  & Nonparametric $4\times4$ & $-$0.064 & 0.832 & 0.173 & 0.182 & 0.931 & 0.185 & 0 \\
\addlinespace[3pt]
(c) & 5,000 & Naive & $-$0.388 & 0.000 & 0.057 & 0.055 & 0.000 & 0.392 & 0 \\
 &  & RMPW, known weights & $-$0.005 & 0.987 & 0.077 & 0.081 & 0.957 & 0.077 & 0 \\
 &  & RMPW, two-step & $-$0.005 & 0.987 & 0.077 & 0.074 & 0.936 & 0.077 & 0 \\
 &  & Nonparametric $3\times3$ & $-$0.069 & 0.823 & 0.062 & 0.071 & 0.861 & 0.092 & 0 \\
 &  & Nonparametric $4\times4$ & $-$0.045 & 0.885 & 0.064 & 0.075 & 0.917 & 0.078 & 0 \\
\addlinespace[3pt]
\end{longtable}\end{footnotesize}
\vspace{-6pt}\begin{footnotesize}\noindent Bias, mean estimate minus the true value; EmpSE, empirical standard error; ModSE, mean model standard error; Cov, coverage of nominal 95\% intervals; RMSE, root mean squared error; Fail, replications without an estimate. \%BR, proportion of the naive estimator's bias removed. Scenario (a): NIE = NDE = 0; (b): NIE = $-0.807$, NDE = 0.787; (c): NIE = $-0.766$, NDE = 0.773. Maximum Monte Carlo standard errors: bias 0.006; EmpSE 0.004; ModSE 0.002; coverage 0.015; RMSE 0.005.\end{footnotesize}

\clearpage
\begin{table}[htbp]
\centering
\caption{Replication of \citet{hongdeutschhill2015}: proportion of naive-estimator bias removed, this replication and (in parentheses) \citet[Table~5]{hongdeutschhill2015}.}
\label{tab:hongcompare}
\footnotesize
\begin{tabular}{lrrrrrr}
\toprule
& \multicolumn{3}{c}{$N = 5{,}000$} & \multicolumn{3}{c}{$N = 800$} \\
\cmidrule(lr){2-4}\cmidrule(lr){5-7}
Scenario & RMPW & NP $3\times3$ & NP $4\times4$ & RMPW & NP $3\times3$ & NP $4\times4$ \\
\midrule
\multicolumn{7}{l}{\textit{Natural indirect effect}}\\
(a) & 0.991 (0.998) & 0.837 (0.856) & 0.896 (0.904) & 0.982 (0.985) & 0.817 (0.856) & 0.887 (0.885) \\
(b) & 0.994 (0.991) & 0.833 (0.865) & 0.892 (0.913) & 0.977 (0.990) & 0.842 (0.864) & 0.778 (0.874) \\
(c) & 0.989 (0.999) & 0.825 (0.856) & 0.887 (0.904) & 0.981 (0.994) & 0.810 (0.823) & 0.786 (0.776) \\
\addlinespace
\multicolumn{7}{l}{\textit{Natural direct effect}}\\
(a) & 0.984 (0.999) & 0.842 (0.857) & 0.901 (0.905) & 0.983 (0.984) & 0.819 (0.843) & 0.885 (0.872) \\
(b) & 0.992 (0.980) & 0.844 (0.875) & 0.904 (0.923) & 0.993 (0.999) & 0.837 (0.854) & 0.855 (0.864) \\
(c) & 0.987 (0.995) & 0.823 (0.859) & 0.885 (0.908) & 0.981 (0.988) & 0.821 (0.818) & 0.832 (0.771) \\
\addlinespace
\bottomrule
\end{tabular}

\vspace{2pt}\begin{minipage}{0.92\textwidth}\footnotesize NP, nonparametric RMPW. Monte Carlo standard errors of the proportions in this replication are at most 0.03.\end{minipage}
\end{table}

\clearpage
\begin{footnotesize}\setlength{\tabcolsep}{4pt}
\begin{longtable}{lllrrrrrrr}
\caption{Mediator-model misspecification: nonlinear and nonadditive departures from the mediator model (scenario (b) of \citet{hongdeutschhill2015} with a nonlinear or nonadditive term of strength $\delta$ omitted from the mediator model); 1,000 replications per cell.}\label{tab:study1b}\\
\toprule
Design & $N$ & Method & Bias & \%BR & EmpSE & ModSE & Cov & RMSE & Fail \\
\midrule
\endfirsthead
\multicolumn{10}{l}{\textit{(continued)}}\\
\toprule
Design & $N$ & Method & Bias & \%BR & EmpSE & ModSE & Cov & RMSE & Fail \\
\midrule
\endhead
\bottomrule
\endfoot
\multicolumn{10}{l}{\textit{A. Natural indirect effect}}\\\addlinespace[2pt]
Nonlinear, $\delta=0.25$ & 800 & Naive & 0.151 & 0.000 & 0.088 & 0.091 & 0.601 & 0.175 & 0 \\
 &  & RMPW, misspecified & 0.071 & 0.530 & 0.120 & 0.122 & 0.920 & 0.139 & 0 \\
 &  & RMPW, correct model & 0.010 & 0.937 & 0.152 & 0.133 & 0.949 & 0.152 & 0 \\
 &  & Nonparametric $3\times3$ & 0.028 & 0.817 & 0.113 & 0.091 & 0.851 & 0.116 & 0 \\
 &  & Nonparametric $4\times4$ & 0.038 & 0.746 & 0.120 & 0.101 & 0.849 & 0.126 & 0 \\
\addlinespace[3pt]
Nonlinear, $\delta=0.25$ & 5,000 & Naive & 0.153 & 0.000 & 0.036 & 0.036 & 0.018 & 0.157 & 0 \\
 &  & RMPW, misspecified & 0.075 & 0.510 & 0.048 & 0.049 & 0.667 & 0.089 & 0 \\
 &  & RMPW, correct model & 0.002 & 0.989 & 0.068 & 0.059 & 0.940 & 0.068 & 0 \\
 &  & Nonparametric $3\times3$ & 0.029 & 0.808 & 0.043 & 0.033 & 0.764 & 0.052 & 0 \\
 &  & Nonparametric $4\times4$ & 0.019 & 0.874 & 0.045 & 0.037 & 0.845 & 0.049 & 0 \\
\addlinespace[3pt]
Nonlinear, $\delta=0.50$ & 800 & Naive & 0.142 & 0.000 & 0.092 & 0.092 & 0.644 & 0.169 & 0 \\
 &  & RMPW, misspecified & 0.108 & 0.237 & 0.115 & 0.117 & 0.852 & 0.158 & 0 \\
 &  & RMPW, correct model & 0.027 & 0.812 & 0.171 & 0.140 & 0.912 & 0.173 & 0 \\
 &  & Nonparametric $3\times3$ & 0.036 & 0.746 & 0.112 & 0.087 & 0.823 & 0.117 & 0 \\
 &  & Nonparametric $4\times4$ & 0.050 & 0.650 & 0.118 & 0.095 & 0.815 & 0.128 & 0 \\
\addlinespace[3pt]
Nonlinear, $\delta=0.50$ & 5,000 & Naive & 0.139 & 0.000 & 0.038 & 0.037 & 0.039 & 0.144 & 0 \\
 &  & RMPW, misspecified & 0.102 & 0.269 & 0.047 & 0.046 & 0.410 & 0.112 & 0 \\
 &  & RMPW, correct model & 0.006 & 0.956 & 0.087 & 0.068 & 0.921 & 0.087 & 0 \\
 &  & Nonparametric $3\times3$ & 0.032 & 0.769 & 0.045 & 0.031 & 0.721 & 0.055 & 0 \\
 &  & Nonparametric $4\times4$ & 0.025 & 0.819 & 0.046 & 0.036 & 0.808 & 0.053 & 0 \\
\addlinespace[3pt]
Nonlinear, $\delta=1$ & 800 & Naive & 0.111 & 0.000 & 0.097 & 0.095 & 0.752 & 0.147 & 0 \\
 &  & RMPW, misspecified & 0.089 & 0.196 & 0.112 & 0.113 & 0.876 & 0.143 & 0 \\
 &  & RMPW, correct model & 0.037 & 0.662 & 0.182 & 0.148 & 0.898 & 0.186 & 0 \\
 &  & Nonparametric $3\times3$ & 0.031 & 0.716 & 0.110 & 0.078 & 0.783 & 0.114 & 0 \\
 &  & Nonparametric $4\times4$ & 0.037 & 0.662 & 0.116 & 0.090 & 0.816 & 0.122 & 0 \\
\addlinespace[3pt]
Nonlinear, $\delta=1$ & 5,000 & Naive & 0.114 & 0.000 & 0.038 & 0.038 & 0.152 & 0.120 & 0 \\
 &  & RMPW, misspecified & 0.092 & 0.196 & 0.046 & 0.045 & 0.471 & 0.102 & 0 \\
 &  & RMPW, correct model & 0.021 & 0.816 & 0.117 & 0.079 & 0.851 & 0.119 & 1 \\
 &  & Nonparametric $3\times3$ & 0.034 & 0.703 & 0.045 & 0.028 & 0.654 & 0.056 & 0 \\
 &  & Nonparametric $4\times4$ & 0.029 & 0.743 & 0.045 & 0.032 & 0.749 & 0.054 & 0 \\
\addlinespace[3pt]
Nonadditive, $\delta=0.25$ & 800 & Naive & 0.152 & 0.000 & 0.091 & 0.090 & 0.604 & 0.177 & 0 \\
 &  & RMPW, misspecified & 0.032 & 0.788 & 0.128 & 0.126 & 0.931 & 0.132 & 0 \\
 &  & RMPW, correct model & 0.009 & 0.942 & 0.140 & 0.134 & 0.936 & 0.140 & 0 \\
 &  & Nonparametric $3\times3$ & 0.030 & 0.800 & 0.113 & 0.090 & 0.843 & 0.117 & 0 \\
 &  & Nonparametric $4\times4$ & 0.038 & 0.746 & 0.119 & 0.099 & 0.850 & 0.125 & 0 \\
\addlinespace[3pt]
Nonadditive, $\delta=0.25$ & 5,000 & Naive & 0.154 & 0.000 & 0.035 & 0.036 & 0.010 & 0.158 & 0 \\
 &  & RMPW, misspecified & 0.030 & 0.807 & 0.050 & 0.051 & 0.906 & 0.058 & 0 \\
 &  & RMPW, correct model & 0.000 & 0.999 & 0.060 & 0.057 & 0.959 & 0.060 & 0 \\
 &  & Nonparametric $3\times3$ & 0.028 & 0.817 & 0.043 & 0.032 & 0.770 & 0.051 & 0 \\
 &  & Nonparametric $4\times4$ & 0.019 & 0.879 & 0.044 & 0.036 & 0.847 & 0.048 & 0 \\
\addlinespace[3pt]
Nonadditive, $\delta=0.50$ & 800 & Naive & 0.144 & 0.000 & 0.090 & 0.092 & 0.648 & 0.170 & 0 \\
 &  & RMPW, misspecified & 0.046 & 0.678 & 0.126 & 0.124 & 0.935 & 0.134 & 0 \\
 &  & RMPW, correct model & 0.011 & 0.926 & 0.162 & 0.139 & 0.933 & 0.163 & 0 \\
 &  & Nonparametric $3\times3$ & 0.035 & 0.757 & 0.114 & 0.086 & 0.820 & 0.119 & 0 \\
 &  & Nonparametric $4\times4$ & 0.037 & 0.741 & 0.121 & 0.097 & 0.823 & 0.126 & 0 \\
\addlinespace[3pt]
Nonadditive, $\delta=0.50$ & 5,000 & Naive & 0.142 & 0.000 & 0.036 & 0.037 & 0.030 & 0.147 & 0 \\
 &  & RMPW, misspecified & 0.045 & 0.686 & 0.049 & 0.049 & 0.847 & 0.066 & 0 \\
 &  & RMPW, correct model & 0.000 & 1.000 & 0.071 & 0.063 & 0.940 & 0.071 & 0 \\
 &  & Nonparametric $3\times3$ & 0.032 & 0.778 & 0.043 & 0.031 & 0.728 & 0.053 & 0 \\
 &  & Nonparametric $4\times4$ & 0.022 & 0.846 & 0.045 & 0.035 & 0.820 & 0.050 & 0 \\
\addlinespace[3pt]
Nonadditive, $\delta=1$ & 800 & Naive & 0.112 & 0.000 & 0.092 & 0.095 & 0.765 & 0.145 & 0 \\
 &  & RMPW, misspecified & 0.040 & 0.647 & 0.118 & 0.120 & 0.943 & 0.125 & 0 \\
 &  & RMPW, correct model & 0.011 & 0.901 & 0.182 & 0.146 & 0.946 & 0.182 & 0 \\
 &  & Nonparametric $3\times3$ & 0.028 & 0.753 & 0.108 & 0.081 & 0.831 & 0.111 & 0 \\
 &  & Nonparametric $4\times4$ & 0.026 & 0.765 & 0.113 & 0.091 & 0.835 & 0.116 & 0 \\
\addlinespace[3pt]
Nonadditive, $\delta=1$ & 5,000 & Naive & 0.117 & 0.000 & 0.036 & 0.038 & 0.121 & 0.122 & 0 \\
 &  & RMPW, misspecified & 0.046 & 0.607 & 0.046 & 0.048 & 0.836 & 0.065 & 0 \\
 &  & RMPW, correct model & 0.003 & 0.973 & 0.127 & 0.075 & 0.926 & 0.127 & 0 \\
 &  & Nonparametric $3\times3$ & 0.031 & 0.734 & 0.043 & 0.029 & 0.706 & 0.053 & 0 \\
 &  & Nonparametric $4\times4$ & 0.024 & 0.799 & 0.043 & 0.033 & 0.806 & 0.049 & 0 \\
\addlinespace[3pt]
\multicolumn{10}{l}{\textit{B. Natural direct effect}}\\\addlinespace[2pt]
Nonlinear, $\delta=0.25$ & 800 & Naive & $-$0.154 & 0.000 & 0.129 & 0.131 & 0.775 & 0.201 & 0 \\
 &  & RMPW, misspecified & $-$0.074 & 0.520 & 0.135 & 0.136 & 0.910 & 0.154 & 0 \\
 &  & RMPW, correct model & $-$0.013 & 0.919 & 0.164 & 0.146 & 0.936 & 0.164 & 0 \\
 &  & Nonparametric $3\times3$ & $-$0.029 & 0.811 & 0.131 & 0.145 & 0.955 & 0.134 & 0 \\
 &  & Nonparametric $4\times4$ & $-$0.020 & 0.867 & 0.141 & 0.152 & 0.958 & 0.142 & 0 \\
\addlinespace[3pt]
Nonlinear, $\delta=0.25$ & 5,000 & Naive & $-$0.154 & 0.000 & 0.051 & 0.052 & 0.162 & 0.163 & 0 \\
 &  & RMPW, misspecified & $-$0.077 & 0.504 & 0.055 & 0.055 & 0.710 & 0.094 & 0 \\
 &  & RMPW, correct model & $-$0.003 & 0.979 & 0.074 & 0.064 & 0.932 & 0.074 & 0 \\
 &  & Nonparametric $3\times3$ & $-$0.031 & 0.799 & 0.052 & 0.055 & 0.927 & 0.061 & 0 \\
 &  & Nonparametric $4\times4$ & $-$0.021 & 0.865 & 0.052 & 0.058 & 0.955 & 0.056 & 0 \\
\addlinespace[3pt]
Nonlinear, $\delta=0.50$ & 800 & Naive & $-$0.144 & 0.000 & 0.133 & 0.131 & 0.798 & 0.196 & 0 \\
 &  & RMPW, misspecified & $-$0.111 & 0.232 & 0.132 & 0.131 & 0.869 & 0.172 & 0 \\
 &  & RMPW, correct model & $-$0.029 & 0.797 & 0.185 & 0.153 & 0.918 & 0.187 & 0 \\
 &  & Nonparametric $3\times3$ & $-$0.038 & 0.739 & 0.132 & 0.142 & 0.947 & 0.137 & 0 \\
 &  & Nonparametric $4\times4$ & $-$0.030 & 0.794 & 0.138 & 0.148 & 0.957 & 0.141 & 0 \\
\addlinespace[3pt]
Nonlinear, $\delta=0.50$ & 5,000 & Naive & $-$0.140 & 0.000 & 0.051 & 0.052 & 0.233 & 0.149 & 0 \\
 &  & RMPW, misspecified & $-$0.102 & 0.268 & 0.052 & 0.052 & 0.499 & 0.115 & 0 \\
 &  & RMPW, correct model & $-$0.007 & 0.952 & 0.093 & 0.073 & 0.912 & 0.093 & 0 \\
 &  & Nonparametric $3\times3$ & $-$0.033 & 0.766 & 0.052 & 0.054 & 0.914 & 0.061 & 0 \\
 &  & Nonparametric $4\times4$ & $-$0.026 & 0.816 & 0.052 & 0.057 & 0.941 & 0.058 & 0 \\
\addlinespace[3pt]
Nonlinear, $\delta=1$ & 800 & Naive & $-$0.111 & 0.000 & 0.131 & 0.131 & 0.855 & 0.172 & 0 \\
 &  & RMPW, misspecified & $-$0.089 & 0.195 & 0.128 & 0.128 & 0.895 & 0.156 & 0 \\
 &  & RMPW, correct model & $-$0.038 & 0.659 & 0.195 & 0.161 & 0.903 & 0.199 & 0 \\
 &  & Nonparametric $3\times3$ & $-$0.031 & 0.718 & 0.129 & 0.136 & 0.957 & 0.133 & 0 \\
 &  & Nonparametric $4\times4$ & $-$0.023 & 0.790 & 0.137 & 0.145 & 0.955 & 0.139 & 0 \\
\addlinespace[3pt]
Nonlinear, $\delta=1$ & 5,000 & Naive & $-$0.114 & 0.000 & 0.054 & 0.052 & 0.413 & 0.127 & 0 \\
 &  & RMPW, misspecified & $-$0.092 & 0.195 & 0.053 & 0.051 & 0.563 & 0.106 & 0 \\
 &  & RMPW, correct model & $-$0.021 & 0.812 & 0.121 & 0.084 & 0.847 & 0.122 & 1 \\
 &  & Nonparametric $3\times3$ & $-$0.034 & 0.699 & 0.054 & 0.052 & 0.888 & 0.064 & 0 \\
 &  & Nonparametric $4\times4$ & $-$0.030 & 0.740 & 0.054 & 0.055 & 0.912 & 0.062 & 0 \\
\addlinespace[3pt]
Nonadditive, $\delta=0.25$ & 800 & Naive & $-$0.147 & 0.000 & 0.130 & 0.130 & 0.795 & 0.196 & 0 \\
 &  & RMPW, misspecified & $-$0.027 & 0.814 & 0.141 & 0.140 & 0.942 & 0.143 & 0 \\
 &  & RMPW, correct model & $-$0.004 & 0.973 & 0.151 & 0.146 & 0.945 & 0.151 & 0 \\
 &  & Nonparametric $3\times3$ & $-$0.025 & 0.829 & 0.130 & 0.144 & 0.964 & 0.133 & 0 \\
 &  & Nonparametric $4\times4$ & $-$0.012 & 0.918 & 0.137 & 0.151 & 0.968 & 0.137 & 0 \\
\addlinespace[3pt]
Nonadditive, $\delta=0.25$ & 5,000 & Naive & $-$0.154 & 0.000 & 0.051 & 0.052 & 0.161 & 0.162 & 0 \\
 &  & RMPW, misspecified & $-$0.030 & 0.804 & 0.057 & 0.057 & 0.912 & 0.064 & 0 \\
 &  & RMPW, correct model & 0.000 & 0.998 & 0.066 & 0.062 & 0.943 & 0.066 & 0 \\
 &  & Nonparametric $3\times3$ & $-$0.029 & 0.815 & 0.052 & 0.055 & 0.933 & 0.059 & 0 \\
 &  & Nonparametric $4\times4$ & $-$0.019 & 0.877 & 0.052 & 0.057 & 0.960 & 0.055 & 0 \\
\addlinespace[3pt]
Nonadditive, $\delta=0.50$ & 800 & Naive & $-$0.142 & 0.000 & 0.126 & 0.131 & 0.806 & 0.190 & 0 \\
 &  & RMPW, misspecified & $-$0.045 & 0.686 & 0.136 & 0.137 & 0.939 & 0.143 & 0 \\
 &  & RMPW, correct model & $-$0.009 & 0.938 & 0.171 & 0.150 & 0.931 & 0.171 & 0 \\
 &  & Nonparametric $3\times3$ & $-$0.032 & 0.772 & 0.129 & 0.141 & 0.962 & 0.133 & 0 \\
 &  & Nonparametric $4\times4$ & $-$0.012 & 0.914 & 0.135 & 0.148 & 0.964 & 0.135 & 0 \\
\addlinespace[3pt]
Nonadditive, $\delta=0.50$ & 5,000 & Naive & $-$0.142 & 0.000 & 0.054 & 0.052 & 0.228 & 0.152 & 0 \\
 &  & RMPW, misspecified & $-$0.045 & 0.685 & 0.057 & 0.055 & 0.851 & 0.073 & 0 \\
 &  & RMPW, correct model & 0.000 & 1.000 & 0.077 & 0.068 & 0.938 & 0.077 & 0 \\
 &  & Nonparametric $3\times3$ & $-$0.032 & 0.778 & 0.053 & 0.054 & 0.913 & 0.061 & 0 \\
 &  & Nonparametric $4\times4$ & $-$0.022 & 0.846 & 0.053 & 0.056 & 0.934 & 0.058 & 0 \\
\addlinespace[3pt]
Nonadditive, $\delta=1$ & 800 & Naive & $-$0.114 & 0.000 & 0.129 & 0.131 & 0.866 & 0.172 & 0 \\
 &  & RMPW, misspecified & $-$0.042 & 0.636 & 0.132 & 0.133 & 0.939 & 0.139 & 0 \\
 &  & RMPW, correct model & $-$0.013 & 0.885 & 0.198 & 0.158 & 0.939 & 0.198 & 0 \\
 &  & Nonparametric $3\times3$ & $-$0.027 & 0.760 & 0.129 & 0.138 & 0.944 & 0.132 & 0 \\
 &  & Nonparametric $4\times4$ & $-$0.003 & 0.977 & 0.137 & 0.144 & 0.960 & 0.137 & 0 \\
\addlinespace[3pt]
Nonadditive, $\delta=1$ & 5,000 & Naive & $-$0.117 & 0.000 & 0.052 & 0.053 & 0.383 & 0.128 & 0 \\
 &  & RMPW, misspecified & $-$0.046 & 0.606 & 0.053 & 0.053 & 0.859 & 0.070 & 0 \\
 &  & RMPW, correct model & $-$0.003 & 0.972 & 0.132 & 0.079 & 0.917 & 0.132 & 0 \\
 &  & Nonparametric $3\times3$ & $-$0.031 & 0.733 & 0.051 & 0.053 & 0.919 & 0.060 & 0 \\
 &  & Nonparametric $4\times4$ & $-$0.024 & 0.798 & 0.052 & 0.055 & 0.941 & 0.057 & 0 \\
\addlinespace[3pt]
\end{longtable}\end{footnotesize}
\vspace{-6pt}\begin{footnotesize}\noindent Bias, mean estimate minus the true value; EmpSE, empirical standard error; ModSE, mean model standard error; Cov, coverage of nominal 95\% intervals; RMSE, root mean squared error; Fail, replications without an estimate. \%BR, proportion of the naive estimator's bias removed. True values: NIE $= -0.807$, NDE $= 0.787$. Misspecified: linear additive mediator model; correct model: mediator model with the omitted term; both with two-step standard errors. The nonparametric standard errors treat the weights as known. Maximum Monte Carlo standard errors: bias 0.006; EmpSE 0.004; coverage 0.016.\end{footnotesize}

\clearpage
\begin{footnotesize}\setlength{\tabcolsep}{3pt}
\begin{longtable}{lllrrrrrrr}
\caption{Alternative mediator, outcome, and treatment types: randomized treatment; 1,000 replications per cell.}\label{tab:study2}\\
\toprule
Design & $N$ & Method & Bias & \%BR & EmpSE & ModSE & Cov & RMSE & Fail \\
\midrule
\endfirsthead
\multicolumn{10}{l}{\textit{(continued)}}\\
\toprule
Design & $N$ & Method & Bias & \%BR & EmpSE & ModSE & Cov & RMSE & Fail \\
\midrule
\endhead
\bottomrule
\endfoot
\multicolumn{10}{l}{\textit{A. Natural indirect effect}}\\\addlinespace[2pt]
Continuous $M$, common & 800 & Naive & 0.213 & 0.000 & 0.109 & 0.050 & 0.089 & 0.239 & 0 \\
 &  & RMPW, known weights & $-$0.003 & 0.988 & 0.067 & 0.058 & 0.909 & 0.067 & 0 \\
 &  & RMPW, two-step & $-$0.003 & 0.988 & 0.067 & 0.066 & 0.950 & 0.067 & 0 \\
 &  & Odds & 0.000 & 1.000 & 0.068 & 0.054 & 0.875 & 0.068 & 0 \\
 &  & Odds, two-step & 0.000 & 1.000 & 0.068 & 0.067 & 0.952 & 0.068 & 0 \\
\addlinespace[3pt]
Continuous $M$, common & 5,000 & Naive & 0.215 & 0.000 & 0.045 & 0.019 & 0.000 & 0.220 & 0 \\
 &  & RMPW, known weights & 0.001 & 0.994 & 0.025 & 0.022 & 0.914 & 0.025 & 0 \\
 &  & RMPW, two-step & 0.001 & 0.994 & 0.025 & 0.024 & 0.933 & 0.025 & 0 \\
 &  & Odds & 0.002 & 0.992 & 0.028 & 0.022 & 0.876 & 0.028 & 0 \\
 &  & Odds, two-step & 0.002 & 0.992 & 0.028 & 0.027 & 0.939 & 0.028 & 0 \\
\addlinespace[3pt]
Continuous $M$, arm-specific & 800 & Naive & 0.109 & 0.000 & 0.083 & 0.041 & 0.315 & 0.137 & 0 \\
 &  & RMPW, known weights & $-$0.003 & 0.970 & 0.104 & 0.084 & 0.893 & 0.104 & 0 \\
 &  & RMPW, two-step & $-$0.003 & 0.970 & 0.104 & 0.096 & 0.944 & 0.104 & 0 \\
 &  & Odds & 0.061 & 0.436 & 0.078 & 0.049 & 0.702 & 0.099 & 0 \\
 &  & Odds, two-step & 0.061 & 0.436 & 0.078 & 0.078 & 0.896 & 0.099 & 0 \\
\addlinespace[3pt]
Continuous $M$, arm-specific & 5,000 & Naive & 0.107 & 0.000 & 0.034 & 0.016 & 0.011 & 0.112 & 0 \\
 &  & RMPW, known weights & $-$0.001 & 0.995 & 0.045 & 0.036 & 0.899 & 0.045 & 0 \\
 &  & RMPW, two-step & $-$0.001 & 0.995 & 0.045 & 0.041 & 0.942 & 0.045 & 0 \\
 &  & Odds & 0.058 & 0.455 & 0.032 & 0.020 & 0.244 & 0.066 & 0 \\
 &  & Odds, two-step & 0.058 & 0.455 & 0.032 & 0.031 & 0.530 & 0.066 & 0 \\
\addlinespace[3pt]
Count $M$ & 800 & Naive & 0.278 & 0.000 & 0.099 & 0.059 & 0.025 & 0.295 & 0 \\
 &  & RMPW, known weights & $-$0.002 & 0.992 & 0.109 & 0.087 & 0.931 & 0.109 & 0 \\
 &  & RMPW, two-step & $-$0.002 & 0.992 & 0.109 & 0.091 & 0.940 & 0.109 & 0 \\
\addlinespace[3pt]
Count $M$ & 5,000 & Naive & 0.282 & 0.000 & 0.039 & 0.024 & 0.000 & 0.284 & 0 \\
 &  & RMPW, known weights & $-$0.002 & 0.994 & 0.041 & 0.038 & 0.940 & 0.041 & 0 \\
 &  & RMPW, two-step & $-$0.002 & 0.994 & 0.041 & 0.039 & 0.955 & 0.041 & 0 \\
\addlinespace[3pt]
Binary $Y$ & 800 & Naive & 0.024 & 0.000 & 0.020 & 0.013 & 0.519 & 0.031 & 0 \\
 &  & RMPW, known weights & $-$0.001 & 0.947 & 0.031 & 0.028 & 0.920 & 0.031 & 0 \\
 &  & RMPW, two-step & $-$0.001 & 0.947 & 0.031 & 0.029 & 0.936 & 0.031 & 0 \\
 &  & Nonparametric $4\times4$ & $-$0.001 & 0.971 & 0.032 & 0.029 & 0.916 & 0.032 & 0 \\
\addlinespace[3pt]
Binary $Y$ & 5,000 & Naive & 0.025 & 0.000 & 0.008 & 0.005 & 0.015 & 0.027 & 0 \\
 &  & RMPW, known weights & 0.000 & 0.985 & 0.012 & 0.011 & 0.919 & 0.012 & 0 \\
 &  & RMPW, two-step & 0.000 & 0.985 & 0.012 & 0.012 & 0.938 & 0.012 & 0 \\
 &  & Nonparametric $4\times4$ & 0.003 & 0.878 & 0.011 & 0.010 & 0.918 & 0.011 & 0 \\
\addlinespace[3pt]
Count $Y$ & 800 & Naive & 0.117 & 0.000 & 0.062 & 0.041 & 0.261 & 0.132 & 0 \\
 &  & RMPW, known weights & 0.001 & 0.995 & 0.093 & 0.078 & 0.922 & 0.093 & 0 \\
 &  & RMPW, two-step & 0.001 & 0.995 & 0.093 & 0.081 & 0.928 & 0.093 & 0 \\
 &  & Nonparametric $4\times4$ & 0.005 & 0.958 & 0.086 & 0.077 & 0.927 & 0.086 & 0 \\
\addlinespace[3pt]
Count $Y$ & 5,000 & Naive & 0.115 & 0.000 & 0.024 & 0.016 & 0.000 & 0.117 & 0 \\
 &  & RMPW, known weights & 0.000 & 0.996 & 0.033 & 0.032 & 0.949 & 0.033 & 0 \\
 &  & RMPW, two-step & 0.000 & 0.996 & 0.033 & 0.033 & 0.955 & 0.033 & 0 \\
 &  & Nonparametric $4\times4$ & 0.012 & 0.893 & 0.030 & 0.028 & 0.907 & 0.032 & 0 \\
\addlinespace[3pt]
Three-level $T$, 1 vs.\ 0 & 800 & Naive & 0.121 & 0.000 & 0.068 & 0.032 & 0.159 & 0.139 & 0 \\
 &  & RMPW, known weights & 0.004 & 0.965 & 0.077 & 0.069 & 0.926 & 0.077 & 0 \\
 &  & RMPW, two-step & 0.004 & 0.965 & 0.077 & 0.073 & 0.944 & 0.077 & 0 \\
 &  & Nonparametric $3\times3$ & 0.020 & 0.838 & 0.074 & 0.065 & 0.919 & 0.076 & 0 \\
\addlinespace[3pt]
Three-level $T$, 1 vs.\ 0 & 5,000 & Naive & 0.119 & 0.000 & 0.028 & 0.013 & 0.000 & 0.123 & 0 \\
 &  & RMPW, known weights & 0.000 & 1.000 & 0.032 & 0.028 & 0.918 & 0.032 & 0 \\
 &  & RMPW, two-step & 0.000 & 1.000 & 0.032 & 0.030 & 0.937 & 0.032 & 0 \\
 &  & Nonparametric $3\times3$ & 0.020 & 0.835 & 0.027 & 0.023 & 0.832 & 0.033 & 0 \\
\addlinespace[3pt]
Three-level $T$, 2 vs.\ 0 & 800 & Naive & 0.228 & 0.000 & 0.092 & 0.065 & 0.107 & 0.246 & 0 \\
 &  & RMPW, known weights & $-$0.004 & 0.981 & 0.084 & 0.075 & 0.926 & 0.084 & 0 \\
 &  & RMPW, two-step & $-$0.004 & 0.981 & 0.084 & 0.077 & 0.940 & 0.084 & 0 \\
 &  & Nonparametric $3\times3$ & 0.030 & 0.870 & 0.086 & 0.079 & 0.911 & 0.091 & 0 \\
\addlinespace[3pt]
Three-level $T$, 2 vs.\ 0 & 5,000 & Naive & 0.231 & 0.000 & 0.036 & 0.026 & 0.000 & 0.233 & 0 \\
 &  & RMPW, known weights & 0.000 & 0.999 & 0.030 & 0.029 & 0.953 & 0.030 & 0 \\
 &  & RMPW, two-step & 0.000 & 0.999 & 0.030 & 0.030 & 0.955 & 0.030 & 0 \\
 &  & Nonparametric $3\times3$ & 0.036 & 0.843 & 0.029 & 0.028 & 0.744 & 0.046 & 0 \\
\addlinespace[3pt]
\multicolumn{10}{l}{\textit{B. Natural direct effect}}\\\addlinespace[2pt]
Continuous $M$, common & 800 & Naive & $-$0.215 & 0.000 & 0.096 & 0.136 & 0.729 & 0.235 & 0 \\
 &  & RMPW, known weights & 0.001 & 0.995 & 0.124 & 0.139 & 0.975 & 0.124 & 0 \\
 &  & RMPW, two-step & 0.001 & 0.995 & 0.124 & 0.127 & 0.956 & 0.124 & 0 \\
 &  & Odds & $-$0.001 & 0.994 & 0.122 & 0.137 & 0.969 & 0.122 & 0 \\
 &  & Odds, two-step & $-$0.001 & 0.994 & 0.122 & 0.128 & 0.953 & 0.122 & 0 \\
\addlinespace[3pt]
Continuous $M$, common & 5,000 & Naive & $-$0.216 & 0.000 & 0.037 & 0.054 & 0.001 & 0.219 & 0 \\
 &  & RMPW, known weights & $-$0.002 & 0.991 & 0.050 & 0.055 & 0.965 & 0.050 & 0 \\
 &  & RMPW, two-step & $-$0.002 & 0.991 & 0.050 & 0.050 & 0.950 & 0.050 & 0 \\
 &  & Odds & $-$0.002 & 0.989 & 0.052 & 0.055 & 0.965 & 0.052 & 0 \\
 &  & Odds, two-step & $-$0.002 & 0.989 & 0.052 & 0.051 & 0.947 & 0.052 & 0 \\
\addlinespace[3pt]
Continuous $M$, arm-specific & 800 & Naive & $-$0.106 & 0.000 & 0.099 & 0.116 & 0.890 & 0.145 & 0 \\
 &  & RMPW, known weights & 0.006 & 0.945 & 0.140 & 0.141 & 0.960 & 0.140 & 0 \\
 &  & RMPW, two-step & 0.006 & 0.945 & 0.140 & 0.131 & 0.946 & 0.140 & 0 \\
 &  & Odds & $-$0.059 & 0.446 & 0.117 & 0.127 & 0.940 & 0.131 & 0 \\
 &  & Odds, two-step & $-$0.059 & 0.446 & 0.117 & 0.115 & 0.915 & 0.131 & 0 \\
\addlinespace[3pt]
Continuous $M$, arm-specific & 5,000 & Naive & $-$0.108 & 0.000 & 0.038 & 0.046 & 0.319 & 0.114 & 0 \\
 &  & RMPW, known weights & 0.000 & 0.998 & 0.058 & 0.058 & 0.966 & 0.058 & 0 \\
 &  & RMPW, two-step & 0.000 & 0.998 & 0.058 & 0.054 & 0.956 & 0.058 & 0 \\
 &  & Odds & $-$0.059 & 0.452 & 0.046 & 0.050 & 0.810 & 0.075 & 0 \\
 &  & Odds, two-step & $-$0.059 & 0.452 & 0.046 & 0.046 & 0.746 & 0.075 & 0 \\
\addlinespace[3pt]
Count $M$ & 800 & Naive & $-$0.281 & 0.000 & 0.104 & 0.117 & 0.295 & 0.299 & 0 \\
 &  & RMPW, known weights & 0.000 & 0.999 & 0.136 & 0.137 & 0.964 & 0.136 & 0 \\
 &  & RMPW, two-step & 0.000 & 0.999 & 0.136 & 0.129 & 0.948 & 0.136 & 0 \\
\addlinespace[3pt]
Count $M$ & 5,000 & Naive & $-$0.281 & 0.000 & 0.042 & 0.047 & 0.000 & 0.284 & 0 \\
 &  & RMPW, known weights & 0.002 & 0.993 & 0.055 & 0.057 & 0.962 & 0.055 & 0 \\
 &  & RMPW, two-step & 0.002 & 0.993 & 0.055 & 0.054 & 0.953 & 0.055 & 0 \\
\addlinespace[3pt]
Binary $Y$ & 800 & Naive & $-$0.026 & 0.000 & 0.035 & 0.038 & 0.921 & 0.043 & 0 \\
 &  & RMPW, known weights & 0.000 & 0.982 & 0.045 & 0.045 & 0.955 & 0.045 & 0 \\
 &  & RMPW, two-step & 0.000 & 0.982 & 0.045 & 0.043 & 0.948 & 0.045 & 0 \\
 &  & Nonparametric $4\times4$ & $-$0.002 & 0.914 & 0.045 & 0.046 & 0.952 & 0.045 & 0 \\
\addlinespace[3pt]
Binary $Y$ & 5,000 & Naive & $-$0.025 & 0.000 & 0.015 & 0.015 & 0.620 & 0.029 & 0 \\
 &  & RMPW, known weights & 0.000 & 0.982 & 0.019 & 0.018 & 0.940 & 0.019 & 0 \\
 &  & RMPW, two-step & 0.000 & 0.982 & 0.019 & 0.017 & 0.928 & 0.019 & 0 \\
 &  & Nonparametric $4\times4$ & $-$0.003 & 0.876 & 0.018 & 0.018 & 0.940 & 0.018 & 0 \\
\addlinespace[3pt]
Count $Y$ & 800 & Naive & $-$0.114 & 0.000 & 0.102 & 0.113 & 0.850 & 0.153 & 0 \\
 &  & RMPW, known weights & 0.002 & 0.984 & 0.136 & 0.137 & 0.960 & 0.136 & 0 \\
 &  & RMPW, two-step & 0.002 & 0.984 & 0.136 & 0.129 & 0.943 & 0.136 & 0 \\
 &  & Nonparametric $4\times4$ & $-$0.010 & 0.911 & 0.131 & 0.136 & 0.960 & 0.132 & 0 \\
\addlinespace[3pt]
Count $Y$ & 5,000 & Naive & $-$0.117 & 0.000 & 0.041 & 0.045 & 0.242 & 0.123 & 0 \\
 &  & RMPW, known weights & $-$0.001 & 0.990 & 0.052 & 0.055 & 0.959 & 0.052 & 0 \\
 &  & RMPW, two-step & $-$0.001 & 0.990 & 0.052 & 0.052 & 0.950 & 0.052 & 0 \\
 &  & Nonparametric $4\times4$ & $-$0.014 & 0.881 & 0.050 & 0.052 & 0.943 & 0.052 & 0 \\
\addlinespace[3pt]
Three-level $T$, 1 vs.\ 0 & 800 & Naive & $-$0.120 & 0.000 & 0.117 & 0.129 & 0.865 & 0.167 & 0 \\
 &  & RMPW, known weights & $-$0.004 & 0.969 & 0.140 & 0.144 & 0.954 & 0.140 & 0 \\
 &  & RMPW, two-step & $-$0.004 & 0.969 & 0.140 & 0.138 & 0.939 & 0.140 & 0 \\
 &  & Nonparametric $3\times3$ & $-$0.006 & 0.949 & 0.160 & 0.142 & 0.911 & 0.160 & 0 \\
\addlinespace[3pt]
Three-level $T$, 1 vs.\ 0 & 5,000 & Naive & $-$0.121 & 0.000 & 0.048 & 0.051 & 0.352 & 0.130 & 0 \\
 &  & RMPW, known weights & $-$0.002 & 0.987 & 0.059 & 0.058 & 0.939 & 0.059 & 0 \\
 &  & RMPW, two-step & $-$0.002 & 0.987 & 0.059 & 0.056 & 0.929 & 0.059 & 0 \\
 &  & Nonparametric $3\times3$ & $-$0.021 & 0.825 & 0.055 & 0.055 & 0.929 & 0.059 & 0 \\
\addlinespace[3pt]
Three-level $T$, 2 vs.\ 0 & 800 & Naive & $-$0.228 & 0.000 & 0.128 & 0.145 & 0.658 & 0.262 & 0 \\
 &  & RMPW, known weights & 0.004 & 0.982 & 0.146 & 0.149 & 0.957 & 0.146 & 0 \\
 &  & RMPW, two-step & 0.004 & 0.982 & 0.146 & 0.140 & 0.947 & 0.146 & 0 \\
 &  & Nonparametric $3\times3$ & $-$0.033 & 0.858 & 0.150 & 0.153 & 0.944 & 0.153 & 0 \\
\addlinespace[3pt]
Three-level $T$, 2 vs.\ 0 & 5,000 & Naive & $-$0.231 & 0.000 & 0.054 & 0.058 & 0.016 & 0.238 & 0 \\
 &  & RMPW, known weights & $-$0.001 & 0.996 & 0.058 & 0.059 & 0.953 & 0.058 & 0 \\
 &  & RMPW, two-step & $-$0.001 & 0.996 & 0.058 & 0.056 & 0.938 & 0.058 & 0 \\
 &  & Nonparametric $3\times3$ & $-$0.037 & 0.841 & 0.058 & 0.059 & 0.910 & 0.068 & 0 \\
\addlinespace[3pt]
\end{longtable}\end{footnotesize}
\vspace{-6pt}\begin{footnotesize}\noindent Bias, mean estimate minus the true value; EmpSE, empirical standard error; ModSE, mean model standard error; Cov, coverage of nominal 95\% intervals; RMSE, root mean squared error; Fail, replications without an estimate. \%BR, proportion of the naive estimator's bias removed. Continuous mediator with common or arm-specific covariate effects and residual variance; binary and count outcomes with a binary mediator; three-level treatment with a binary mediator. Maximum Monte Carlo standard errors: bias 0.005; EmpSE 0.004; ModSE 0.001; coverage 0.016; RMSE 0.006.\end{footnotesize}

\clearpage
\begin{footnotesize}\setlength{\tabcolsep}{4pt}
\begin{longtable}{lllrrrrrrr}
\caption{Nonrandom treatment assignment: alternative treatment-selection adjustments; 1,000 replications per cell.}\label{tab:study3a}\\
\toprule
Design & $N$ & Method & Bias & \%BR & EmpSE & ModSE & Cov & RMSE & Fail \\
\midrule
\endfirsthead
\multicolumn{10}{l}{\textit{(continued)}}\\
\toprule
Design & $N$ & Method & Bias & \%BR & EmpSE & ModSE & Cov & RMSE & Fail \\
\midrule
\endhead
\bottomrule
\endfoot
\multicolumn{10}{l}{\textit{A. Natural indirect effect}}\\\addlinespace[2pt]
Binary $M$ & 800 & Naive & 0.322 & 0.000 & 0.072 & 0.050 & 0.000 & 0.330 & 0 \\
 &  & Mediator selection only & $-$0.016 & 0.949 & 0.083 & 0.072 & 0.920 & 0.085 & 0 \\
 &  & IPTW, known weights & $-$0.011 & 0.967 & 0.097 & 0.084 & 0.935 & 0.097 & 0 \\
 &  & IPTW, two-step & $-$0.011 & 0.967 & 0.097 & 0.082 & 0.934 & 0.097 & 0 \\
 &  & MMWS 5 strata & $-$0.009 & 0.973 & 0.094 & 0.081 & 0.933 & 0.095 & 0 \\
 &  & MMWS 10 strata & $-$0.009 & 0.972 & 0.096 & 0.083 & 0.934 & 0.096 & 0 \\
 &  & IPTW, $W_2$ omitted & $-$0.007 & 0.977 & 0.094 & 0.082 & 0.932 & 0.094 & 0 \\
\addlinespace[3pt]
Binary $M$ & 5,000 & Naive & 0.321 & 0.000 & 0.029 & 0.020 & 0.000 & 0.322 & 0 \\
 &  & Mediator selection only & $-$0.015 & 0.953 & 0.031 & 0.029 & 0.921 & 0.035 & 0 \\
 &  & IPTW, known weights & $-$0.003 & 0.992 & 0.035 & 0.034 & 0.961 & 0.035 & 0 \\
 &  & IPTW, two-step & $-$0.003 & 0.992 & 0.035 & 0.034 & 0.960 & 0.035 & 0 \\
 &  & MMWS 5 strata & $-$0.002 & 0.993 & 0.034 & 0.033 & 0.955 & 0.034 & 0 \\
 &  & MMWS 10 strata & $-$0.002 & 0.994 & 0.035 & 0.034 & 0.956 & 0.035 & 0 \\
 &  & IPTW, $W_2$ omitted & $-$0.002 & 0.995 & 0.034 & 0.033 & 0.958 & 0.034 & 0 \\
\addlinespace[3pt]
Continuous $M$ & 800 & Naive & 0.129 & 0.000 & 0.028 & 0.025 & 0.000 & 0.132 & 0 \\
 &  & Mediator selection only & $-$0.003 & 0.980 & 0.033 & 0.031 & 0.938 & 0.033 & 0 \\
 &  & IPTW, known weights & $-$0.006 & 0.952 & 0.034 & 0.034 & 0.951 & 0.034 & 0 \\
 &  & IPTW, two-step & $-$0.006 & 0.952 & 0.034 & 0.033 & 0.937 & 0.034 & 0 \\
 &  & MMWS 5 strata & $-$0.005 & 0.964 & 0.033 & 0.034 & 0.950 & 0.034 & 0 \\
 &  & MMWS 10 strata & $-$0.005 & 0.959 & 0.034 & 0.034 & 0.949 & 0.034 & 0 \\
 &  & IPTW, $W_2$ omitted & $-$0.005 & 0.959 & 0.034 & 0.034 & 0.949 & 0.034 & 0 \\
\addlinespace[3pt]
Continuous $M$ & 5,000 & Naive & 0.130 & 0.000 & 0.011 & 0.010 & 0.000 & 0.130 & 0 \\
 &  & Mediator selection only & 0.000 & 0.998 & 0.012 & 0.013 & 0.959 & 0.012 & 0 \\
 &  & IPTW, known weights & $-$0.001 & 0.993 & 0.013 & 0.014 & 0.964 & 0.013 & 0 \\
 &  & IPTW, two-step & $-$0.001 & 0.993 & 0.013 & 0.013 & 0.952 & 0.013 & 0 \\
 &  & MMWS 5 strata & 0.000 & 0.999 & 0.013 & 0.014 & 0.965 & 0.013 & 0 \\
 &  & MMWS 10 strata & 0.000 & 0.996 & 0.013 & 0.014 & 0.970 & 0.013 & 0 \\
 &  & IPTW, $W_2$ omitted & $-$0.001 & 0.996 & 0.013 & 0.014 & 0.963 & 0.013 & 0 \\
\addlinespace[3pt]
Three-level $T$, 1 vs.\ 0 & 800 & Naive & 0.212 & 0.000 & 0.075 & 0.040 & 0.011 & 0.225 & 0 \\
 &  & Mediator selection only & $-$0.038 & 0.822 & 0.076 & 0.063 & 0.837 & 0.085 & 0 \\
 &  & IPTW, known weights & $-$0.008 & 0.964 & 0.104 & 0.092 & 0.932 & 0.104 & 0 \\
 &  & IPTW, two-step & $-$0.008 & 0.964 & 0.104 & 0.091 & 0.935 & 0.104 & 0 \\
 &  & MMWS 5 strata & $-$0.008 & 0.961 & 0.100 & 0.088 & 0.920 & 0.100 & 0 \\
 &  & MMWS 10 strata & $-$0.009 & 0.956 & 0.103 & 0.091 & 0.927 & 0.104 & 0 \\
 &  & IPTW, $W_2$ omitted & 0.002 & 0.991 & 0.100 & 0.086 & 0.913 & 0.100 & 0 \\
\addlinespace[3pt]
Three-level $T$, 1 vs.\ 0 & 5,000 & Naive & 0.213 & 0.000 & 0.030 & 0.016 & 0.000 & 0.215 & 0 \\
 &  & Mediator selection only & $-$0.042 & 0.805 & 0.030 & 0.025 & 0.610 & 0.051 & 0 \\
 &  & IPTW, known weights & 0.000 & 0.999 & 0.045 & 0.043 & 0.940 & 0.045 & 0 \\
 &  & IPTW, two-step & 0.000 & 0.999 & 0.045 & 0.041 & 0.928 & 0.045 & 0 \\
 &  & MMWS 5 strata & $-$0.004 & 0.980 & 0.042 & 0.039 & 0.934 & 0.042 & 0 \\
 &  & MMWS 10 strata & $-$0.001 & 0.995 & 0.044 & 0.041 & 0.943 & 0.044 & 0 \\
 &  & IPTW, $W_2$ omitted & 0.002 & 0.993 & 0.042 & 0.039 & 0.928 & 0.042 & 0 \\
\addlinespace[3pt]
Three-level $T$, 2 vs.\ 0 & 800 & Naive & 0.411 & 0.000 & 0.107 & 0.085 & 0.001 & 0.425 & 0 \\
 &  & Mediator selection only & 0.040 & 0.902 & 0.099 & 0.091 & 0.894 & 0.106 & 0 \\
 &  & IPTW, known weights & $-$0.003 & 0.994 & 0.113 & 0.096 & 0.918 & 0.113 & 0 \\
 &  & IPTW, two-step & $-$0.003 & 0.994 & 0.113 & 0.102 & 0.933 & 0.113 & 0 \\
 &  & MMWS 5 strata & 0.001 & 0.998 & 0.106 & 0.094 & 0.924 & 0.106 & 0 \\
 &  & MMWS 10 strata & $-$0.001 & 0.997 & 0.113 & 0.097 & 0.911 & 0.113 & 0 \\
 &  & IPTW, $W_2$ omitted & $-$0.005 & 0.987 & 0.100 & 0.089 & 0.919 & 0.100 & 0 \\
\addlinespace[3pt]
Three-level $T$, 2 vs.\ 0 & 5,000 & Naive & 0.413 & 0.000 & 0.043 & 0.034 & 0.000 & 0.416 & 0 \\
 &  & Mediator selection only & 0.041 & 0.901 & 0.037 & 0.035 & 0.768 & 0.055 & 0 \\
 &  & IPTW, known weights & $-$0.001 & 0.998 & 0.039 & 0.035 & 0.924 & 0.039 & 0 \\
 &  & IPTW, two-step & $-$0.001 & 0.998 & 0.039 & 0.038 & 0.945 & 0.039 & 0 \\
 &  & MMWS 5 strata & 0.004 & 0.991 & 0.037 & 0.035 & 0.930 & 0.037 & 0 \\
 &  & MMWS 10 strata & 0.001 & 0.997 & 0.038 & 0.035 & 0.925 & 0.038 & 0 \\
 &  & IPTW, $W_2$ omitted & $-$0.001 & 0.997 & 0.035 & 0.032 & 0.929 & 0.035 & 0 \\
\addlinespace[3pt]
\multicolumn{10}{l}{\textit{B. Natural direct effect}}\\\addlinespace[2pt]
Binary $M$ & 800 & Naive & 0.164 & 0.000 & 0.107 & 0.118 & 0.737 & 0.196 & 0 \\
 &  & Mediator selection only & 0.502 & $-$2.059 & 0.131 & 0.131 & 0.033 & 0.519 & 0 \\
 &  & IPTW, known weights & 0.009 & 0.945 & 0.131 & 0.164 & 0.990 & 0.132 & 0 \\
 &  & IPTW, two-step & 0.009 & 0.945 & 0.131 & 0.120 & 0.928 & 0.132 & 0 \\
 &  & MMWS 5 strata & 0.058 & 0.646 & 0.122 & 0.153 & 0.982 & 0.135 & 0 \\
 &  & MMWS 10 strata & 0.029 & 0.823 & 0.125 & 0.159 & 0.983 & 0.128 & 0 \\
 &  & IPTW, $W_2$ omitted & 0.087 & 0.471 & 0.124 & 0.156 & 0.960 & 0.152 & 0 \\
\addlinespace[3pt]
Binary $M$ & 5,000 & Naive & 0.167 & 0.000 & 0.042 & 0.047 & 0.041 & 0.172 & 0 \\
 &  & Mediator selection only & 0.502 & $-$2.014 & 0.053 & 0.052 & 0.000 & 0.505 & 0 \\
 &  & IPTW, known weights & 0.001 & 0.997 & 0.052 & 0.066 & 0.987 & 0.052 & 0 \\
 &  & IPTW, two-step & 0.001 & 0.997 & 0.052 & 0.050 & 0.951 & 0.052 & 0 \\
 &  & MMWS 5 strata & 0.053 & 0.684 & 0.049 & 0.061 & 0.910 & 0.072 & 0 \\
 &  & MMWS 10 strata & 0.021 & 0.873 & 0.050 & 0.063 & 0.980 & 0.054 & 0 \\
 &  & IPTW, $W_2$ omitted & 0.078 & 0.529 & 0.049 & 0.063 & 0.814 & 0.093 & 0 \\
\addlinespace[3pt]
Continuous $M$ & 800 & Naive & $-$0.016 & 0.000 & 0.035 & 0.041 & 0.963 & 0.039 & 0 \\
 &  & Mediator selection only & 0.116 & $-$6.473 & 0.044 & 0.046 & 0.293 & 0.124 & 0 \\
 &  & IPTW, known weights & 0.007 & 0.563 & 0.048 & 0.054 & 0.972 & 0.048 & 0 \\
 &  & IPTW, two-step & 0.007 & 0.563 & 0.048 & 0.046 & 0.948 & 0.048 & 0 \\
 &  & MMWS 5 strata & 0.016 & $-$0.059 & 0.046 & 0.052 & 0.972 & 0.049 & 0 \\
 &  & MMWS 10 strata & 0.010 & 0.353 & 0.048 & 0.054 & 0.976 & 0.049 & 0 \\
 &  & IPTW, $W_2$ omitted & 0.019 & $-$0.222 & 0.046 & 0.053 & 0.967 & 0.050 & 0 \\
\addlinespace[3pt]
Continuous $M$ & 5,000 & Naive & $-$0.017 & 0.000 & 0.014 & 0.016 & 0.862 & 0.022 & 0 \\
 &  & Mediator selection only & 0.113 & $-$5.822 & 0.018 & 0.019 & 0.000 & 0.114 & 0 \\
 &  & IPTW, known weights & 0.001 & 0.933 & 0.019 & 0.022 & 0.985 & 0.019 & 0 \\
 &  & IPTW, two-step & 0.001 & 0.933 & 0.019 & 0.019 & 0.953 & 0.019 & 0 \\
 &  & MMWS 5 strata & 0.012 & 0.260 & 0.018 & 0.021 & 0.949 & 0.022 & 0 \\
 &  & MMWS 10 strata & 0.005 & 0.689 & 0.018 & 0.022 & 0.981 & 0.019 & 0 \\
 &  & IPTW, $W_2$ omitted & 0.015 & 0.120 & 0.018 & 0.021 & 0.935 & 0.023 & 0 \\
\addlinespace[3pt]
Three-level $T$, 1 vs.\ 0 & 800 & Naive & 0.284 & 0.000 & 0.126 & 0.136 & 0.446 & 0.311 & 0 \\
 &  & Mediator selection only & 0.534 & $-$0.879 & 0.152 & 0.146 & 0.045 & 0.556 & 0 \\
 &  & IPTW, known weights & 0.002 & 0.993 & 0.164 & 0.196 & 0.983 & 0.164 & 0 \\
 &  & IPTW, two-step & 0.002 & 0.993 & 0.164 & 0.146 & 0.921 & 0.164 & 0 \\
 &  & MMWS 5 strata & 0.044 & 0.846 & 0.155 & 0.189 & 0.977 & 0.161 & 0 \\
 &  & MMWS 10 strata & 0.014 & 0.952 & 0.163 & 0.198 & 0.983 & 0.163 & 0 \\
 &  & IPTW, $W_2$ omitted & 0.074 & 0.740 & 0.151 & 0.186 & 0.972 & 0.168 & 0 \\
\addlinespace[3pt]
Three-level $T$, 1 vs.\ 0 & 5,000 & Naive & 0.294 & 0.000 & 0.048 & 0.055 & 0.000 & 0.298 & 0 \\
 &  & Mediator selection only & 0.549 & $-$0.866 & 0.055 & 0.059 & 0.000 & 0.552 & 0 \\
 &  & IPTW, known weights & 0.001 & 0.995 & 0.072 & 0.084 & 0.986 & 0.072 & 0 \\
 &  & IPTW, two-step & 0.001 & 0.995 & 0.072 & 0.065 & 0.927 & 0.072 & 0 \\
 &  & MMWS 5 strata & 0.047 & 0.840 & 0.063 & 0.078 & 0.953 & 0.078 & 0 \\
 &  & MMWS 10 strata & 0.016 & 0.944 & 0.066 & 0.082 & 0.981 & 0.068 & 0 \\
 &  & IPTW, $W_2$ omitted & 0.077 & 0.739 & 0.063 & 0.078 & 0.885 & 0.099 & 0 \\
\addlinespace[3pt]
Three-level $T$, 2 vs.\ 0 & 800 & Naive & 0.066 & 0.000 & 0.149 & 0.161 & 0.945 & 0.163 & 0 \\
 &  & Mediator selection only & 0.438 & $-$5.590 & 0.159 & 0.165 & 0.239 & 0.465 & 0 \\
 &  & IPTW, known weights & 0.000 & 0.999 & 0.167 & 0.203 & 0.984 & 0.167 & 0 \\
 &  & IPTW, two-step & 0.000 & 0.999 & 0.167 & 0.154 & 0.934 & 0.167 & 0 \\
 &  & MMWS 5 strata & 0.037 & 0.443 & 0.157 & 0.195 & 0.986 & 0.161 & 0 \\
 &  & MMWS 10 strata & 0.008 & 0.884 & 0.167 & 0.204 & 0.983 & 0.168 & 0 \\
 &  & IPTW, $W_2$ omitted & $-$0.066 & 0.005 & 0.153 & 0.191 & 0.984 & 0.167 & 0 \\
\addlinespace[3pt]
Three-level $T$, 2 vs.\ 0 & 5,000 & Naive & 0.070 & 0.000 & 0.058 & 0.064 & 0.836 & 0.091 & 0 \\
 &  & Mediator selection only & 0.443 & $-$5.291 & 0.062 & 0.065 & 0.000 & 0.447 & 0 \\
 &  & IPTW, known weights & 0.000 & 0.994 & 0.067 & 0.081 & 0.989 & 0.067 & 0 \\
 &  & IPTW, two-step & 0.000 & 0.994 & 0.067 & 0.063 & 0.929 & 0.067 & 0 \\
 &  & MMWS 5 strata & 0.039 & 0.452 & 0.060 & 0.077 & 0.965 & 0.071 & 0 \\
 &  & MMWS 10 strata & 0.014 & 0.808 & 0.063 & 0.079 & 0.984 & 0.064 & 0 \\
 &  & IPTW, $W_2$ omitted & $-$0.070 & 0.007 & 0.059 & 0.076 & 0.908 & 0.092 & 0 \\
\addlinespace[3pt]
\end{longtable}\end{footnotesize}
\vspace{-6pt}\begin{footnotesize}\noindent Bias, mean estimate minus the true value; EmpSE, empirical standard error; ModSE, mean model standard error; Cov, coverage of nominal 95\% intervals; RMSE, root mean squared error; Fail, replications without an estimate. \%BR, proportion of the naive estimator's bias removed. Binary mediator: binary treatment and mediator; continuous mediator: binary treatment, continuous mediator, binary outcome; three-level: three-level treatment, binary mediator. MMWS and the misspecified model use standard errors that treat the weights as known. Maximum Monte Carlo standard errors: bias 0.005; EmpSE 0.004; ModSE 0.001; coverage 0.016; RMSE 0.005.\end{footnotesize}

\clearpage
\begin{footnotesize}\setlength{\tabcolsep}{4pt}
\begin{longtable}{lllrrrrrr}
\caption{Ordinal treatment: alternative treatment and mediator-weight models, with the treatment generated by an ordered logit model (two-step standard errors); 1,000 replications per cell.}\label{tab:study3c}\\
\toprule
Design & $N$ & Method & Bias & EmpSE & ModSE & Cov & RMSE & Fail \\
\midrule
\endfirsthead
\multicolumn{9}{l}{\textit{(continued)}}\\
\toprule
Design & $N$ & Method & Bias & EmpSE & ModSE & Cov & RMSE & Fail \\
\midrule
\endhead
\bottomrule
\endfoot
\multicolumn{9}{l}{\textit{A. Natural indirect effect}}\\\addlinespace[2pt]
1 vs.\ 0 & 800 & IPTW, mlogit & 0.000 & 0.098 & 0.083 & 0.934 & 0.098 & 0 \\
 &  & IPTW, ologit & 0.001 & 0.099 & 0.085 & 0.941 & 0.099 & 0 \\
 &  & IPTW, oprobit & 0.001 & 0.099 & 0.085 & 0.942 & 0.099 & 0 \\
 &  & Odds, mlogit & 0.003 & 0.052 & 0.052 & 0.947 & 0.052 & 0 \\
 &  & Odds, ologit & 0.017 & 0.043 & 0.043 & 0.951 & 0.046 & 0 \\
\addlinespace[3pt]
1 vs.\ 0 & 5,000 & IPTW, mlogit & 0.000 & 0.034 & 0.034 & 0.950 & 0.034 & 0 \\
 &  & IPTW, ologit & 0.000 & 0.034 & 0.035 & 0.956 & 0.034 & 0 \\
 &  & IPTW, oprobit & 0.000 & 0.034 & 0.035 & 0.954 & 0.034 & 0 \\
 &  & Odds, mlogit & 0.000 & 0.020 & 0.020 & 0.953 & 0.020 & 0 \\
 &  & Odds, ologit & 0.016 & 0.017 & 0.017 & 0.859 & 0.023 & 0 \\
\addlinespace[3pt]
2 vs.\ 0 & 800 & IPTW, mlogit & $-$0.005 & 0.097 & 0.091 & 0.933 & 0.097 & 0 \\
 &  & IPTW, ologit & $-$0.004 & 0.095 & 0.090 & 0.938 & 0.095 & 0 \\
 &  & IPTW, oprobit & $-$0.005 & 0.096 & 0.091 & 0.935 & 0.096 & 0 \\
 &  & Odds, mlogit & 0.020 & 0.089 & 0.086 & 0.945 & 0.091 & 0 \\
 &  & Odds, ologit & 0.012 & 0.087 & 0.085 & 0.955 & 0.088 & 0 \\
\addlinespace[3pt]
2 vs.\ 0 & 5,000 & IPTW, mlogit & $-$0.001 & 0.036 & 0.035 & 0.947 & 0.036 & 0 \\
 &  & IPTW, ologit & $-$0.001 & 0.036 & 0.034 & 0.946 & 0.036 & 0 \\
 &  & IPTW, oprobit & $-$0.001 & 0.036 & 0.035 & 0.948 & 0.036 & 0 \\
 &  & Odds, mlogit & 0.020 & 0.034 & 0.034 & 0.920 & 0.039 & 0 \\
 &  & Odds, ologit & 0.013 & 0.033 & 0.034 & 0.954 & 0.036 & 0 \\
\addlinespace[3pt]
\multicolumn{9}{l}{\textit{B. Natural direct effect}}\\\addlinespace[2pt]
1 vs.\ 0 & 800 & IPTW, mlogit & $-$0.008 & 0.136 & 0.123 & 0.931 & 0.137 & 0 \\
 &  & IPTW, ologit & $-$0.005 & 0.163 & 0.150 & 0.938 & 0.163 & 0 \\
 &  & IPTW, oprobit & $-$0.003 & 0.163 & 0.150 & 0.943 & 0.163 & 0 \\
 &  & Odds, mlogit & $-$0.010 & 0.107 & 0.106 & 0.947 & 0.107 & 0 \\
 &  & Odds, ologit & $-$0.021 & 0.138 & 0.137 & 0.947 & 0.140 & 0 \\
\addlinespace[3pt]
1 vs.\ 0 & 5,000 & IPTW, mlogit & $-$0.007 & 0.050 & 0.050 & 0.948 & 0.051 & 0 \\
 &  & IPTW, ologit & $-$0.002 & 0.062 & 0.061 & 0.947 & 0.062 & 0 \\
 &  & IPTW, oprobit & 0.001 & 0.062 & 0.061 & 0.945 & 0.062 & 0 \\
 &  & Odds, mlogit & $-$0.007 & 0.043 & 0.042 & 0.950 & 0.043 & 0 \\
 &  & Odds, ologit & $-$0.018 & 0.055 & 0.055 & 0.938 & 0.058 & 0 \\
\addlinespace[3pt]
2 vs.\ 0 & 800 & IPTW, mlogit & $-$0.011 & 0.141 & 0.128 & 0.934 & 0.141 & 0 \\
 &  & IPTW, ologit & 0.003 & 0.141 & 0.128 & 0.930 & 0.141 & 0 \\
 &  & IPTW, oprobit & $-$0.008 & 0.142 & 0.128 & 0.932 & 0.142 & 0 \\
 &  & Odds, mlogit & $-$0.036 & 0.140 & 0.130 & 0.935 & 0.144 & 0 \\
 &  & Odds, ologit & $-$0.014 & 0.139 & 0.130 & 0.936 & 0.140 & 0 \\
\addlinespace[3pt]
2 vs.\ 0 & 5,000 & IPTW, mlogit & $-$0.013 & 0.054 & 0.051 & 0.937 & 0.055 & 0 \\
 &  & IPTW, ologit & 0.001 & 0.054 & 0.051 & 0.941 & 0.054 & 0 \\
 &  & IPTW, oprobit & $-$0.011 & 0.054 & 0.051 & 0.943 & 0.055 & 0 \\
 &  & Odds, mlogit & $-$0.035 & 0.053 & 0.052 & 0.897 & 0.063 & 0 \\
 &  & Odds, ologit & $-$0.013 & 0.053 & 0.052 & 0.949 & 0.054 & 0 \\
\addlinespace[3pt]
\end{longtable}\end{footnotesize}
\vspace{-6pt}\begin{footnotesize}\noindent Bias, mean estimate minus the true value; EmpSE, empirical standard error; ModSE, mean model standard error; Cov, coverage of nominal 95\% intervals; RMSE, root mean squared error; Fail, replications without an estimate. True values: 1 vs.\ 0, NIE = 0.16, NDE = 0.38; 2 vs.\ 0, NIE = 0.35, NDE = 0.66. Maximum Monte Carlo standard errors: bias 0.005; EmpSE 0.004; ModSE 0.001; coverage 0.011; RMSE 0.004.\end{footnotesize}

\clearpage
\begin{footnotesize}\setlength{\tabcolsep}{4pt}
\begin{longtable}{lllrrrrrrrr}
\caption{Treatment-model specification and weight stability: alternative fitted treatment models (IPTW, two-step standard errors); rows by data-generating treatment model; 1,000 replications per cell.}\label{tab:study3b}\\
\toprule
Design & $N$ & Method & Bias & EmpSE & ModSE & Cov & RMSE & ESS & ESS$<$5\% & Fail \\
\midrule
\endfirsthead
\multicolumn{11}{l}{\textit{(continued)}}\\
\toprule
Design & $N$ & Method & Bias & EmpSE & ModSE & Cov & RMSE & ESS & ESS$<$5\% & Fail \\
\midrule
\endhead
\bottomrule
\endfoot
\multicolumn{11}{l}{\textit{A. Natural indirect effect}}\\\addlinespace[2pt]
Logit & 800 & Logit & $-$0.008 & 0.096 & 0.082 & 0.939 & 0.097 & -- & -- & 0 \\
 &  & Probit & $-$0.010 & 0.099 & 0.084 & 0.940 & 0.100 & -- & -- & 0 \\
 &  & Het.\ probit & $-$0.010 & 0.099 & 0.083 & 0.940 & 0.100 & -- & -- & 0 \\
\addlinespace[3pt]
Logit & 5,000 & Logit & $-$0.003 & 0.035 & 0.034 & 0.952 & 0.035 & -- & -- & 0 \\
 &  & Probit & $-$0.004 & 0.036 & 0.035 & 0.950 & 0.036 & -- & -- & 0 \\
 &  & Het.\ probit & $-$0.004 & 0.036 & 0.035 & 0.955 & 0.036 & -- & -- & 0 \\
\addlinespace[3pt]
Probit & 800 & Logit & $-$0.008 & 0.094 & 0.082 & 0.936 & 0.095 & -- & -- & 0 \\
 &  & Probit & $-$0.009 & 0.096 & 0.083 & 0.936 & 0.096 & -- & -- & 0 \\
 &  & Het.\ probit & $-$0.009 & 0.096 & 0.082 & 0.933 & 0.096 & -- & -- & 0 \\
\addlinespace[3pt]
Probit & 5,000 & Logit & $-$0.004 & 0.034 & 0.034 & 0.953 & 0.034 & -- & -- & 0 \\
 &  & Probit & $-$0.004 & 0.036 & 0.035 & 0.951 & 0.036 & -- & -- & 0 \\
 &  & Het.\ probit & $-$0.004 & 0.035 & 0.035 & 0.952 & 0.035 & -- & -- & 0 \\
\addlinespace[3pt]
Het.\ probit, $g=0.50$ & 800 & Logit & $-$0.007 & 0.098 & 0.083 & 0.923 & 0.099 & 29.6 & 0.001 & 0 \\
 &  & Probit & $-$0.007 & 0.099 & 0.083 & 0.921 & 0.099 & 28.0 & 0.004 & 0 \\
 &  & Het.\ probit & $-$0.018 & 0.119 & 0.092 & 0.914 & 0.120 & 26.1 & 0.021 & 1 \\
\addlinespace[3pt]
Het.\ probit, $g=0.50$ & 5,000 & Logit & $-$0.003 & 0.037 & 0.035 & 0.937 & 0.037 & 29.9 & 0.001 & 0 \\
 &  & Probit & $-$0.003 & 0.038 & 0.035 & 0.935 & 0.038 & 26.2 & 0.010 & 0 \\
 &  & Het.\ probit & $-$0.009 & 0.060 & 0.044 & 0.928 & 0.061 & 19.2 & 0.075 & 0 \\
\addlinespace[3pt]
Het.\ probit, $g=0.25$ & 800 & Logit & $-$0.004 & 0.090 & 0.080 & 0.938 & 0.090 & 29.8 & 0.000 & 0 \\
 &  & Probit & $-$0.004 & 0.091 & 0.081 & 0.935 & 0.091 & 28.2 & 0.006 & 0 \\
 &  & Het.\ probit & $-$0.013 & 0.099 & 0.086 & 0.931 & 0.100 & 27.3 & 0.009 & 0 \\
\addlinespace[3pt]
Het.\ probit, $g=0.25$ & 5,000 & Logit & $-$0.003 & 0.036 & 0.034 & 0.944 & 0.036 & 30.0 & 0.001 & 0 \\
 &  & Probit & $-$0.003 & 0.036 & 0.035 & 0.948 & 0.036 & 26.5 & 0.011 & 1 \\
 &  & Het.\ probit & $-$0.008 & 0.045 & 0.039 & 0.951 & 0.046 & 22.3 & 0.027 & 0 \\
\addlinespace[3pt]
Het.\ probit, $g=0.10$ & 800 & Logit & $-$0.010 & 0.090 & 0.082 & 0.938 & 0.091 & 30.1 & 0.000 & 0 \\
 &  & Probit & $-$0.011 & 0.091 & 0.082 & 0.938 & 0.092 & 28.9 & 0.001 & 0 \\
 &  & Het.\ probit & $-$0.014 & 0.095 & 0.084 & 0.935 & 0.096 & 28.7 & 0.003 & 0 \\
\addlinespace[3pt]
Het.\ probit, $g=0.10$ & 5,000 & Logit & 0.000 & 0.036 & 0.034 & 0.945 & 0.036 & 30.3 & 0.000 & 0 \\
 &  & Probit & 0.000 & 0.036 & 0.035 & 0.946 & 0.036 & 28.2 & 0.001 & 0 \\
 &  & Het.\ probit & $-$0.003 & 0.039 & 0.036 & 0.940 & 0.039 & 25.7 & 0.008 & 0 \\
\addlinespace[3pt]
\multicolumn{11}{l}{\textit{B. Natural direct effect}}\\\addlinespace[2pt]
Logit & 800 & Logit & 0.012 & 0.134 & 0.119 & 0.927 & 0.135 & -- & -- & 0 \\
 &  & Probit & 0.001 & 0.142 & 0.122 & 0.920 & 0.142 & -- & -- & 0 \\
 &  & Het.\ probit & 0.007 & 0.142 & 0.120 & 0.919 & 0.142 & -- & -- & 0 \\
\addlinespace[3pt]
Logit & 5,000 & Logit & 0.003 & 0.052 & 0.050 & 0.936 & 0.052 & -- & -- & 0 \\
 &  & Probit & $-$0.010 & 0.056 & 0.053 & 0.929 & 0.057 & -- & -- & 0 \\
 &  & Het.\ probit & $-$0.009 & 0.057 & 0.052 & 0.930 & 0.057 & -- & -- & 0 \\
\addlinespace[3pt]
Probit & 800 & Logit & 0.020 & 0.130 & 0.118 & 0.928 & 0.132 & -- & -- & 0 \\
 &  & Probit & 0.011 & 0.136 & 0.121 & 0.925 & 0.136 & -- & -- & 0 \\
 &  & Het.\ probit & 0.014 & 0.137 & 0.120 & 0.920 & 0.137 & -- & -- & 0 \\
\addlinespace[3pt]
Probit & 5,000 & Logit & 0.017 & 0.049 & 0.050 & 0.940 & 0.052 & -- & -- & 0 \\
 &  & Probit & 0.004 & 0.054 & 0.053 & 0.959 & 0.054 & -- & -- & 0 \\
 &  & Het.\ probit & 0.004 & 0.053 & 0.052 & 0.953 & 0.053 & -- & -- & 0 \\
\addlinespace[3pt]
Het.\ probit, $g=0.50$ & 800 & Logit & 0.003 & 0.140 & 0.126 & 0.931 & 0.140 & 29.6 & 0.001 & 0 \\
 &  & Probit & $-$0.013 & 0.166 & 0.131 & 0.927 & 0.166 & 28.0 & 0.004 & 0 \\
 &  & Het.\ probit & 0.069 & 0.167 & 0.129 & 0.876 & 0.180 & 26.1 & 0.021 & 1 \\
\addlinespace[3pt]
Het.\ probit, $g=0.50$ & 5,000 & Logit & $-$0.002 & 0.054 & 0.053 & 0.949 & 0.054 & 29.9 & 0.001 & 0 \\
 &  & Probit & $-$0.023 & 0.068 & 0.060 & 0.949 & 0.072 & 26.2 & 0.010 & 0 \\
 &  & Het.\ probit & 0.041 & 0.091 & 0.064 & 0.806 & 0.100 & 19.2 & 0.075 & 0 \\
\addlinespace[3pt]
Het.\ probit, $g=0.25$ & 800 & Logit & 0.006 & 0.132 & 0.121 & 0.935 & 0.132 & 29.8 & 0.000 & 0 \\
 &  & Probit & $-$0.009 & 0.149 & 0.127 & 0.936 & 0.149 & 28.2 & 0.006 & 0 \\
 &  & Het.\ probit & 0.030 & 0.144 & 0.121 & 0.913 & 0.147 & 27.3 & 0.009 & 0 \\
\addlinespace[3pt]
Het.\ probit, $g=0.25$ & 5,000 & Logit & 0.006 & 0.051 & 0.052 & 0.954 & 0.052 & 30.0 & 0.001 & 0 \\
 &  & Probit & $-$0.016 & 0.072 & 0.060 & 0.958 & 0.073 & 26.5 & 0.011 & 1 \\
 &  & Het.\ probit & 0.017 & 0.070 & 0.058 & 0.913 & 0.072 & 22.3 & 0.027 & 0 \\
\addlinespace[3pt]
Het.\ probit, $g=0.10$ & 800 & Logit & 0.023 & 0.129 & 0.119 & 0.922 & 0.131 & 30.1 & 0.000 & 0 \\
 &  & Probit & 0.013 & 0.136 & 0.121 & 0.927 & 0.136 & 28.9 & 0.001 & 0 \\
 &  & Het.\ probit & 0.026 & 0.133 & 0.118 & 0.918 & 0.135 & 28.7 & 0.003 & 0 \\
\addlinespace[3pt]
Het.\ probit, $g=0.10$ & 5,000 & Logit & 0.012 & 0.051 & 0.050 & 0.939 & 0.053 & 30.3 & 0.000 & 0 \\
 &  & Probit & $-$0.003 & 0.057 & 0.054 & 0.949 & 0.057 & 28.2 & 0.001 & 0 \\
 &  & Het.\ probit & 0.003 & 0.061 & 0.054 & 0.937 & 0.061 & 25.7 & 0.008 & 0 \\
\addlinespace[3pt]
\end{longtable}\end{footnotesize}
\vspace{-6pt}\begin{footnotesize}\noindent Bias, mean estimate minus the true value; EmpSE, empirical standard error; ModSE, mean model standard error; Cov, coverage of nominal 95\% intervals; RMSE, root mean squared error; Fail, replications without an estimate. Design: data-generating treatment model (Het. probit, heteroskedastic probit with standard deviation $\exp(gW_1)$); Method: fitted treatment model. ESS, mean of the smallest Kish effective sample size of any copy (\% of its observations); ESS$<$5\%, proportion of replications in which some copy's ESS fell below 5\% (recorded for the heteroskedastic probit designs only). True NIE = 0.242, NDE = 0.418. Maximum Monte Carlo standard errors: bias 0.005; EmpSE 0.004; ModSE 0.002; coverage 0.013; RMSE 0.015.\end{footnotesize}

\clearpage
\begin{footnotesize}\setlength{\tabcolsep}{4pt}
\begin{longtable}{lllrrrrrrrr}
\caption{Heteroskedastic continuous mediator: alternative mediator-weight models, $\ln\text{var}(M\mid T,X) = 1.5cX_1 + cTX_2$ (IPTW, two-step standard errors); 1,000 replications per cell.}\label{tab:study4}\\
\toprule
Design & $N$ & Method & Bias & EmpSE & ModSE & Cov & RMSE & ESS & ESS$<$5\% & Fail \\
\midrule
\endfirsthead
\multicolumn{11}{l}{\textit{(continued)}}\\
\toprule
Design & $N$ & Method & Bias & EmpSE & ModSE & Cov & RMSE & ESS & ESS$<$5\% & Fail \\
\midrule
\endhead
\bottomrule
\endfoot
\multicolumn{11}{l}{\textit{A. Natural indirect effect}}\\\addlinespace[2pt]
$c=0.40$ & 800 & Constant variance & $-$0.235 & 0.207 & 0.170 & 0.478 & 0.313 & 18.5 & 0.129 & 11 \\
 &  & Heteroskedastic & $-$0.095 & 0.234 & 0.185 & 0.758 & 0.252 & 19.3 & 0.049 & 2 \\
 &  & Heteroskedastic ($X_1, X_2$) & $-$0.090 & 0.245 & 0.179 & 0.753 & 0.261 & 19.6 & 0.056 & 2 \\
 &  & Odds & $-$0.144 & 0.264 & 0.168 & 0.636 & 0.301 & 16.0 & 0.125 & 2 \\
\addlinespace[3pt]
$c=0.40$ & 5,000 & Constant variance & $-$0.201 & 0.103 & 0.079 & 0.172 & 0.226 & 11.7 & 0.198 & 2 \\
 &  & Heteroskedastic & $-$0.036 & 0.146 & 0.096 & 0.751 & 0.150 & 15.5 & 0.073 & 1 \\
 &  & Heteroskedastic ($X_1, X_2$) & $-$0.038 & 0.145 & 0.097 & 0.746 & 0.150 & 15.5 & 0.074 & 1 \\
 &  & Odds & $-$0.085 & 0.266 & 0.124 & 0.544 & 0.280 & 9.6 & 0.272 & 1 \\
\addlinespace[3pt]
$c=0.20$ & 800 & Constant variance & $-$0.204 & 0.155 & 0.130 & 0.554 & 0.256 & 23.0 & 0.026 & 1 \\
 &  & Heteroskedastic & $-$0.067 & 0.190 & 0.161 & 0.837 & 0.201 & 22.5 & 0.030 & 2 \\
 &  & Heteroskedastic ($X_1, X_2$) & $-$0.069 & 0.184 & 0.151 & 0.839 & 0.196 & 22.8 & 0.024 & 1 \\
 &  & Odds & $-$0.158 & 0.173 & 0.139 & 0.648 & 0.234 & 21.4 & 0.022 & 0 \\
\addlinespace[3pt]
$c=0.20$ & 5,000 & Constant variance & $-$0.166 & 0.059 & 0.054 & 0.173 & 0.176 & 19.6 & 0.022 & 0 \\
 &  & Heteroskedastic & $-$0.010 & 0.129 & 0.073 & 0.875 & 0.129 & 20.8 & 0.018 & 0 \\
 &  & Heteroskedastic ($X_1, X_2$) & $-$0.011 & 0.102 & 0.073 & 0.883 & 0.102 & 21.1 & 0.018 & 0 \\
 &  & Odds & $-$0.142 & 0.075 & 0.067 & 0.368 & 0.161 & 19.0 & 0.022 & 0 \\
\addlinespace[3pt]
$c=0.10$ & 800 & Constant variance & $-$0.128 & 0.145 & 0.130 & 0.751 & 0.193 & 24.2 & 0.009 & 0 \\
 &  & Heteroskedastic & $-$0.043 & 0.179 & 0.152 & 0.897 & 0.184 & 23.2 & 0.027 & 1 \\
 &  & Heteroskedastic ($X_1, X_2$) & $-$0.047 & 0.171 & 0.144 & 0.878 & 0.177 & 23.6 & 0.017 & 0 \\
 &  & Odds & $-$0.094 & 0.169 & 0.143 & 0.788 & 0.194 & 23.7 & 0.014 & 0 \\
\addlinespace[3pt]
$c=0.10$ & 5,000 & Constant variance & $-$0.101 & 0.056 & 0.053 & 0.471 & 0.116 & 22.4 & 0.006 & 0 \\
 &  & Heteroskedastic & $-$0.012 & 0.060 & 0.057 & 0.912 & 0.061 & 22.8 & 0.005 & 0 \\
 &  & Heteroskedastic ($X_1, X_2$) & $-$0.012 & 0.063 & 0.059 & 0.909 & 0.064 & 23.3 & 0.004 & 0 \\
 &  & Odds & $-$0.085 & 0.075 & 0.066 & 0.640 & 0.114 & 21.8 & 0.006 & 0 \\
\addlinespace[3pt]
$c=0.00$ & 800 & Constant variance & $-$0.040 & 0.153 & 0.138 & 0.908 & 0.158 & 25.0 & 0.008 & 0 \\
 &  & Heteroskedastic & $-$0.038 & 0.193 & 0.150 & 0.905 & 0.197 & 23.6 & 0.025 & 2 \\
 &  & Heteroskedastic ($X_1, X_2$) & $-$0.041 & 0.162 & 0.143 & 0.906 & 0.167 & 23.9 & 0.019 & 2 \\
 &  & Odds & $-$0.011 & 0.196 & 0.155 & 0.888 & 0.196 & 24.9 & 0.012 & 0 \\
\addlinespace[3pt]
$c=0.00$ & 5,000 & Constant variance & $-$0.011 & 0.062 & 0.057 & 0.912 & 0.063 & 24.1 & 0.003 & 0 \\
 &  & Heteroskedastic & $-$0.011 & 0.061 & 0.054 & 0.917 & 0.062 & 23.6 & 0.003 & 0 \\
 &  & Heteroskedastic ($X_1, X_2$) & $-$0.013 & 0.058 & 0.054 & 0.908 & 0.060 & 24.0 & 0.002 & 0 \\
 &  & Odds & $-$0.004 & 0.075 & 0.070 & 0.921 & 0.075 & 23.6 & 0.002 & 0 \\
\addlinespace[3pt]
\multicolumn{11}{l}{\textit{B. Natural direct effect}}\\\addlinespace[2pt]
$c=0.40$ & 800 & Constant variance & 0.237 & 0.228 & 0.193 & 0.609 & 0.329 & 18.5 & 0.129 & 11 \\
 &  & Heteroskedastic & 0.097 & 0.250 & 0.198 & 0.773 & 0.268 & 19.3 & 0.049 & 2 \\
 &  & Heteroskedastic ($X_1, X_2$) & 0.093 & 0.263 & 0.193 & 0.785 & 0.279 & 19.6 & 0.056 & 2 \\
 &  & Odds & 0.147 & 0.285 & 0.192 & 0.704 & 0.321 & 16.0 & 0.125 & 2 \\
\addlinespace[3pt]
$c=0.40$ & 5,000 & Constant variance & 0.201 & 0.110 & 0.087 & 0.212 & 0.229 & 11.7 & 0.198 & 2 \\
 &  & Heteroskedastic & 0.036 & 0.150 & 0.102 & 0.776 & 0.154 & 15.5 & 0.073 & 1 \\
 &  & Heteroskedastic ($X_1, X_2$) & 0.039 & 0.148 & 0.103 & 0.774 & 0.153 & 15.5 & 0.074 & 1 \\
 &  & Odds & 0.085 & 0.268 & 0.130 & 0.586 & 0.281 & 9.6 & 0.272 & 1 \\
\addlinespace[3pt]
$c=0.20$ & 800 & Constant variance & 0.203 & 0.179 & 0.157 & 0.658 & 0.271 & 23.0 & 0.026 & 1 \\
 &  & Heteroskedastic & 0.069 & 0.213 & 0.178 & 0.851 & 0.223 & 22.5 & 0.030 & 2 \\
 &  & Heteroskedastic ($X_1, X_2$) & 0.070 & 0.202 & 0.168 & 0.849 & 0.214 & 22.8 & 0.024 & 1 \\
 &  & Odds & 0.158 & 0.217 & 0.168 & 0.735 & 0.269 & 21.4 & 0.022 & 0 \\
\addlinespace[3pt]
$c=0.20$ & 5,000 & Constant variance & 0.167 & 0.069 & 0.067 & 0.272 & 0.181 & 19.6 & 0.022 & 0 \\
 &  & Heteroskedastic & 0.012 & 0.131 & 0.081 & 0.899 & 0.132 & 20.8 & 0.018 & 0 \\
 &  & Heteroskedastic ($X_1, X_2$) & 0.013 & 0.105 & 0.081 & 0.896 & 0.106 & 21.1 & 0.018 & 0 \\
 &  & Odds & 0.144 & 0.083 & 0.078 & 0.460 & 0.166 & 19.0 & 0.022 & 0 \\
\addlinespace[3pt]
$c=0.10$ & 800 & Constant variance & 0.131 & 0.165 & 0.154 & 0.811 & 0.211 & 24.2 & 0.009 & 0 \\
 &  & Heteroskedastic & 0.046 & 0.191 & 0.171 & 0.902 & 0.196 & 23.2 & 0.027 & 1 \\
 &  & Heteroskedastic ($X_1, X_2$) & 0.050 & 0.185 & 0.163 & 0.898 & 0.192 & 23.6 & 0.017 & 0 \\
 &  & Odds & 0.097 & 0.196 & 0.170 & 0.834 & 0.218 & 23.7 & 0.014 & 0 \\
\addlinespace[3pt]
$c=0.10$ & 5,000 & Constant variance & 0.103 & 0.067 & 0.064 & 0.591 & 0.122 & 22.4 & 0.006 & 0 \\
 &  & Heteroskedastic & 0.013 & 0.069 & 0.067 & 0.930 & 0.070 & 22.8 & 0.005 & 0 \\
 &  & Heteroskedastic ($X_1, X_2$) & 0.014 & 0.071 & 0.068 & 0.928 & 0.072 & 23.3 & 0.004 & 0 \\
 &  & Odds & 0.087 & 0.084 & 0.077 & 0.700 & 0.121 & 21.8 & 0.006 & 0 \\
\addlinespace[3pt]
$c=0.00$ & 800 & Constant variance & 0.046 & 0.180 & 0.159 & 0.898 & 0.186 & 25.0 & 0.008 & 0 \\
 &  & Heteroskedastic & 0.045 & 0.213 & 0.168 & 0.894 & 0.217 & 23.6 & 0.025 & 2 \\
 &  & Heteroskedastic ($X_1, X_2$) & 0.048 & 0.187 & 0.163 & 0.899 & 0.193 & 23.9 & 0.019 & 2 \\
 &  & Odds & 0.017 & 0.221 & 0.180 & 0.891 & 0.222 & 24.9 & 0.012 & 0 \\
\addlinespace[3pt]
$c=0.00$ & 5,000 & Constant variance & 0.009 & 0.071 & 0.066 & 0.929 & 0.071 & 24.1 & 0.003 & 0 \\
 &  & Heteroskedastic & 0.009 & 0.070 & 0.064 & 0.923 & 0.071 & 23.6 & 0.003 & 0 \\
 &  & Heteroskedastic ($X_1, X_2$) & 0.011 & 0.068 & 0.064 & 0.927 & 0.068 & 24.0 & 0.002 & 0 \\
 &  & Odds & 0.002 & 0.085 & 0.080 & 0.936 & 0.085 & 23.6 & 0.002 & 0 \\
\addlinespace[3pt]
\end{longtable}\end{footnotesize}
\vspace{-6pt}\begin{footnotesize}\noindent Bias, mean estimate minus the true value; EmpSE, empirical standard error; ModSE, mean model standard error; Cov, coverage of nominal 95\% intervals; RMSE, root mean squared error; Fail, replications without an estimate. ESS, mean of the smallest Kish effective sample size of any copy (\% of its observations); ESS$<$5\%, proportion of replications in which some copy's ESS fell below 5\%. True NIE = 0.70, NDE = 0.30. Maximum Monte Carlo standard errors: bias 0.009; EmpSE 0.006; ModSE 0.007; coverage 0.016; RMSE 0.029.\end{footnotesize}

\clearpage
\begin{footnotesize}\setlength{\tabcolsep}{4pt}
\begin{longtable}{lllrrrrrr}
\caption{Heteroskedastic continuous mediator: two-step versus bootstrap standard errors (same simulated data sets for each pair); 500 replications per cell, 200 bootstrap resamples.}\label{tab:study4boot}\\
\toprule
Design & $N$ & Method & Bias & EmpSE & ModSE & Cov & RMSE & Fail \\
\midrule
\endfirsthead
\multicolumn{9}{l}{\textit{(continued)}}\\
\toprule
Design & $N$ & Method & Bias & EmpSE & ModSE & Cov & RMSE & Fail \\
\midrule
\endhead
\bottomrule
\endfoot
\multicolumn{9}{l}{\textit{A. Natural indirect effect}}\\\addlinespace[2pt]
$c=0.40$ & 800 & Heteroskedastic, two-step & $-$0.088 & 0.241 & 0.189 & 0.784 & 0.256 & 0 \\
 &  & Heteroskedastic, bootstrap & $-$0.088 & 0.241 & 0.225 & 0.862 & 0.256 & 0 \\
\addlinespace[3pt]
$c=0.40$ & 5,000 & Heteroskedastic, two-step & $-$0.024 & 0.213 & 0.104 & 0.766 & 0.214 & 0 \\
 &  & Heteroskedastic, bootstrap & $-$0.024 & 0.213 & 0.114 & 0.786 & 0.214 & 0 \\
\addlinespace[3pt]
$c=0.20$ & 800 & Heteroskedastic, two-step & $-$0.057 & 0.185 & 0.163 & 0.844 & 0.193 & 0 \\
 &  & Heteroskedastic, bootstrap & $-$0.057 & 0.185 & 0.202 & 0.896 & 0.193 & 0 \\
\addlinespace[3pt]
$c=0.20$ & 5,000 & Heteroskedastic, two-step & $-$0.003 & 0.126 & 0.073 & 0.876 & 0.126 & 0 \\
 &  & Heteroskedastic, bootstrap & $-$0.003 & 0.126 & 0.081 & 0.898 & 0.126 & 0 \\
\addlinespace[3pt]
$c=0.10$ & 800 & Heteroskedastic, two-step & $-$0.048 & 0.168 & 0.152 & 0.882 & 0.175 & 0 \\
 &  & Heteroskedastic, bootstrap & $-$0.048 & 0.168 & 0.194 & 0.932 & 0.175 & 0 \\
\addlinespace[3pt]
$c=0.10$ & 5,000 & Heteroskedastic, two-step & $-$0.010 & 0.063 & 0.058 & 0.898 & 0.064 & 0 \\
 &  & Heteroskedastic, bootstrap & $-$0.010 & 0.063 & 0.066 & 0.922 & 0.064 & 0 \\
\addlinespace[3pt]
$c=0.00$ & 800 & Heteroskedastic, two-step & $-$0.034 & 0.183 & 0.153 & 0.892 & 0.186 & 0 \\
 &  & Heteroskedastic, bootstrap & $-$0.034 & 0.183 & 0.191 & 0.954 & 0.186 & 0 \\
 &  & Constant variance, two-step & $-$0.041 & 0.173 & 0.137 & 0.880 & 0.177 & 0 \\
 &  & Constant variance, bootstrap & $-$0.041 & 0.173 & 0.155 & 0.906 & 0.177 & 0 \\
\addlinespace[3pt]
$c=0.00$ & 5,000 & Heteroskedastic, two-step & $-$0.008 & 0.058 & 0.054 & 0.906 & 0.059 & 0 \\
 &  & Heteroskedastic, bootstrap & $-$0.008 & 0.058 & 0.061 & 0.928 & 0.059 & 0 \\
 &  & Constant variance, two-step & $-$0.007 & 0.062 & 0.058 & 0.922 & 0.062 & 0 \\
 &  & Constant variance, bootstrap & $-$0.007 & 0.062 & 0.060 & 0.924 & 0.062 & 0 \\
\addlinespace[3pt]
\multicolumn{9}{l}{\textit{B. Natural direct effect}}\\\addlinespace[2pt]
$c=0.40$ & 800 & Heteroskedastic, two-step & 0.084 & 0.255 & 0.203 & 0.802 & 0.269 & 0 \\
 &  & Heteroskedastic, bootstrap & 0.084 & 0.255 & 0.237 & 0.862 & 0.269 & 0 \\
\addlinespace[3pt]
$c=0.40$ & 5,000 & Heteroskedastic, two-step & 0.024 & 0.215 & 0.110 & 0.794 & 0.216 & 0 \\
 &  & Heteroskedastic, bootstrap & 0.024 & 0.215 & 0.120 & 0.816 & 0.216 & 0 \\
\addlinespace[3pt]
$c=0.20$ & 800 & Heteroskedastic, two-step & 0.055 & 0.202 & 0.181 & 0.872 & 0.209 & 0 \\
 &  & Heteroskedastic, bootstrap & 0.055 & 0.202 & 0.217 & 0.920 & 0.209 & 0 \\
\addlinespace[3pt]
$c=0.20$ & 5,000 & Heteroskedastic, two-step & 0.005 & 0.130 & 0.081 & 0.922 & 0.130 & 0 \\
 &  & Heteroskedastic, bootstrap & 0.005 & 0.130 & 0.090 & 0.936 & 0.130 & 0 \\
\addlinespace[3pt]
$c=0.10$ & 800 & Heteroskedastic, two-step & 0.048 & 0.177 & 0.169 & 0.900 & 0.183 & 0 \\
 &  & Heteroskedastic, bootstrap & 0.048 & 0.177 & 0.208 & 0.938 & 0.183 & 0 \\
\addlinespace[3pt]
$c=0.10$ & 5,000 & Heteroskedastic, two-step & 0.011 & 0.072 & 0.068 & 0.926 & 0.073 & 0 \\
 &  & Heteroskedastic, bootstrap & 0.011 & 0.072 & 0.075 & 0.932 & 0.073 & 0 \\
\addlinespace[3pt]
$c=0.00$ & 800 & Heteroskedastic, two-step & 0.043 & 0.194 & 0.171 & 0.906 & 0.198 & 0 \\
 &  & Heteroskedastic, bootstrap & 0.043 & 0.194 & 0.206 & 0.934 & 0.198 & 0 \\
 &  & Constant variance, two-step & 0.051 & 0.188 & 0.157 & 0.892 & 0.195 & 0 \\
 &  & Constant variance, bootstrap & 0.051 & 0.188 & 0.174 & 0.910 & 0.195 & 0 \\
\addlinespace[3pt]
$c=0.00$ & 5,000 & Heteroskedastic, two-step & 0.007 & 0.070 & 0.064 & 0.914 & 0.070 & 0 \\
 &  & Heteroskedastic, bootstrap & 0.007 & 0.070 & 0.070 & 0.916 & 0.070 & 0 \\
 &  & Constant variance, two-step & 0.006 & 0.073 & 0.068 & 0.922 & 0.074 & 0 \\
 &  & Constant variance, bootstrap & 0.006 & 0.073 & 0.069 & 0.922 & 0.074 & 0 \\
\addlinespace[3pt]
\end{longtable}\end{footnotesize}
\vspace{-6pt}\begin{footnotesize}\noindent Bias, mean estimate minus the true value; EmpSE, empirical standard error; ModSE, mean model standard error; Cov, coverage of nominal 95\% intervals; RMSE, root mean squared error; Fail, replications without an estimate. True NIE = 0.70, NDE = 0.30. Maximum Monte Carlo standard errors: bias 0.011; EmpSE 0.008; ModSE 0.007; coverage 0.019; RMSE 0.037.\end{footnotesize}

\clearpage
\begin{footnotesize}\setlength{\tabcolsep}{3pt}
\begin{longtable}{lllrrrrrrr}
\caption{Variance estimation: two-step and bootstrap inference across weighting methods (200 bootstrap resamples), $N = 800$; 500 replications per cell.}\label{tab:study6}\\
\toprule
Design & $N$ & Method & Bias & EmpSE & ModSE & Cov & RMSE & Failed res. & Fail \\
\midrule
\endfirsthead
\multicolumn{10}{l}{\textit{(continued)}}\\
\toprule
Design & $N$ & Method & Bias & EmpSE & ModSE & Cov & RMSE & Failed res. & Fail \\
\midrule
\endhead
\bottomrule
\endfoot
\multicolumn{10}{l}{\textit{A. Natural indirect effect}}\\\addlinespace[2pt]
Binary $M$ & 800 & IPTW, two-step & $-$0.009 & 0.091 & 0.082 & 0.940 & 0.091 & 0.00 & 0 \\
 &  & IPTW, bootstrap & $-$0.009 & 0.091 & 0.090 & 0.958 & 0.091 & 0.00 & 0 \\
 &  & MMWS 5, bootstrap & $-$0.007 & 0.086 & 0.088 & 0.966 & 0.086 & 0.00 & 0 \\
 &  & MMWS 10, bootstrap & $-$0.008 & 0.091 & 0.091 & 0.960 & 0.091 & 0.00 & 0 \\
 &  & Nonparametric, bootstrap & 0.026 & 0.079 & 0.088 & 0.970 & 0.083 & 0.00 & 0 \\
 &  & Odds, bootstrap & $-$0.004 & 0.056 & 0.056 & 0.944 & 0.056 & 0.00 & 0 \\
\addlinespace[3pt]
Three-level $T$, 1 vs.\ 0 & 800 & IPTW, two-step & $-$0.004 & 0.105 & 0.091 & 0.920 & 0.105 & 0.00 & 0 \\
 &  & IPTW, bootstrap & $-$0.004 & 0.105 & 0.104 & 0.952 & 0.105 & 0.04 & 0 \\
 &  & MMWS 5, bootstrap & $-$0.008 & 0.099 & 0.103 & 0.960 & 0.099 & 0.01 & 0 \\
 &  & MMWS 10, bootstrap & $-$0.010 & 0.104 & 0.106 & 0.954 & 0.104 & 0.02 & 0 \\
 &  & Nonparametric, bootstrap & 0.011 & 0.094 & 0.111 & 0.982 & 0.094 & 0.00 & 0 \\
 &  & Odds, bootstrap & $-$0.041 & 0.053 & 0.052 & 0.810 & 0.067 & 0.01 & 0 \\
\addlinespace[3pt]
Three-level $T$, 2 vs.\ 0 & 800 & IPTW, two-step & $-$0.006 & 0.111 & 0.103 & 0.954 & 0.111 & 0.00 & 0 \\
 &  & IPTW, bootstrap & $-$0.006 & 0.111 & 0.121 & 0.974 & 0.111 & 0.04 & 0 \\
 &  & MMWS 5, bootstrap & 0.000 & 0.106 & 0.123 & 0.980 & 0.106 & 0.01 & 0 \\
 &  & MMWS 10, bootstrap & $-$0.006 & 0.112 & 0.125 & 0.974 & 0.112 & 0.02 & 0 \\
 &  & Nonparametric, bootstrap & 0.046 & 0.110 & 0.131 & 0.982 & 0.119 & 0.00 & 0 \\
 &  & Odds, bootstrap & $-$0.011 & 0.092 & 0.093 & 0.952 & 0.092 & 0.01 & 0 \\
\addlinespace[3pt]
\multicolumn{10}{l}{\textit{B. Natural direct effect}}\\\addlinespace[2pt]
Binary $M$ & 800 & IPTW, two-step & 0.004 & 0.131 & 0.119 & 0.928 & 0.131 & 0.00 & 0 \\
 &  & IPTW, bootstrap & 0.004 & 0.131 & 0.126 & 0.948 & 0.131 & 0.00 & 0 \\
 &  & MMWS 5, bootstrap & 0.052 & 0.122 & 0.121 & 0.936 & 0.133 & 0.00 & 0 \\
 &  & MMWS 10, bootstrap & 0.020 & 0.125 & 0.127 & 0.942 & 0.127 & 0.00 & 0 \\
 &  & Nonparametric, bootstrap & $-$0.031 & 0.126 & 0.125 & 0.946 & 0.130 & 0.00 & 0 \\
 &  & Odds, bootstrap & $-$0.002 & 0.114 & 0.108 & 0.928 & 0.114 & 0.00 & 0 \\
\addlinespace[3pt]
Three-level $T$, 1 vs.\ 0 & 800 & IPTW, two-step & 0.011 & 0.168 & 0.147 & 0.934 & 0.168 & 0.00 & 0 \\
 &  & IPTW, bootstrap & 0.011 & 0.168 & 0.160 & 0.954 & 0.168 & 0.04 & 0 \\
 &  & MMWS 5, bootstrap & 0.051 & 0.154 & 0.157 & 0.950 & 0.162 & 0.01 & 0 \\
 &  & MMWS 10, bootstrap & 0.025 & 0.165 & 0.165 & 0.956 & 0.166 & 0.02 & 0 \\
 &  & Nonparametric, bootstrap & 0.034 & 0.189 & 0.211 & 0.964 & 0.192 & 0.00 & 0 \\
 &  & Odds, bootstrap & 0.048 & 0.146 & 0.145 & 0.928 & 0.154 & 0.01 & 0 \\
\addlinespace[3pt]
Three-level $T$, 2 vs.\ 0 & 800 & IPTW, two-step & $-$0.001 & 0.174 & 0.156 & 0.930 & 0.174 & 0.00 & 0 \\
 &  & IPTW, bootstrap & $-$0.001 & 0.174 & 0.172 & 0.952 & 0.174 & 0.04 & 0 \\
 &  & MMWS 5, bootstrap & 0.035 & 0.157 & 0.169 & 0.964 & 0.161 & 0.01 & 0 \\
 &  & MMWS 10, bootstrap & 0.010 & 0.170 & 0.177 & 0.958 & 0.170 & 0.02 & 0 \\
 &  & Nonparametric, bootstrap & $-$0.049 & 0.177 & 0.206 & 0.974 & 0.184 & 0.00 & 0 \\
 &  & Odds, bootstrap & 0.004 & 0.181 & 0.164 & 0.934 & 0.181 & 0.01 & 0 \\
\addlinespace[3pt]
\end{longtable}\end{footnotesize}
\vspace{-6pt}\begin{footnotesize}\noindent Bias, mean estimate minus the true value; EmpSE, empirical standard error; ModSE, mean model standard error; Cov, coverage of nominal 95\% intervals; RMSE, root mean squared error; Fail, replications without an estimate. Failed res., mean number of failed bootstrap resamples per replication. Designs as in Table~\ref{tab:study3a}. Maximum Monte Carlo standard errors: bias 0.008; EmpSE 0.006; ModSE 0.002; coverage 0.018; RMSE 0.007.\end{footnotesize}

\clearpage
\begin{footnotesize}\setlength{\tabcolsep}{4pt}
\begin{longtable}{lllrrrrrr}
\caption{Continuous treatment: GPS weighting under alternative bandwidths, with bootstrap standard errors (200 resamples); 1,000 replications per cell.}\label{tab:study5}\\
\toprule
Design & $N$ & Method & Bias & EmpSE & ModSE & Cov & RMSE & Fail \\
\midrule
\endfirsthead
\multicolumn{9}{l}{\textit{(continued)}}\\
\toprule
Design & $N$ & Method & Bias & EmpSE & ModSE & Cov & RMSE & Fail \\
\midrule
\endhead
\bottomrule
\endfoot
\multicolumn{9}{l}{\textit{A. Natural indirect effect}}\\\addlinespace[2pt]
dose 2 vs.\ 1 & 800 & Default bandwidth & 0.005 & 0.128 & 0.118 & 0.926 & 0.128 & 0 \\
 &  & Half bandwidth & $-$0.040 & 0.147 & 0.138 & 0.886 & 0.153 & 0 \\
\addlinespace[3pt]
dose 2 vs.\ 1 & 5,000 & Default bandwidth & 0.005 & 0.079 & 0.073 & 0.934 & 0.079 & 0 \\
 &  & Half bandwidth & $-$0.013 & 0.098 & 0.092 & 0.893 & 0.099 & 0 \\
\addlinespace[3pt]
dose 0 vs.\ 1 & 800 & Default bandwidth & $-$0.021 & 0.098 & 0.094 & 0.950 & 0.101 & 0 \\
 &  & Half bandwidth & 0.027 & 0.113 & 0.110 & 0.912 & 0.116 & 0 \\
\addlinespace[3pt]
dose 0 vs.\ 1 & 5,000 & Default bandwidth & $-$0.008 & 0.066 & 0.058 & 0.924 & 0.066 & 0 \\
 &  & Half bandwidth & 0.008 & 0.078 & 0.072 & 0.908 & 0.078 & 0 \\
\addlinespace[3pt]
\multicolumn{9}{l}{\textit{B. Natural direct effect}}\\\addlinespace[2pt]
dose 2 vs.\ 1 & 800 & Default bandwidth & $-$0.048 & 0.260 & 0.259 & 0.934 & 0.265 & 0 \\
 &  & Half bandwidth & 0.150 & 0.325 & 0.330 & 0.906 & 0.358 & 0 \\
\addlinespace[3pt]
dose 2 vs.\ 1 & 5,000 & Default bandwidth & $-$0.042 & 0.155 & 0.153 & 0.953 & 0.160 & 0 \\
 &  & Half bandwidth & 0.027 & 0.200 & 0.199 & 0.934 & 0.201 & 0 \\
\addlinespace[3pt]
dose 0 vs.\ 1 & 800 & Default bandwidth & 0.039 & 0.233 & 0.225 & 0.941 & 0.236 & 0 \\
 &  & Half bandwidth & $-$0.124 & 0.286 & 0.288 & 0.914 & 0.312 & 0 \\
\addlinespace[3pt]
dose 0 vs.\ 1 & 5,000 & Default bandwidth & 0.028 & 0.133 & 0.130 & 0.951 & 0.136 & 0 \\
 &  & Half bandwidth & $-$0.025 & 0.171 & 0.170 & 0.928 & 0.173 & 0 \\
\addlinespace[3pt]
\end{longtable}\end{footnotesize}
\vspace{-6pt}\begin{footnotesize}\noindent Bias, mean estimate minus the true value; EmpSE, empirical standard error; ModSE, mean model standard error; Cov, coverage of nominal 95\% intervals; RMSE, root mean squared error; Fail, replications without an estimate. True values: 2 vs.\ 1, NIE = 0.54, NDE = 0.52; 0 vs.\ 1, NIE = $-0.30$, NDE = $-0.52$. Maximum Monte Carlo standard errors: bias 0.010; EmpSE 0.007; ModSE 0.002; coverage 0.010; RMSE 0.007.\end{footnotesize}

\end{document}